%% file: main.tex
\documentclass[a4paper,11pt]{article}
\pdfoutput=1 

\usepackage{jheppub} 

\usepackage{amsmath}%
\usepackage{amssymb}%
\usepackage{bm}%
\usepackage{caption}%
\usepackage{cleveref}%
\usepackage{makecell}%
\usepackage{mathrsfs}%
\usepackage{MnSymbol}%
\usepackage{mathtools}%
\usepackage{multirow}%
\usepackage{physics}%
\usepackage{amsthm}%
\usepackage{simpler-wick}%
\usepackage[subrefformat=parens]{subcaption}%
\usepackage{subfiles}%
\usepackage{tabularx}%
\usepackage{xspace}%

\graphicspath{{figures/}}%
\newcommand{\apriori}{\textit{a priori}}%
\newcommand{\cf}{\textit{cf.\@\xspace}}%
\newcommand{\eg}{\textit{e.g.\@\xspace}}%
\newcommand{\ie}{\textit{i.e.\@\xspace}}%
\DeclareMathOperator{\diag}{diag}%
\DeclareMathOperator{\dom}{dom}%
\DeclareMathOperator{\He}{He}%
\DeclareMathOperator{\Image}{im}%
\DeclareMathOperator{\Span}{span}%
\DeclareMathOperator{\Sym}{Sym}%
\DeclareMathSymbol{\varkappa}{\mathord}{AMSb}{"7B}%

\theoremstyle{remark}%

\title{\boldmath Group-averaged baby universe field theory}

\author{Hong Zhe (Vincent) Chen}
\affiliation{Department of Physics, University of California,\\
Santa Barbara, CA 93106, U.S.A.}

\emailAdd{hzchen@ucsb.edu}

\abstract{Gravitational sums over spacetime topology suggest a baby-universe
  field theory whose quanta are entire universes. We develop such a description
  from group-averaged amplitudes that impose gravitational constraints and
  define physical inner products in the single-universe theory. To sharpen the
  contrast with ordinary QFT, we start with gravity on one-dimensional
  worldlines, which can arise from reducing higher-dimensional cylinders. For
  multiple noninteracting worldlines that pair boundary components, the
  baby-universe Hilbert space is the symmetric Fock space over the physical
  one-universe Hilbert space, and the universe field expands in the on-shell
  modes spanning the latter. Each mode multiplies the sum of creation and
  annihilation operators, so the mode coefficients commute. The mode expansion
  thus makes explicit the field's commutativity and its Wheeler–DeWitt linear
  field equation. Diagonalizing the coefficients recovers a statistical
  description by a Gaussian path integral over on-shell field configurations
  labelling superselection (\(\alpha\)-)sectors.

  We take exploratory steps toward more general topologies. Disk-like topologies
  coherently shift the no-boundary state, turning on the Hartle--Hawking
  wavefunction as a one-point function. Even with nontrivial topologies
  represented by interacting Feynman diagrams, we argue that group averaging
  near each boundary still enforces the same linear field equation, while
  interactions instead enter through a non-Gaussian measure over
  \(\alpha\)-sectors. In the topological Marolf--Maxfield model, we promote a
  disk-and-cylinder truncation to an exact Fock-space representation of the full
  theory incorporating all topologies. We also relate the truncated and exact
  inner products by an insertion of interactions, whose perturbative expansion
  produces a sum over time evolutions generated by pairs of pants, evoking an
  analogy to group averaging.}

\begin{document}
\maketitle
\flushbottom

\subfile{sections/section01_introduction}

\subfile{sections/section02_worldline}

\subfile{sections/section03_universeFieldTheory}

\subfile{sections/section04_discussion}

\appendix

\subfile{sections/appendix01_minisuperspace}

\subfile{sections/appendix02_qftwightman}

\subfile{sections/appendix03_alphapathint}

\subfile{sections/appendix04_gaugingstar}

\subfile{sections/appendix05_generalizedFock}

\subfile{sections/appendix06_diskCylinder}

\acknowledgments


The author is especially grateful to Donald Marolf for guidance on this project.
The author also thanks Xi Dong, Steven Giddings, Maciej Kolanowski, Henry
Maxfield, and Victor Rodriguez for interesting discussions. The author is
supported by a DeBenedictis Postdoctoral Fellowship from the University of
California, Santa Barbara. ChatGPT (OpenAI) was used to assist in preparing and
editing all figures; obtaining acceptable final results required extensive and
repeated author guidance, correction, and revision.










\bibliographystyle{JHEP.bst}
\bibliography{references.bib}

\end{document}

%% file: sections/section01_introduction.tex
\section{Introduction}
\label{sec:intro}

Some of the most striking recent successes of gravitational path integrals hinge
on contributions from nontrivial spacetime topology, from replica-wormhole
calculations of fine-grained entropy
\cite{Penington:2019npb,Almheiri:2019psf,Penington:2019kki,Almheiri:2019qdq}
to the relation between low-dimensional gravity and random-matrix statistics \cite{Saad:2019lba}. These developments have renewed
interest in an older idea: that a gravitational sum over topologies might admit
a field-theoretic description in which entire universes are the quanta
\cite{Coleman:1988cy,Giddings:1988cx,Giddings:1988wv}.

There is a suggestive analogy with quantum field theory. Upon dimensional
reduction in space, gravity on higher-dimensional cylinder-like topologies,
\((\text{time})\times(\text{spatial topology})\), are described by gravitational
path integrals on worldlines, while more general spacetime topologies can give
rise schematically to Feynman-like diagrams built from such worldlines. A sum
over spacetime topologies in the gravitational path integral can therefore
resemble a Feynman-diagram expansion. This resemblance, however, does not by
itself determine the resulting field theory.

Two questions are especially important for making this analogy precise: (i) how
is the gravitational path integral for a single worldline defined, and hence
which single-worldline amplitude does it produce; and (ii) which two-point
correlation function in the corresponding many-worldline theory is identified
with that amplitude? For example, different choices of contour in the
gravitational path integral on a worldline can lead to different single-worldline
amplitudes, while in the many-worldline theory one must decide whether the
chosen amplitude is to reproduce a time-ordered correlator or a Wightman
function with operators ordered as written. Casali, Marolf, Maxfield, and
Rangamani
\cite{Casali:2021ewu} analyze a broader set of choices involved in passing from
gravity on a worldline to theories describing multiple worldlines or universes.
Here we use the two questions above as a convenient organizing framework for the
comparison between ordinary QFT and the baby-universe theory developed below.

To make these choices concrete, consider a single gravitational worldline with
target space \(\mathscr{M}\). From the perspective of a dimensional reduction of
a higher-dimensional gravitational theory, \(\mathscr{M}\) is the relevant
minisuperspace: a point \(x\in\mathscr{M}\) specifies the induced field data
retained on a spatial slice labelled by the universe time coordinate \(t\), and
a trajectory \(x(t)\) describes how these data vary from slice to slice. In
Lorentzian signature, the worldline metric may be written
\(\dd{s}^2=-N^2(t)\dd{t}^2\). The lapse \(N\) enters the worldline Lagrangian as
a Lagrange multiplier for the Wheeler--DeWitt Hamiltonian \(H\), which generates
evolution along the worldline. Since this evolution is a gauge redundancy in
gravity, physical states satisfy the Wheeler--DeWitt constraint \(H=0\). After
gauge fixing the local time-reparametrization freedom, there remains the
invariant total lapse
\begin{align}
  T
  &= \int \dd{t}\, N(t)
    \;.
\end{align}
At fixed \(T\), the path integral on a worldline interval whose endpoints are
fixed at \(x,x'\in\mathscr{M}\) computes the kernel
\begin{align}
  \mel{x'}{e^{-iT H}}{x}_{\mathrm{wl}}
  \;.
\end{align}
Since the worldline metric is itself integrated over in the gravitational path
integral, one must subsequently integrate this kernel over \(T\). The choice of
contour for this remaining integral is therefore one of the ingredients
specifying the single-worldline amplitude.

The standard worldline construction of QFT provides a familiar point of
comparison. Taking \(T\) over the positive half-line, with the usual
\(i\epsilon\) prescription, gives
\begin{align}
  \int_0^\infty \dd{T}\, e^{-iT(H-i\epsilon)}
  &= \frac{1}{iH+\epsilon}
    \;.
\end{align}
In many minisuperspace models of interest, \(H\) is a Klein--Gordon-like
differential operator on \(\mathscr{M}\), so the right-hand side is a Green's
function and, when the requisite analyticity conditions hold, can be identified
with the Feynman two-point function of an ordinary QFT. In this familiar
construction, a particular choice of worldline contour therefore produces a
Green's function, which is used as a time-ordered two-point function in the
corresponding field theory.

We instead integrate the residual lapse over the full real line,
\begin{align}
  \eta
  &= \int_{-\infty}^{\infty} \dd{T}\, e^{-iT H}
    = 2\pi\delta(H)
    \;.
  \label{eq:introgroupproj}
\end{align}
This is the group average over the evolution generated by \(H\), which is gauged
in the gravitational theory. The expression \(2\pi\delta(H)\) makes manifest
that \(\eta\) maps onto solutions of the Wheeler--DeWitt constraint, while its
group-average form makes gauge invariance manifest. Its matrix elements
\begin{align}
  \mel{x'}{\eta}{x}_{\mathrm{wl}}
\end{align}
define the group-averaged physical inner product and hence the physical
one-universe Hilbert space \cite{Marolf:1994ae,Casali:2021ewu,Held:2025mai}. The
full-line amplitude is therefore natural for the purpose of describing
gravitational universes subject to physical gravitational constraints.

The remaining choice concerns how this one-universe amplitude enters the
many-universe theory. In the modern baby-universe framework, multi-boundary
gravitational amplitudes are interpreted as inner products between states
prepared by boundary conditions; through the GNS construction, adding a boundary
is represented by acting with a corresponding boundary-inserting operator on the
baby-universe Hilbert space \cite{Marolf:2020xie}. We denote the no-boundary
state by \(\ket{\varnothing}_{\mathrm{BU}}\) and by \(\Phi(x)\) the operator
associated with a boundary labelled by \(x\in\mathscr{M}\). The two-boundary
gravitational amplitude is then identified with\footnote{For oriented theories where
  the antilinear orientation reversal \(\star\) acts nontrivially on induced
  field configurations \(x\) --- see \cref{sec:orientrev} --- one should
  actually identify \(\mel{x'}{\eta}{x}_{\mathrm{wl}}\) with
  \(\mel{\varnothing}{\Phi(x')\Phi(x^\star)}{\varnothing}_{\mathrm{BU}}\), up to
  normalization.}
\begin{align}
  \mel{\varnothing}{\Phi(x')\Phi(x)}{\varnothing}_{\mathrm{BU}}
  \;,
\end{align}
with the operator product left in the order written rather than time ordered.
Thus the gravitational amplitudes are identified with Wightman correlators
rather than time-ordered correlators.

We refer to the resulting construction as \textbf{group-averaged baby universe
  field theory (GABUFT)}. \Cref{tab:qftvsuft} summarizes the immediate contrast
between the standard worldline construction of QFT (Approach 1) and GABUFT. The
group-averaged construction underlying GABUFT is among the possibilities
considered in \cite{Casali:2021ewu}. Here we take this branch as our starting
point and further develop its universe-field-theoretic realization and
consequences, before asking how its structure might extend beyond the free
worldline topology sum.

\begin{table}
  \centering
  \begin{tabularx}{\linewidth}{|X||c|c|c|}
    \hline
    & \multicolumn{2}{|c|}{QFT}
    & \multirow{2}{*}{GABUFT}
    \\
    & \multicolumn{1}{|c}{\scriptsize Approach 1}
    & \multicolumn{1}{c|}{\scriptsize Approach 2}
    &
    \\
    \hline
    \hline
    \makecell[l]{Worldline\\amplitude}
    & \(\mel{x'}{\frac{1}{iH+\epsilon}}{x}\)
    & \(\mel{x'}{P_+\eta}{x}\)
    & \(\mel{x'}{\eta}{x}\)
    \\
    \hline
    \makecell[l]{FT 2-pt.\\function}
    & \(\mel{0}{\mathrm{T}\Phi
      (x')\Phi
      (x)}{0}\)
    & \(\mel{0}{\Phi
      (x')\Phi
      (x)}{0}\)
    & \(\mel{\varnothing}{\Phi(x')\Phi(x)}{\varnothing}\)
    \\
    \hline
    \makecell[l]{Mode\\expansion}
    & \multicolumn{2}{|c|}{\(\Phi(x) = \int_{\mathscr{K}_+} \kappa(k) (a_k u_k(x) + a_k^\dagger u_k(x)^*)\)}
    & \(\Phi(x) = \int_{\mathscr{K}} \kappa(k) (a_k + a_{k^\star}^\dagger) u_k(x)\)
    \\
    \hline
    \makecell[l]{FT Hilbert\\space}
    & \multicolumn{2}{|c|}{\(\overline{\bigoplus_{i=0}^\infty \Sym^i(L^2(\mathscr{K}_+,\kappa))}\)}
    & \(\aleph^{1/2}\overline{\bigoplus_{i=0}^\infty \Sym^i(L^2(\mathscr{K},\kappa))}\)
      \\
      \hline
  \end{tabularx}
  \caption{Comparison of QFT and group-averaged baby universe field theory
    (GABUFT). (For ease of comparison in this table, expressions here are
    written for \emph{unoriented} worldlines --- see \cref{sec:reversal} --- and
    \emph{real} scalar fields.)}
  \label{tab:qftvsuft}
\end{table}

We start by considering free GABUFTs based on worldline pairings. In this class
of theories, the gravitational path integral sums over arbitrary collections of
disjoint smooth worldlines, with no processes that join or split them. In one
dimension, each connected smooth spacetime with nonempty boundary is an
interval. Thus, apart from boundary-independent closed
components,\footnote{These will be absorbed into the normalization
  \(\aleph\equiv\braket{\varnothing}{\varnothing}_{\mathrm{BU}}\) of the
  no-boundary state \(\ket{\varnothing}_{\mathrm{BU}}\).} for an even number of
specified boundaries the gravitational path integral sums over all ways of
pairing those boundaries by worldline intervals, with each interval contributing
the group-averaged one-universe inner product. This is precisely the
combinatorics of Wick's theorem. The boundary-inserting operators can therefore
be represented in terms of universe creation and annihilation operators, with
each Wick contraction corresponding geometrically to the worldline connecting
the contracted pair of boundaries.

This construction also makes precise the sense in which the boundary-inserting
operator \(\Phi(x)\) is a field on \(\mathscr{M}\). The group-averaged two-point
function admits a spectral decomposition
\begin{align}
  \mel{x'}{\eta}{x}_{\mathrm{wl}}
  &= \int_{\mathscr{K}} \kappa(k)\,
    u_k(x') u_k(x)^*
    \;,
\end{align}
where \(k\in\mathscr{K}\) labels an orthogonal basis of physical one-universe
modes and \(\kappa\) is the corresponding measure. To reproduce the gravitational
path integral through Wick contractions, the universe field may then be expanded
in creation and annihilation operators associated with these same modes,
\begin{align}
  \Phi(x)
  &= \int_{\mathscr{K}} \kappa(k)\,
    \bigl(a_k+a_{k^\star}^\dagger\bigr)u_k(x)
    \equiv
    \int_{\mathscr{K}} \kappa(k)\,\phi(k)u_k(x)
    \;.
\end{align}
The involution \(k\mapsto k^\star\), inherited from the operation relating
boundary conditions that prepare bras and kets \cite{Witten:2025ayw}, is
realized as antilinear time reversal in unoriented theories and as antilinear
orientation reversal when orientation is retained as a degree of freedom. Since
the modes appearing in the spectral decomposition satisfy the Wheeler--DeWitt
constraint \(H u_k=0\), the mode expansion immediately implies
\begin{align}
  H\Phi(x)
  &=0
    \;.
\end{align}

The difference from ordinary QFT is especially transparent in the mode
expansion. When the solution space admits an appropriate positive-frequency
decomposition, a real QFT field can be expressed as
\begin{align}
  \Phi_{\mathrm{QFT}}(x)
  &= \int_{\mathscr{K}_+} \kappa(k)
    \left[
      a_k u_k(x)
      +a_k^\dagger u_k(x)^*
    \right]
    \;,
\end{align}
where annihilation operators multiply only positive-frequency modes \(u_k\),
\(k\in\mathscr{K}_+\), while creation operators multiply their
negative-frequency complex conjugates. GABUFT requires no such split. Its modes
are fixed instead by the spectral decomposition of the group average, and each
mode \(u_k\), \(k\in\mathscr{K}\), appears with the combined coefficient
\begin{align}
  \phi(k)
  &= a_k+a_{k^\star}^\dagger
    \;.
\end{align}
These coefficients mutually commute, and hence
\begin{align}
  [\Phi(x),\Phi(x')]
  &=0
    \qquad
    \text{for all }x,x'\in\mathscr{M}
    \;.
\end{align}
The commutativity expected abstractly from the baby-universe boundary algebra is
therefore visible directly in the creation--annihilation structure of the field.
By contrast, ordinary QFT separates creation and annihilation through a choice
of positive/negative-frequency decomposition, and its fields need not commute at
timelike separation.

The mode expansion also makes the lower entries of \cref{tab:qftvsuft}
transparent. In the standard worldline construction of QFT, the half-line lapse
contour produces a Green's function identified with the Feynman two-point
function, and the resulting free QFT is a Fock space over its chosen
positive-frequency one-particle Hilbert space. A second route to ordinary QFT
starts with the group average but then projects onto the chosen
positive-frequency part,
\begin{align}
  \mel{x'}{P_+\eta}{x}_{\mathrm{wl}}
  \;,
\end{align}
before constructing the many-particle theory. This gives the QFT Wightman
function and again leads to a Fock space built only from the chosen
positive-frequency modes. From the perspective of the gravitational theory,
however, \(P_+\) is additional structure that need not be naturally available.
Free GABUFT instead retains the full physical one-universe Hilbert space
\(L^2(\mathscr{K},\kappa)\) and forms the corresponding symmetric Fock space, up
to the overall normalization \(\aleph\) of the no-boundary state
\(\ket{\varnothing}_{\mathrm{BU}}\).

The commutativity of the GABUFT field \(\Phi(x)\) also permits a very different
description of the same baby-universe Hilbert space. By definition, an
\(\alpha\)-sector is a simultaneous eigenspace of the commuting algebra of
boundary-inserting operators. In the modern baby-universe framework these
eigenspaces are one-dimensional superselection sectors \cite{Marolf:2020xie}. In
free GABUFT, all boundary insertions are generated by the commuting mode
operators \(\phi(k)=a_k+a_{k^\star}^\dagger\), so an \(\alpha\)-sector can
equivalently be specified by their simultaneous eigenvalues,
\begin{align}
  \phi(k)\ket{\alpha}_{\mathrm{BU}}
  &= \phi_\alpha(k)\ket{\alpha}_{\mathrm{BU}}
    \;.
\end{align}
Let \(\mathcal{A}\) denote the joint spectrum of the boundary-inserting algebra;
in the present free theory, a point \(\alpha\in\mathcal{A}\) is therefore
parametrized by the complete set of eigenvalues \(\{\phi_\alpha(k)\}\). Through
the universe-field mode expansion, the same data determine a definite field
configuration
\begin{align}
  \Phi_\alpha(x)
  &= \int_{\mathscr{K}} \kappa(k)\,
    \phi_\alpha(k)u_k(x)
    \;,
  &
    H\Phi_\alpha(x)
  &=0
    \;.
\end{align}
Thus the same \(\alpha\)-sector may equivalently be labelled by an on-shell
configuration of the universe field.

Because correlation functions in the no-boundary state
\(\ket{\varnothing}_{\mathrm{BU}}\) of free GABUFT are generated entirely by Wick
contractions, the joint moments of the commuting variables \(\phi(k)\) are
Gaussian. In the simultaneous eigenbasis, \(\ket{\varnothing}_{\mathrm{BU}}\)
therefore induces a Gaussian measure
\begin{align}
  \mathcal{D}\phi_\alpha\,e^{-I[\phi_\alpha]}
\end{align}
on \(\mathcal{A}\). The baby-universe Hilbert space can consequently be
represented as
\begin{align}
  \mathcal{H}_{\mathrm{BU}}
  &=
  L^2\!\left(
    \mathcal{A},
    \mathcal{D}\phi_\alpha\,e^{-I[\phi_\alpha]}
  \right)
  \;,
\end{align}
and its correlation functions can be written as path integrals over the
eigenvalue configurations \(\phi_\alpha(k)\), or equivalently over the associated
on-shell fields \(\Phi_\alpha(x)\).

Despite its formal resemblance to an ordinary QFT path integral, this path
integral has a different Hilbert-space interpretation. As emphasized by
Casali--Marolf--Maxfield--Rangamani \cite{Casali:2021ewu}, it is naturally
interpreted as a classical statistical field theory: each field configuration
\(\Phi_\alpha(x)\), equivalently each set of eigenvalues \(\{\phi_\alpha(k)\}\),
specifies a microscopic state of the statistical theory --- an \(\alpha\)-state.
Distinct \(\alpha\)-states lie in superselection sectors, so a superposition of
them is, with respect to the boundary observables, indistinguishable from a
classical statistical ensemble over \(\alpha\). For the no-boundary state
\(\ket{\varnothing}_{\mathrm{BU}}\), this ensemble is weighted by
\(\mathcal{D}\phi_\alpha e^{-I[\phi_\alpha]}\). Moreover, the field
configurations \(\Phi_\alpha\) are on shell, satisfying \(H\Phi_\alpha=0\). In
an ordinary QFT path integral, by contrast, the field configurations being
integrated over are generally off shell and do not themselves label states or
superselection sectors of the QFT Hilbert space. QFT states are instead
represented by wavefunctionals of field data on suitable hypersurfaces in
\(\mathscr{M}\); in a path-integral representation, they can enter through
initial, final, or asymptotic conditions in \(\mathscr{M}\).

These structural differences become physically consequential already in simple
cosmological models and also shape how the construction might extend beyond the
free topology sum. Minisuperspace models of recollapsing universes provide a
simple example where the GABUFT and QFT approaches qualitatively disagree.
GABUFT does not require choosing a preferred time on \(\mathscr{M}\) or
splitting solutions into positive and negative frequencies. This becomes
important when the scale factor, although often associated with the timelike
direction of the DeWitt metric, fails to be a global clock because a classical
universe expands, turns around, and recollapses. In such models, evolving
positive- and negative-frequency branches independently can produce
exponentially growing solutions beyond the classical turning point
\cite{Casali:2021ewu}. The spectral modes selected by the group average instead
behave much like energy eigenfunctions for a quantum-mechanical particle
reflected from a potential barrier: they decay in the classically forbidden
region and combine incoming and reflected components with equal magnitude.
Semiclassically, the group-averaged inner product then tracks the complete
classical trajectory through expansion, turnaround, and recollapse.

We next ask which features of GABUFT should survive when more general topology
change is included. Two properties of the universe field appear robust: a
reality condition and its linear Wheeler--DeWitt equation. Positivity of the
gravitational path integral, already required for the baby-universe inner
product, implies
\begin{align}
  \Phi(x)^\dagger
  &= \Phi(x^\star)
    \;,
\end{align}
so this reality condition need not be imposed independently. We also expect
\begin{align}
  H\Phi(x)
  &=0
    \;
\end{align}
to continue to hold. The basic reason is that, however complicated the topology
in the bulk interior, each boundary should still have a locally cylinder-like
neighborhood. Defining the lapse contour there as in the cylinder theory should
again group-average over the time evolution generated by the same
Wheeler--DeWitt Hamiltonian \(H\), thereby projecting the boundary data onto
physical one-universe modes. If this expectation is correct, topology-changing
interactions cannot enter through nonlinear terms in the target-space field
equation. The interacting theory should instead remain describable in terms of
\(\star\)-real on-shell configurations,
\begin{align}
  \Phi_\alpha(x)
  &= \int_{\mathscr{K}} \kappa(k)\,
    \phi_\alpha(k)u_k(x)
    \;,
\end{align}
with topology change encoded by replacing the Gaussian weight in the path
integral over \(\alpha\)-sectors by a generally non-Gaussian one.

A simple extension of the Gaussian construction based on worldline pairings is
to include disk-like topologies in which universes can cap off. This produces a
nonzero one-point function
\begin{align}
  \Phi_{\varnothing}(x)
  &\equiv
  \frac{
    \mel{\varnothing}{\Phi(x)}{\varnothing}_{\mathrm{BU}}
  }{
    \braket{\varnothing}{\varnothing}_{\mathrm{BU}}
  }
  \;,
\end{align}
which we identify with the Hartle--Hawking wavefunction \cite{Hartle:1983ai}.
The arguments above imply that \(\Phi_{\varnothing}\) is on shell and obeys the
appropriate \(\star\)-reality condition. If only disk- and cylinder-like
connected topologies are retained, the measure on the space \(\mathcal{A}\) of
\(\alpha\)-sectors remains Gaussian but is shifted away from zero mean,
equivalently describing a coherent displacement of the no-boundary state
\(\ket{\varnothing}_{\mathrm{BU}}\). Under the same assumptions, the Vilenkin
tunneling wavefunction \cite{Vilenkin:1984wp,Vilenkin:1986cy} fails the relevant
reality condition and therefore cannot arise as such a GABUFT one-point
function. Moreover, the Vilenkin wavefunction also fails to satisfy boundary
conditions corresponding to any self-adjoint Wheeler-DeWitt Hamiltonian \(H\),
so we do not expect it to describe any correlation function
\(\mel{\varnothing}{\Phi(x)\cdots}{\varnothing}_{\mathrm{BU}}\).

The two-dimensional topological Marolf--Maxfield model \cite{Marolf:2020xie}
provides an exactly solvable setting in which we can demonstrate explicitly how
the disk-and-cylinder approximation is corrected to reproduce the complete
topology sum. Remarkably, the same oscillator Fock space used in the
disk-and-cylinder approximation can still be used for the exact theory: higher
topologies renormalize the coefficients already present in the approximation,
and only one additional term is required to obtain the exact boundary-inserting
operator,\footnote{In Marolf--Maxfield, this operator is just called \(Z\). In
  this paper, we will use \(Z\) to denote boundary-inserting operators obtained
  from smearing the universe field \(\Phi\) in target space \(\mathscr{M}\). For
  the Marolf--Maxfield model, target space can be taken to be comprised of a
  single element \(\mathscr{M}=\{S^1\}\), so the distinction is moot.}
\begin{align}
  \Phi(S^1)
  = \lambda + \sqrt{\lambda}\,(a+a^\dagger)+a^\dagger a
  \;,
\end{align}
where \(\lambda>0\) is the genus-summed value of the connected gravitational
path integral. We give each term in this expression a direct geometric
interpretation and show that repeated insertions of \(\Phi(S^1)\) generate all
spacetimes contributing to the full gravitational path integral. After a shift
of the oscillator, this representation is equivalent to the pure number-operator
representation already given by Marolf and Maxfield. The simplicity of this
exact construction relies strongly on special features of the model and is not
expected to persist generically.

The same model also lets us explore a logically distinct route to the full
theory, one intended to remain useful when no simple exact Fock-space
construction is available. Starting from the Gaussian disk-and-cylinder theory,
we construct an insertion \(\eta_{\mathrm{pants}}\) that converts its inner
product into that of the full topology-summing theory. From one viewpoint,
\(\eta_{\mathrm{pants}}\) supplies the interaction terms needed to promote the
Gaussian path integral over \(\alpha\)-sectors to the full theory. Repackaging
\(\eta_{\mathrm{pants}}\) as an exponential of interaction vertices shows that a
naive cubic pair-of-pants vertex is insufficient: tadpole and higher-order terms
in the topology-suppressing parameter \(e^{-S_0}\) are required so that the
combined Feynman diagrams count each two-geometry exactly once. In suggestive
analogy with \cref{eq:introgroupproj}, the expansion of
\(\eta_{\mathrm{pants}}\) in \(e^{-S_0}\) may also be viewed as summing over
topology-changing evolution with discrete time steps counted by the number of
pairs-of-pants. In the exact Marolf--Maxfield theory, the spectrum of the
boundary-inserting operator \(\Phi(S^1)\) consists of the nonnegative integers
\cite{Marolf:2020xie}, whereas its disk-and-cylinder counterpart has continuous
spectrum \(\mathbb{R}\). It is as though, evoking group averaging,
\(\eta_{\mathrm{pants}}\) projects the disk-and-cylinder theory onto
integer-valued \(\alpha\)-sectors. In this exactly solvable model, this gives a
concrete realization of the idea that a sum over topology-changing evolution can
impose additional null relations or constraints \cite{Marolf:2020xie}.

In richer gravitational theories, defining the topology-changing amplitudes
needed for such an interacting construction remains an important open problem.
We discuss a Lorentzian proposal in which the gravitational path integral
includes almost-Lorentzian geometries with controlled conical singularities
\cite{Louko:1995jw,Marolf:2022ybi}. Such prescriptions have successfully
reproduced several results conventionally obtained from Euclidean gravitational
path integrals, including partition functions
\cite{Marolf:2022ybi,Dittrich:2024awu}, R\'enyi entropies
\cite{Colin-Ellerin:2020mva,Colin-Ellerin:2021jev,Marolf:2021mbi,Held:2024qcl},
and spectral form factors \cite{Blommaert:2023vbz}, at least at leading,
classical order. Extending these results systematically even to one-loop order
can be subtle, and it would be valuable to understand more precisely the
relation between such Lorentzian prescriptions and conventional ``Euclidean''
gravitational path integrals beyond leading order.

A broader future problem is to find a useful field-theoretic organization of
general topology-changing amplitudes. In the Gaussian theory, the space
\(\mathcal{A}\) of \(\alpha\)-sectors is parametrized by coefficients
\(\phi_\alpha(k)\) of on-shell modes \(u_k(x)\), \(k\in\mathscr{K}\), but
interactions need not be simple or local in the label \(k\) for an arbitrary
choice of basis. It would be interesting to determine whether topology change
selects preferred variables in which the interactions simplify, perhaps even
truncating the hierarchy of vertices required to cover gravitational moduli
space correctly. JT gravity offers a concrete testing ground: its multi-boundary
amplitudes admit a remarkably local cubic Kodaira--Spencer QFT description
\cite{Post:2022dfi}, but it remains to understand how the Kodaira--Spencer
fields relate to closed JT-universe states and how the QFT itself fits into the
baby-universe/GABUFT framework, for example whether its path integral can be
interpreted as a path integral over \(\alpha\)-sectors. In richer theories, the
interacting path integral over \(\alpha\)-sectors may also develop divergences
and require renormalization. This suggests asking whether the Wilsonian logic
familiar from QFT has a gravitational counterpart in integrating out microscopic
wormholes. Such wormholes are known to renormalize the effective couplings
appearing in the gravitational path integral
\cite{Coleman:1988cy,Giddings:1988cx,Giddings:1988wv}; it would be interesting
to understand how this is related to renormalization of the action governing the
path integral over \(\alpha\)-sectors.

The remainder of the paper is organized as follows. In \cref{sec:worldline}, we
develop the single-universe gravitational worldline theory, construct its
group-averaged physical inner product and Hilbert space, and introduce the
\(\star\)-operation relating bra- and ket-boundary conditions. In
\cref{sec:uft}, we construct free GABUFT based on the sum over worldline
pairings. We first define the theory abstractly by applying the general
baby-universe GNS construction to the worldline topology sum and then realize it
explicitly in field-theoretic terms using Wick contractions, universe creation
and annihilation operators, the universe-field mode expansion, the Fock-space
Hilbert space, and the path integral over \(\alpha\)-sectors. In
\cref{sec:discussqftvsuft}, we compare QFT and GABUFT in recollapsing
minisuperspace models. In \cref{sec:interactions}, we ask which structural
properties of GABUFT should survive more general topology change, then study the
effects of allowing disk-like topologies and the resulting Hartle--Hawking
one-point function, and finally use the exactly solvable Marolf--Maxfield model
to demonstrate explicitly how the disk-and-cylinder approximation can be
promoted to an exact Fock-space representation of the full theory.
\Cref{sec:nextsteps} discusses future directions, including Lorentzian
prescriptions for gravitational path integrals that include topology change,
group averaging over topology-changing evolution, and perturbative interacting
GABUFT.

\Cref{app:minisuperspace} provides minisuperspace examples supporting several
discussions in the main text. \Cref{app:qftwightman} explains how QFT Wightman
functions can be represented by worldline amplitudes. \Cref{app:alphapathint}
gives a mathematically precise construction of the path integral over
\(\alpha\)-sectors for the free GABUFT developed in \cref{sec:uft} and proves
its identification with the GNS baby-universe Hilbert space.
\Cref{app:pathintgauging} reviews how the gravitational path integral gauges
certain linear\footnote{In upcoming work \cite{Chen:2026wip2}, we hope to return
  to the distinct question of whether antilinear transformations can be gauged in
  the same sense and the ramifications such gauging would have on the baby
  universe framework. To be clear, in the present paper, \(\star\) is
  \emph{not} gauged in the sense defined by
  \cref{eq:ggaugestate,eq:zetaggauge}.} transformations such as time
evolution and spatial diffeomorphisms, including reflections which are
disconnected from the identity, as well as fermion parity.
\Cref{app:generalizedfock} develops a Sch\"urmann Fock-space representation for
the class of baby-universe theories with Hermitian, conditionally positive
connected gravitational path integrals, which includes the Marolf--Maxfield
model but not, for example, JT gravity. \Cref{app:dckinematics} derives the
disk-and-cylinder Marolf--Maxfield theory from a more primitive kinematic
Hilbert space in which even the elementary processes underlying that truncated
theory are initially forbidden.

%% file: sections/section02_worldline.tex
\section{Gravitational worldline theory}
\label{sec:worldline}

In this section, we develop a gravitational worldline theory, contrasting it to
the more usual particle worldline theory which leads to QFT. The main difference
at the level of the worldline theory will lie in the amplitude computed by the
worldline path integral: in the gravitational case, the worldline path integral
computes a group-averaged inner product projecting onto solutions of the
gravitational constraint; in the particle case, the worldline path integral is
usually used to produce a Green's function for the worldline Hamiltonian. In
\cref{sec:uft}, we will also see how the universe field theory arising from the
gravitational worldline theory differs from QFT.

At an abstract level, what we refer to as a gravitational worldline theory is
defined by a target space \(\mathscr{M}\) with a measure \(\mu\), a kinematic
Hilbert space \(\mathcal{H}_{\mathrm{kin}}=L^2(\mathscr{M},\mu)\), and a Hamiltonian
constraint \(H\) that acts on \(\mathcal{H}_{\mathrm{kin}}\). We will assume
that there is a natural choice of boundary conditions, if needed, which renders
\(H\) self-adjoint. For the most part, we will assume that zero is in the
\emph{continuous} spectrum of \(H\).\footnote{This simplifying assumption rules
  out the possibility of a normalizable \(H\)-invariant state in
  \(\mathcal{H}_{\mathrm{kin}}\), which alleviates certain subtleties. See
  \cref{foot:invariantga} for further comments.}

In \cref{sec:motivation}, we first review why we might be interested in such a
system and what it has to do with gravity \cite{Casali:2021ewu,Held:2025mai}. In
\cref{sec:groupaverage}, we discuss the physical inner product and Hilbert space
of the theory. In \cref{sec:reversal}, we introduce an additional structure to
the gravitational worldline theory --- a \(\star\)-involution interpretable as
antilinear time or orientation reversal --- which will be needed when we
construct its baby universe theory in \cref{sec:uft}.

\subsection{Gravity on a worldline}
\label{sec:motivation}

The most straightforward interpretation of the gravitational worldline theory is
just as a gravitational theory on a one-dimensional spacetime (with just time
and no space) --- \ie{} a worldline. On each time slice of the worldline --- a
point --- \(\mathscr{M}\) is the classical configuration space of matter degrees
of freedom \(x\in\mathscr{M}\). In one dimension, gravity has no propagating
degrees of freedom and the only remnant of gravity is a Wheeler--DeWitt
Hamiltonian \(H\) which annihilates time-reparametrization invariant matter
states. We will describe in \cref{sec:groupaverage} how this constraint is
imposed quantum mechanically by the path integral of the lapse function \(N(t)\)
determining the one remaining metric component \(g_{tt}= -N^2\) on a
one-dimensional worldline.

A second viewpoint is that we are considering a spatial dimensional reduction of
a higher-dimensional gravitational theory on a cylinder-like topology,
\((\text{time interval})\times(\text{spatial topology})\), where \(H\)
is the Wheeler--DeWitt Hamiltonian generating evolution along the sole surviving
direction, namely the time direction.

For example, we may have in mind a minisuperspace model where we restrict to
spatially homogeneous spacetimes that are invariant under some spatial
homogeneity group. Practically speaking, we restrict to spacetime metrics of the
form\footnote{The goal here is to illustrate the quantization of the Hamiltonian
  constraint which is present both in (simplified form in) the minisuperspace
  model and in the full theory. To make the analysis as clean as possible, we
  would like to do away with spatial dependence, momentum constraints, and shift
  components of the metric (which would appear as Lagrange multipliers for the
  momentum constraints in \cref{eq:lagrangian1}).

  The quantization of the restricted set of metrics \labelcref{eq:confmetric}
  will not lead to a Hilbert space that is the naive subsector of the Hilbert
  space of the full theory (still on a fixed topology) that one might have
  expected. For example, in the minisuperspace model, note that the inner
  product \labelcref{eq:gainnerprod} between a generic coherent state centred on
  on-shell initial data and another coherent state that is a \(90^\circ\)
  spatial rotation of the first vanishes in the semiclassical limit.
  However, in the unrestricted theory, the two states are equivalent to each
  other. This difference arises because we have neglected to quantize the
  momentum constraints in the minisuperspace model. \label{foot:diffoned}}

\begin{align}
  g_{\mu\nu} \dd{X}^\mu \dd{X}^\nu
  &= -N^2(t)\dd{t}^2 + h_{ij}(t) \sigma^i \sigma^j
    \;,
    \label{eq:confmetric}
\end{align}
where \(h_{ij}(t)\) is drawn from some space \(\mathscr{M}\) of
positive-definite matrices, and \(\sigma^i\) are (fixed) time-invariant
one-forms such that \(h_{ij}\sigma^i\sigma^j\) is also invariant under the
homogeneity group for all \(h_{ij}\in\mathscr{M}\). Different models correspond
to different choices of the homogeneity group, and often also place further
restrictions on the minisuperspace \(\mathscr{M}\), \eg{} to be a space of
diagonal \(h_{ij}(t)\). We will focus on models for which the momentum
constraints (time-space components of the Einstein equations) are identically
satisfied by \cref{eq:confmetric}. Consolidating notation, it will be useful to
use \(x^A\) to denote coordinates on \(\mathscr{M}\), \ie{} the variable \(x\)
parametrizes the space of allowed spatial metrics \(h_{ij}\). As this notation
suggests, under the minisuperspace dimensional reduction to just the time
direction, the propagating gravitational degrees of freedom become what we
previously called the ``matter'' of the gravitational worldline theory.

Adopting the spatially homogeneous metric ansatz \cref{eq:confmetric}, let us
further focus on models where the spatial directions are compact (or have been
compactified) so that we can set
\begin{align}
  \frac{1}{16\pi{}G_{\mathrm{N}}\,\hbar}\int \sigma^1 \wedge \cdots \wedge \sigma^{D-1}
  =1
  \;,
  \label{eq:spacevol}
\end{align}
where \(D\) is the spacetime dimension.
The Einstein--Hilbert Lagrangian, supplemented by a Gibbons--Hawking--York term for
boundaries in time, can be expressed in canonical form as
\begin{align}
  L &= N \sqrt{h} \left(
      R(g) - 2\Lambda
      \right)
      - 2 \partial_t(\sqrt{h} K)
      = 
      \pi^{ij} \dot{h}_{ij} - N H
      \;,
      \label{eq:lagrangian1}
  \\
  \pi^{ij}
    &= \sqrt{h}(K^{ij}- K h^{ij})
      \;,
\end{align}
where \(h\) denotes the determinant of \(h_{ij}\), \(R(g)\) is the Ricci
scalar of the metric \(g_{\mu\nu}\), the dot over a symbol denotes
differentiation with respect to \(t\),
\begin{align}
  H
  &= \frac{1}{\sqrt{h}}
    (\mathcal{G}^{-1})_{ijk\ell} \pi^{ij} \pi^{k\ell} + \sqrt{h}\,\left(
    2\Lambda -
    R(h)
    \right)
\end{align}
is the Wheeler--DeWitt Hamiltonian, and
\begin{align}
  (\mathcal{G}^{-1})_{ijk\ell}
    &= \frac{1}{2} \left(
      h_{ik} h_{j\ell}
      + h_{i\ell} h_{jk}
      - \frac{2}{D-2}
      h_{ij} h_{k\ell}
      \right)
      \label{eq:dewittmet}
\end{align}
is the inverse DeWitt metric. We see from \cref{eq:lagrangian1} that the lapse
\(N\) is a Lagrange multiplier imposing the constraint \(H=0\). To proceed, it
will be helpful to write everything in terms of \(x^A\) and its conjugate
momentum
\begin{align}
      p_A
  &= \pi^{ij} \partial_A h_{ij}
    \;,
\end{align}
giving:
\begin{align}
  L &= p_A \dot{x}^A - N H
      \;,
      \label{eq:lagrangian2}
  \\
  H
    &= \frac{1}{\sqrt{h(x)}}
      \mathcal{G}^{AB}(x) p_A p_B + \sqrt{h(x)}\,(2\Lambda-R(x))
      \;,
      \label{eq:miniham}
  \\
  \mathcal{G}_{AB}
    &= \mathcal{G}^{ijk\ell}
      \partial_A h_{ij}
      \partial_B h_{k\ell}
      \;.
      \label{eq:minimetric}
\end{align}
There is an obvious analogy between the dynamics of minisuperspace gravity and a
particle propagating in minisuperspace with metric \(\sqrt{h}\mathcal{G}_{AB}\)
and potential
\begin{align}
  V(x)
  &= \sqrt{h(x)}\,(2\Lambda-R(x))
    \;.
    \label{eq:wdwpot}
\end{align}

In \cref{app:minisuperspace}, we provide a brief review of some illustrative
examples of minisuperspace models. Our focus will be on diagonal minisuperspace
models in \(D=4\) where the spatial metric can be expressed in terms of a
conformal factor \(e^{2 x^0}\) and anisotropies \(x^1\) and \(x^2\):
\begin{align}
  h_{ij}
  &= e^{2 x^0} \diag\left(
    e^{2x^1 + 2\sqrt{3} x^2},
    e^{2x^1 - 2\sqrt{3} x^2},
    e^{-4x^1}
    \right)
    \;,
    \label{eq:misnermetric}
\end{align}
and \(\mathcal{G}_{AB}\) is the flat Minkowski metric
\begin{align}
  \mathcal{G}_{AB} \dd{x^A}\dd{x^B}
  &= 24\left(
    -(\dd{x^0})^2 + (\dd{x^1})^2 + (\dd{x^2})^2
    \right)
    \;.
    \label{eq:misnersupermetric}
\end{align}
Even with this flat metric or the conformally flat metric
\(\sqrt{h}\mathcal{G}_{AB} = e^{3 x^0}\mathcal{G}_{AB}\) in these
minisuperspace models, the dynamics encoded in the Hamiltonian constraint can be
nontrivial due to the potential term. Somewhat unusually (from the particle
perspective) this potential can behave like a reflective wall in the timelike
direction \(x^0\) in minisuperspace. In \cref{app:minisuperspace}, we will use
examples to illustrate how these potentials lead to certain features in the
gravitational inner product, which we now define.

\paragraph{In summary,} we have described how a gravitational worldline theory
can be viewed either as an intrinsically one-dimensional gravitational theory or
as a minisuperspace reduction of a higher-dimensional theory. In any case, we
have some set of variables \((x^A, p_A)\) parameterizing a kinematic phase space
that is the cotangent bundle of a target space \(\mathscr{M}\). We also have the
lapse \(N\) --- giving the only metric component \(-N^2\dd{t}^2\) in one
dimension --- which appears as a Lagrange multiplier for the Wheeler--DeWitt
Hamiltonian \(H\).

\subsection{Group-averaged inner product and the physical Hilbert space}
\label{sec:groupaverage}

Let us turn now to the quantum theory. If we ignore the Wheeler--DeWitt
constraint \(H\), the kinematic Hilbert space is simply
\(\mathcal{H}_{\mathrm{kin}}=L^2(\mathscr{M},\mu)\) with an inner product
\begin{align}
  \braket{\psi'}{\psi}
  &= \int_{\mathscr{M}} \mu(x) (\psi'(x))^* \psi(x)
    \;,
\end{align}
involving the volume measure \(\mu\) on \(\mathscr{M}\) (\eg{} associated with
the metric over minisuperspace \(\mathscr{M}\)). Equivalently, we may write
\begin{align}
  \ket{\psi}
  &= \int_{\mathscr{M}} \mu(x) \psi(x) \ket{x}
    \;,
  &
  \braket{x'}{x}
  &= \delta_\mu(x',x)
    \;,
  &
  1
  &= \int_{\mathscr{M}} \mu(x) \ketbra{x}{x}
    \;,
    \label{eq:psismear}
\end{align}
introducing an orthonormal position basis \(\{\ket{x}:x\in\mathscr{M}\}\),
\(\delta\)-function normalized with respect to the measure \(\mu\). Below, we
will review how the gravitational worldline path integral naturally defines a
different inner product \(\llangle\bullet|\bullet\rrangle\) and we will explain
in \cref{sec:physhilb} how this defines a physical Hilbert space
\(\mathcal{H}_{\mathrm{co}}\cong\mathcal{H}_{\mathrm{inv}}\) that accounts for
the constraint \(H=0\) associated with time-reparametrization invariance.

Specifically, let us define the physical inner product
\(\llangle x'|x\rrangle\) using the (canonical) gravitational path integral
\begin{align}
  \llangle x'|x\rrangle
  &= \int_{x(t=0)=x}^{x(t=1)=x'} \frac{\mathcal{D}x\,\mathcal{D}p\,\mathcal{D}N}{\text{diff}}
    e^{i \int_0^1 \dd{t} (p_A \dot{x}^A - N H)}
\end{align}
over histories of \(x^A(t)\), \(p_A(t)\), and \(N(t)\) with boundary conditions
\(x^A(t=0)=x^A\) and \(x^A(t=1) = (x')^A\).\footnote{If the worldline theory is
  oriented, the boundary condition for the bra-state should more generally be
  \(x^A(t=1)=(x^{\prime\star})^A\) --- see \cref{sec:orientrev}.} The ``diff''
in the denominator reminds us to gauge-fix the integration to
diffeomorphism-inequivalent (\ie{} time-reparametrization-inequivalent)
histories. We have left the precise path integration measure
\(\frac{\mathcal{D}x\,\mathcal{D}p\,\mathcal{D}N}{\text{diff}}\) unspecified ---
we will take it such that the calculation of the path integral proceeds as we
now describe:
\begin{itemize}
\item{From the functional integral over \(N\), we seem to obtain
    \(\delta\)-functionals setting \(H=0\) throughout history --- that would be
    the case, if it were not for gauge-fixing conditions that can depend on
    \(N\). Actually, using time-reparametrization, we can gauge fix away most of
    the freedom in the function \(N(t)\) (\eg{} with the gauge-fixing condition
    \(\dot{N}(t)=0\)). We are nonetheless left with the invariant
    \begin{align}
      T &= \int_0^1 \dd{t} N(t)
          \;,
    \end{align}
    which must be integrated over.}
\item{The path integral at a fixed \(T\) is just the usual one computing the
    matrix element \(\mel{x'}{e^{-i T H}}{x}\) for the time evolution generated
    by \(H\) as an operator on the
    \(\mathcal{H}_{\mathrm{kin}}=L^2(\mathscr{M},\mu)\). When promoting the
    Hamiltonian \(H\) to a quantum operator, there is an implicit choice of
    operator ordering, with different orderings corresponding possibly to
    corrections to the Lagrangian (of order at least \(\hbar\)) or measure used
    to define the path integral. To give one example of a natural choice, we can
    take the kinetic term in the Hamiltonian to be the Laplacian
    \(\Box_{\sqrt{h}\mathcal{G}}\), associated with the minisuperspace metric
    \(\sqrt{h}\mathcal{G}_{AB}\), acting on
    \(\mathcal{H}_{\mathrm{kin}}=L^2(\mathscr{M},\mu)\):
    \begin{align}
      H
      &= \Box_{\sqrt{h}\mathcal{G}} + \sqrt{h(x)}\,(2\Lambda-R(x)) \;.
        \label{eq:quantumH}
    \end{align}
    The main text of this paper will not rely on any particular form for the
    Hamiltonian \(H\), and in \cref{app:minisuperspace}, we will also consider a
    simplifying choice differing slightly from the above.}
\item{Altogether, including the final integral over \(T\), the worldline path
    integral gives rise to the inner product
    \begin{align}
      \llangle x' | x\rrangle
      &= \mel{x'}{\eta}{x}
        \label{eq:gainnerprod}
      \\
      \eta
      &= \int_{-\infty}^\infty \dd{T}
        e^{-i T H}
        = 2\pi\delta(H)
        \;,
        \label{eq:groupavg}
    \end{align}
    the interpretation of which we will turn to shortly in
    \cref{sec:physhilb}.}
\end{itemize}

Let us briefly reflect on the choice of integration contour for \(T\). Other
than the contour \(\mathbb{R}\) we selected in \cref{eq:groupavg}, what choices
are there? To respect locality, we note that the contour can only terminate at
\(0\) or \(\infty\) --- any other endpoint would mean placing a nonlocal
restriction on the path integral over \(N\). Let us consider some obvious
possibilities for the \(T\)-contour.

First, let us consider the candidate integration contours \(\mathbb{R}_{<0}\)
and \(\mathbb{R}_{>0}\). Because the spectrum of \(H\) includes zero, these
integrals are perhaps best written with a regulator, \eg{}
\begin{align}
  \int_0^\infty \dd{T} \mel{x'}{e^{-i T (H-i\epsilon)}}{x}
  &= \mel{x'}{\frac{1}{iH+\epsilon}}{x}
    \;,
    \label{eq:feynman}
  \\
    \int_{-\infty}^0 \dd{T} \mel{x'}{e^{-i T (H+i\epsilon)}}{x}
  &= \mel{x'}{\frac{1}{-iH+\epsilon}}{x}
    \;.
    \label{eq:antifeynman}
\end{align}
These amplitudes are Green's functions for the differential operator \(H\). If we
want to identify this worldline amplitude with some two-point function in a
(free) universe ``field theory'' description of arbitrary numbers of universes
(worldlines), it is important to recall that the Wheeler--DeWitt Hamiltonian
\(H\) generates a redundancy of the gravitational worldline theory which ought
to act trivially on physical worldline states. Worldline states correspond to
field theory operators, so this physical requirement will translate to an
equation of motion \(H\Phi(x)=0\) satisfied by the field operator. If we want to
identify a \emph{Green's function} for \(H\) (rather than a solution) with a
two-point function for a universe field theory, that two-point function ought to
come with some nontrivial ordering prescription. Happily, the \(i\epsilon\)
prescription of the amplitude \labelcref{eq:feynman} (respectively
\labelcref{eq:antifeynman}) can sometimes\footnote{The relation between an
  amplitude like \cref{eq:feynman} and a QFT two-point function is perhaps most
  easily established on a Euclidean section of complexified \(\mathscr{M}\)
  where \(H\) becomes positive and has a well-defined inverse. Assuming analytic
  continuation back to the original section \(\mathscr{M}\) gives a well-defined
  vacuum state \(\ket{0}_{\mathrm{QFT}}\) and \cref{eq:feynman} is the analytic
  continuation of its Euclidean counterpart, then it can be identified with a
  time-ordered correlation function in the state \(\ket{0}_{\mathrm{QFT}}\).}
give it precisely the right analyticity properties appropriate for a QFT
(anti-)time-ordered two-point function
\(\mel{0}{\mathrm{T}\Phi_{\mathrm{QFT}}(x')\Phi_{\mathrm{QFT}}(x)}{0}_{\mathrm{QFT}}\)
(\(\mel{0}{\overline{\mathrm{T}}\Phi_{\mathrm{QFT}}(x')\Phi_{\mathrm{QFT}}(x)}{0}_{\mathrm{QFT}}\))
in some preferred vacuum \(\ket{0}_{\mathrm{QFT}}\). If we are able to make such
an identification, then that would be one approach --- \cref{tab:qftvsuft}'s
``Approach 1'' --- to second-quantization that lands on a QFT with field
equation \(H\Phi_{\mathrm{QFT}}(x)=0\).

As an aside, let us note that \cref{eq:groupavg} can be expressed as a sum of
the propagators \labelcref{eq:feynman,eq:antifeynman} \cite{Casali:2021ewu},
\begin{align}
  \eta
  &= 2\pi\delta(H)
    = \frac{1}{iH + \epsilon} + \frac{1}{-iH+\epsilon}
    \;.
    \label{eq:etaisfeynplusafeyn}
\end{align}
So from the QFT perspective, \(\llangle x'|x\rrangle=\mel{x'}{\eta}{x}\) can
sometimes be given the interpretation of a symmetrized two-point function
\(\mel{0}{\{\Phi_{\mathrm{QFT}}(x'),\Phi_{\mathrm{QFT}}(x)\}}{0}_{\mathrm{QFT}}\).

Next, we note that the Wheeler--DeWitt Hamiltonian \(H\) for gravity is unbounded
from above and below, similar to the field equation of motion of a Lorentzian
QFT. As the theory stands, the \(T\)-contours that stray from \(\mathbb{R}\) are
thus ruled out. This is a manifestation of the conformal factor problem
\cite{Gibbons:1978ac} which prevents the gravitational path integral from being
defined over Euclidean geometries --- here, worldlines with imaginary \(T\).
(Note that the conformal mode \(x^0\) is time-like in minisuperspace, as seen in
\cref{eq:misnersupermetric}.) In QFT, one way to make the field equation of
motion bounded from below is to do a Wick rotation to Euclidean signature on the
space \(\mathscr{M}\) where particles propagate. After Wick rotation, one can
then integrate the worldline lapse \(T\) over \(i\mathbb{R}_{<0}\) and obtain a
Euclidean QFT two-point function. From the worldline perspective,
\(\mathscr{M}\) is target space, so this strategy involves choosing some complex
contour for ``fields'' on the worldline (in particular the conformal mode
\cite{Gibbons:1978ac}). But at present, there is no general consensus on what
the right contour should be to define a ``Euclidean'' gravitational path
integral, prompting suggestions to retreat back to Lorentzian signature
(\(T\in\mathbb{R}\)) where the path integral might in principle be defined over
real Lorentzian fields.

To summarize, we have described how choosing integration contours for \(T\)
running between \(0\) and \(\infty\) leads to Green's functions for the
Wheeler--DeWitt Hamiltonian \(H\) that can play the role of time-ordered
two-point functions in QFT. In this paper, we will focus instead on the other
choice made in \cref{eq:groupavg} of integrating \(T\) over the contour
\(\mathbb{R}\). The resulting worldline amplitude is naturally interpreted as a
physical inner product, as we review in \cref{sec:physhilb}. When we construct a
universe field theory using this worldline amplitude as input in \cref{sec:uft},
we will take care to highlight differences with standard QFT that arise. This is
assuming, of course, that a standard QFT approach makes sense at all --- in
\cref{sec:discussqftvsuft}, however, we will review how reasonable worldline
models of gravity can make the QFT approach quite pathological or at least
unnatural. This provides some justification for the alternative approach taken
in this paper, which we refer to as group-averaged baby universe field theory
(GABUFT).

\subsubsection{Constructing the physical worldline Hilbert space}
\label{sec:physhilb}

The amplitude \labelcref{eq:gainnerprod} resulting from integrating \(T\) over
\(\mathbb{R}\) defines a so-called group-averaged inner product, in this case,
with the group of time translations. Going beyond the minisuperspace truncation,
the more complete path integral on higher-dimensional spacetimes of fixed
cylinder-like \((\text{time interval})\times(\text{spatial topology})\) topology
will involve path integrals over both lapse \(N\) and shift \(N^i\) --- the
time-space components of the metric. After gauge-fixing, the integrals over the
residual freedoms in \(N\) and \(N^i\) are expected to similarly give rise to an
inner product \labelcref{eq:gainnerprod} involving a group average
\begin{align}
  \eta
  &= \int_G \dd{g} U(g)
    \label{eq:generalgav}
\end{align}
over the group \(G\) of space-dependent time translations and spatial
diffeomorphisms, where \(U(g)\) denotes the unitary action of a group element \(g\in G\)
on states. In this paper, our focus will be on worldline models so that we only
have the group average \labelcref{eq:groupavg} over time translations
\(U(g)=e^{-i T H}\). The purpose of this section will be to explain the general
conceptual interpretation of the group-averaged inner product
\(\llangle\bullet|\bullet\rrangle=\mel{\bullet}{\eta}{\bullet}\) and the
physical Hilbert space it constructs.\footnote{The review presented here is
  largely inspired by \cite{Alonso-Monsalve:2025lvt}.}

To make the following discussion a bit more precise, we start, however, with a
technical remark: the group average \(\eta\) is not defined on all of
\(\mathcal{H}_{\mathrm{kin}}=L^2(\mathscr{M},\mu)\), but only on a dense domain
\(\mathcal{D}_{\mathrm{test}}\subset L^2(\mathscr{M},\mu)\) of test functions
\(j(x)\). For the following discussion to make sense,
\(\mathcal{D}_{\mathrm{test}}\) must be preserved by the action \(U(g)\) of
\(g\in G\) and \(\eta\) must map into the conjugate dual
\(\mathcal{D}_{\mathrm{test}}^\times\) of \(\mathcal{D}_{\mathrm{test}}\),
\begin{align} \eta : \mathcal{D}_{\mathrm{test}} \to
  \mathcal{D}_{\mathrm{test}}^\times \;,
  \label{eq:etadomain}
\end{align}
so that \(\mel{j'}{\eta}{j}\) is well-defined.\footnote{Here, the
conjugate \emph{algebraic} dual \(\mathcal{D}_{\mathrm{test}}^\times\) is the
space of \(\mathbb{C}\)-valued antilinear functionals on
\(\mathcal{D}_{\mathrm{test}}\). The construction of the physical Hilbert space
\(\mathcal{H}_{\mathrm{co}}\cong\mathcal{H}_{\mathrm{inv}}\), as described in
the main text, does not require a topology on \(\mathcal{D}_{\mathrm{test}}\).
(And if we required \cref{eq:etadomain} to hold instead with
\(\mathcal{D}_{\mathrm{test}}^\times\) replaced by the \emph{continuous} dual
with respect to some topology on \(\mathcal{D}_{\mathrm{test}}\) --- as some
authors do --- then the constructed
\(\mathcal{H}_{\mathrm{co}}\cong\mathcal{H}_{\mathrm{inv}}\) will be unchanged
if we choose a finer topology on \(\mathcal{D}_{\mathrm{test}}\); in particular,
we can choose an extremely fine topology such that the continuous dual is equal
to the algebraic dual \(\mathcal{D}_{\mathrm{test}}^\times\).)


Nonetheless, introducing a meaningful topology on
\(\mathcal{D}_{\mathrm{test}}\) may be useful for other purposes. For example,
one might want certain bounded (with respect to \(\braket{\bullet}{\bullet}\))
\(G\)-invariant operators
\(O:\mathcal{D}_{\mathrm{test}}\to\mathcal{D}_{\mathrm{test}}\) of interest to give
rise to bounded operators on
\(\mathcal{H}_{\mathrm{co}}\cong\mathcal{H}_{\mathrm{inv}}\), \ie{} the latter
operators are continuous with respect to the
\(\llangle\bullet|\bullet\rrangle\)- (\(\lsem\bullet|\bullet\rsem\)-)norm
topology on \(\mathcal{H}_{\mathrm{co}}\cong\mathcal{H}_{\mathrm{inv}}\). This
can be checked by the equivalent condition for the former operators
\(O:\mathcal{D}_{\mathrm{test}}\to\mathcal{D}_{\mathrm{test}}\) to be continuous
on \(\mathcal{D}_{\mathrm{test}}\) with respect to the
\(\mel{\bullet}{\eta}{\bullet}\) seminorm-topology on
\(\mathcal{D}_{\mathrm{test}}\).

The author thanks Donald Marolf for explaining this point.} Expressions like
\(\mel{x'}{\eta}{x}\) should be understood as distributions to be smeared
against test functions \(j'(x')^*,j(x)\),
\begin{align}
  \ket{j}
  &= \int_{\mathscr{M}} \mu(x) j(x) \ket{x}
    \;.
    \label{eq:jsmear}
\end{align}
In \cref{sec:gennotation}, we will
describe a practical (but perhaps non-rigorous) characterization of the space \(\mathcal{D}_{\mathrm{test}}\)
of allowed test functions \(j(x)\).

Conceptually, the most important property of the group average \(\eta\) is its
invariance under the group action:
\begin{align}
  U(g)^\dagger \eta
  &= \eta
    = \eta U(g)
    \;.
    \label{eq:gainvariance}
\end{align}
The latter equality implies that the difference of two gauge-equivalent states
\(\ket{j},U(g)\ket{j}\in\mathcal{D}_{\mathrm{test}}\) lies in the kernel of
\(\eta\),
\begin{align}
  \eta (1-U(g))\ket{j}
  &= 0
    \;,
  &
    (1-U(g))\ket{j}\in\ker\eta
    \;.
\end{align}
The group-averaged inner product
\begin{align}
  \llangle j' | j\rrangle
  &\equiv \mel{j'}{\eta}{j}
\end{align}
is really a positive-definite inner product on equivalence classes
\begin{align}
  |j\rrangle
  &= \{\ket{j} + \ket{j_0} | \ket{j_0}\in\ker\eta\}
    \in \frac{\mathcal{D}_{\mathrm{test}}}{\ker\eta}
               \;.
\end{align}
In particular, gauge-equivalent states \(\ket{j}, U(g)\ket{j}\) fall into
the same equivalence class \(|j\rrangle\) so their difference gives a null
state \(|j\rrangle-|U(g)j\rrangle=0\). Taking the completion of
\(\mathcal{D}_{\mathrm{test}}/\ker\eta\) in its
\(\llangle\bullet|\bullet\rrangle\)-norm topology defines what we will refer to
as the coinvariant Hilbert space
\begin{align}
  \mathcal{H}_{\mathrm{co}}
  &\equiv \overline{\left( \frac{\mathcal{D}_{\mathrm{test}}}{\ker\eta} \right)}
    \;.
\end{align}
The coinvariant Hilbert space \(\mathcal{H}_{\mathrm{co}}\) provides one
description of the physical Hilbert space of the gravitational theory ---
roughly speaking, physical states are gauge-equivalence classes of kinematic
states.

An alternative description is as a Hilbert space \(\mathcal{H}_{\mathrm{inv}}\)
of invariant states \(|k\rsem\). More precisely, the first equality of
\cref{eq:gainvariance} says that the image \(\Image\eta\) of \(\eta\) contains
only \(U(g)\)-invariant functionals, and we define
\begin{align}
  \mathcal{H}_{\mathrm{inv}}
  &\equiv \overline{\Image\eta}
    \;,
\end{align}
as the completion of \(\Image\eta\) with inner product
\begin{align}
  \psi &= \eta j\;, \quad
      \psi' = \eta j'
  &
  &\implies
      &
  \lsem \psi' | \psi \rsem
  &\equiv \llangle j' | j\rrangle
    = \mel{j'}{\eta}{j}
    = \braket{j'}{\psi}
    = \braket{\psi'}{j}
    \;.
    \label{eq:invinnerprod}
\end{align}
The invariant and coinvariant Hilbert spaces are equivalent descriptions of the
physical Hilbert space:
\begin{align}
  \Image \eta
  &\cong \frac{\mathcal{D}_{\mathrm{test}}}{\ker\eta}
    \;,
    &
    \mathcal{H}_{\mathrm{inv}}
  &\cong \mathcal{H}_{\mathrm{co}}
    \;.
\end{align}

We sketched earlier in \cref{sec:groupaverage} how the gravitational path
integral naturally gives a group-averaged inner product, and we have just
reviewed how this inner product is associated with a Hilbert space
\(\mathcal{H}_{\mathrm{co}}\cong\mathcal{H}_{\mathrm{inv}}\) of physical states.
One might wonder how this is related to simply taking the \(U(g)\)-invariant
subspace of \(\mathcal{H}_{\mathrm{kin}}\). The problem is that, when the group
under consideration is noncompact, the invariant ``states in
\(\mathcal{H}_{\mathrm{kin}}\)'' often have infinite norms under the kinematic
inner product \(\braket{\bullet}{\bullet}\) of \(\mathcal{H}_{\mathrm{kin}}\)
(in which case they aren't truly ``states'' of this Hilbert space).\footnote{In
  fact, this is guaranteed by our assumption that zero is in the continuous
  spectrum of \(H\), as stated at the start of \cref{sec:worldline}.

  Relaxing this assumption, if a (normalizable)
  \(\ket{\psi}\in\mathcal{H}_{\mathrm{kin}}\) happens to be invariant under a
  noncompact subgroup, that noncompact subgroup should be excluded from the
  group average \(\eta\ket{\psi}\) to avoid a divergence in the physical inner
  product \cref{eq:invinnerprod} \cite{Alonso-Monsalve:2025lvt}. (We expect the
  gravitational path integral to automatically give us such a
  prescription.) \label{foot:invariantga}} This is because the kinematic inner
product of invariant states contains a factor of the volume of the gauge group
\(G\). The inner product \(\lsem\bullet|\bullet\rsem\) of the invariant Hilbert
space \(\mathcal{H}_{\mathrm{inv}}\) essentially divides by the volume of \(G\).

Let us now give a practically description of the group-averaged inner product
\(\llangle\bullet|\bullet\rrangle\) in terms of an expansion in invariant modes.

\subsubsection{Mode decomposition of the group-averaged inner product}
\label{sec:gennotation}

We retreat now to a physicist's level of rigour, to give a practical
description of the group-averaged inner product as a projection onto invariant
\(\delta\)-function normalized wavefunctions.

As stated at the outset of \cref{sec:worldline}, we will assume that \(H\) is
self-adjoint on \(\mathcal{H}_{\mathrm{kin}}\), and can therefore be diagonalized. We will
denote (possibly \(\delta\)-function normalizable) eigenstates, eigenfunctions,
and eigenvalues in the spectral decomposition of \(H\) by \(\ket{n}=\ket{u_n}\),
\(u_n(x)\), and \(h_n\), satisfying
\begin{align}
  H \ket{n}
  &= h_n \ket{n}
    \;.
\end{align}
We will use labels \(n\) belonging to some set \(\mathscr{N}\) to index a
complete orthogonal basis of such eigenstates. Since the spectrum of \(H\) is
continuous at least around zero, the label set \(\mathscr{N}\) is a space with a
correspondingly continuous coordinate \(h_n\) near \(h_n=0\). (For example, in
the Bianchi I model with vanishing cosmological constant \(\Lambda=0\) and
constraint Hamiltonian \(\tilde{H}\) discussed in \cref{sec:ksBianchiI},
\(\mathscr{N}\) can be taken to be the momentum space of eigenvalues for
\((p^0,p^1,p^2)\).) The choice of normalization for the eigenstates \(\ket{n}\),
or equivalently eigenfunctions \(u_n\), corresponds to a choice of measure
\(\nu(n)\) on \(\mathscr{N}\):
\begin{align}
  \braket{n'}{n}
  &= \int_{\mathscr{M}} \mu(x) u_{n'}(x)^* u_n(x)
    = \delta_\nu(n,n')
    \;,
    \\
    1 &= \int_{\mathscr{N}} \nu(n)
        \ketbra{n}{n}
        \;,
    \\
    H &= \int_{\mathscr{N}} \nu(n)
        h_n \ketbra{n}{n}
        \;.
\end{align}
(While \(\nu(n)\) and \(\ket{n}\) depend on the choice of normalization, the
combination \(\nu(n)\ketbra{n}{n}\) does not.)

Let us consider now the group-averaged inner product
\(\llangle\bullet|\bullet\rrangle=\mel{\bullet}{\eta}{\bullet}\) which, as we
have described, gives the physical inner product of the gravitational theory.
With just the group of time translations generated by a Hamiltonian \(H\), the
group average can be expressed as a distributional projection
\begin{align}
  \eta
    &= \int_{-\infty}^\infty \dd{T}
        e^{-i T H}
  =2\pi\delta(H)
\end{align}
onto invariant states annihilated by \(H\).
Let us label a basis \(\{\ket{k}\}\) of such states by \(k\) taking values in
the subspace
\begin{align}
  \mathscr{K}
  = \{
  k\in\mathscr{N}
  \mid
  h_k=0
  \}
  \subset\mathscr{N}
  \;.
\end{align}
Then, we can express \(\eta\) as
\begin{align}
  \eta
  &=
    2\pi \int_{\mathscr{N}} \nu(n)
    \delta(h_n)
    \ketbra{n}{n}
    \label{eq:specinnerprod}
    \\
  &= \int_{\mathscr{K}} \kappa(k) \ketbra{k}{k}
    \;,
    \label{eq:etaexpansion}
\end{align}
where the measure \(\kappa(k)\) on \(\mathscr{K}\subset\mathscr{N}\) is the one
equivalent to \(2\pi\delta(h_n)\nu(n)\) on \(\mathscr{N}\).

The physical Hilbert space
\(\mathcal{H}_{\mathrm{co}}\cong\mathcal{H}_{\mathrm{inv}}\) described in
\cref{sec:physhilb} can be thought of as
\begin{align}
  \mathcal{H}_{\mathrm{co}}
  \cong\mathcal{H}_{\mathrm{inv}}
  = L^2(\mathscr{K},\kappa)
  \label{eq:physicall2}
\end{align}
with measure \(\kappa\):
\begin{align}
  \llangle j' | j\rrangle
  &= \int_{\mathscr{K}} \kappa(k) \braket{j'}{k} \braket{k}{j}
    \;.
\end{align}
In hindsight, we see that the test subspace
\(\mathcal{D}_{\mathrm{test}}\subset\mathcal{H}_{\mathrm{kin}}\) should be
chosen so that test states \(\ket{j}\) have well-defined overlaps
\(\braket{k}{j}\) belonging to \(L^2(\mathscr{K},\kappa)\). This requires an
appropriate degree of smoothness of \(\braket{n}{j}\) in \(n\) near the
constraint surface \(\mathscr{K}\subset\mathscr{N}\).

We have essentially described above the ``spectral analysis construction''
\cite{Marolf:1994ae} of the inner product
\(\llangle x'|x\rrangle=\mel{x'}{\eta}{x}=2\pi\mel{x'}{\delta(H)}{x}\). Going
beyond worldline theory, gravitational constraints in higher dimensions will
generate a non-abelian group \(G\) of diffeomorphisms to be averaged over in
\(\eta\), as in \cref{eq:generalgav}. The group-averaging \(\eta\) will still
take the form \labelcref{eq:etaexpansion} of a projection onto states
\(\ket{k}\) annihilated by all constraints; the measure \(\kappa(k)\), defined
by equating \cref{eq:etaexpansion} with \(\eta\), might just lack as obvious a
construction as above. The remainder of this paper will mostly bypass
\(\mathscr{N}\) and make reference to an orthogonal basis
\(\{\ket{k}:k\in\mathscr{K}\}\) and a measure \(\kappa(k)\) for which the
physical gravitational inner product \(\llangle x'|x\rrangle=\mel{x'}{\eta}{x}\)
is given by matrix elements of \cref{eq:etaexpansion}.\footnote{Even with the
  possible modified definition of \(\eta\) described in \cref{foot:invariantga},
  we expect it to be expressible in the form \labelcref{eq:etaexpansion}. (We
  have in mind a broad interpretation of notation like
  \(\int_{\mathscr{K}}\kappa(k)\), which could represent \eg{} an integral, a
  sum, or some mix of both.)}

\subsection{Antilinear time and orientation reversals}
\label{sec:reversal}

To develop the universe field theory framework in \cref{sec:uft}, we will
require an additional structure which is not necessarily part of our earlier
definition of a gravitational worldline theory, but is naturally implied by the
existence of a path integral description. Namely, given boundary conditions ---
call them \(j\) --- preparing a (test) ket-state \(\ket{j}\), there are
corresponding boundary conditions --- call them \(j^\star\) --- that prepare the
corresponding (test) bra-state \(\bra{j}\).\footnote{Most of what we will say
  here should also hold for arbitrary kinematic states. Our interest, though,
  will be mostly in test states
  \(\ket{j},\ket{j'}\in\mathcal{D}_{\mathrm{test}}\subset\mathcal{H}_{\mathrm{kin}}\)
  for which the gravitational path integral evaluates to a well-defined answer
  \(\llangle j'|j\rrangle\).} In the following, we will discuss this
\(\star\)-involution structure, which can be viewed as antilinear time reversal,
\(\ket{j^\star}=\mathsf{T}\ket{j}\), or antilinear orientation reversal,
\(\ket{j^\star}=\mathcal{T}\ket{j}\), depending on whether spacetime (worldline)
configurations are oriented \cite{Witten:2025ayw}. Although \(\mathsf{T}\) and
\(\mathcal{T}\) might be meaningfully distinguished in non-gravitational QFT,
the distinction becomes somewhat ambiguous once orientation is viewed as a
degree of freedom in gravity, as we will see by the end of \cref{sec:orientrev}.

\subsubsection{Time reversal}
\label{sec:timerev}

Let us first consider gravitational theories where spacetime (worldline)
configurations are not assigned orientation, \ie{} are \emph{unoriented}. To be
clear, one-dimensional worldlines are always \emph{orientable} meaning that they
could be assigned orientation in principle; currently, we are considering
theories where each worldline configuration in the path integral is not assigned
an orientation. In higher dimensions, it is natural to consider similar
gravitational theories on non-orientable spacetimes as well.

We expect gravitational theories with unoriented spacetimes (worldlines) to
possess a time-reversal symmetry \(\mathsf{T}\). In
gravitational theories, the most obvious way to break time-reversal symmetry
would have been to require spacetime configurations be oriented and give the
Lagrangian or Hamiltonian orientation-odd terms. Without reference to
orientation, we expect there to be a time-reversal symmetry,
\begin{align}
  \mathsf{T} H \mathsf{T}
  &= H
    \;.
    \label{eq:timerevsym}
\end{align}

For such gravitational theories where spacetime (worldline) configurations are
unoriented, the boundary conditions \(j^\star\) preparing the bra-state
\(\bra{j}\) are also the ones used to prepare the time-reversed ket-state
\(\ket{j^\star}= \mathsf{T}\ket{j}\in\mathcal{D}_{\mathrm{test}}\)
\cite{Witten:2025ayw}. The time-reversal symmetry \(\mathsf{T}\) maps the
Hilbert space \(\mathcal{H}_{\mathrm{kin}}\) of the worldline theory back to
itself (and we assume that it also preserves the test subspace
\(\mathcal{D}_{\mathrm{test}}\subset\mathcal{H}_{\mathrm{kin}}\)). In general, an
orthonormal basis, \eg{} \(\{\ket{x} \mid x\in\mathscr{M}\}\), is mapped by
\(\mathsf{T}\) to an (\apriori{} possibly different) orthonormal basis, \eg{}
\(\{\ket{x^\star}:x^\star\in\mathscr{M}^\star\}\) with a label set that we will
call \(\mathscr{M}^\star\). It is often natural, however, to require (worldline)
time-reversal invariance as part of the definition of this ``position basis'':
\begin{align}
  \mathscr{M}
  &= \mathscr{M}^\star
    \;,
  &
    x
  &= x^\star
    \;.
    \label{eq:startimerev}
\end{align}
This is the case, for example, if \(x\) parametrizes the induced metric and
matter tensor field configurations on a time slice, as opposed to their
conjugate momenta. Our focus in this paper will be on Euclidean boundaries
where, unless otherwise specified, ``induced field configurations'' refers to
ones satisfying the same reality conditions as the bulk fields.\footnote{One
  could also consider Lorentzian boundaries as would be relevant for Lorentzian
  open universes (\eg{} in standard Lorentzian AdS/CFT); real induced field
  configurations on Lorentzian boundaries will generically not satisfy
  \(x^\star\stackrel{?}{=}x\). We will refrain from talking about spinor fields to
  sidestep the discussion of (s)pin structures; see \cite{Witten:2025ayw} for
  details.} We will also focus on tensor fields and avoid discussing spinor
fields (apart from a quick note about gauging fermion parity in
\cref{app:pathintgauging}). Specifically, we expect the second equality of
\cref{eq:startimerev} to hold for the induced configurations of ordinary
(non-pseudo) tensor fields; a short discussion of axions will be provided in the
summary at the end of \cref{sec:orientrev}.

Given \cref{eq:startimerev}, \(\star\) can be extended antilinearly to
\(\mathcal{D}_{\mathrm{test}}\subset\mathcal{H}_{\mathrm{kin}}\). In terms of
the wavefunction \(j(x)\) describing an arbitrary test state \(\ket{j}\) --- see
\cref{eq:jsmear},
\begin{align}
  j^\star(x)
  &= j(x)^*
    \;.
\end{align}
Note the distinction between the five-pointed star \(\star\) denoting
(worldline) time reversal here on the LHS and the six-pointed asterisk \(*\) on
the RHS used to denote complex conjugation in general.

Since time reversal is a symmetry \labelcref{eq:timerevsym}, we can similarly
require
\begin{align}
    \mathscr{K}
  &=\mathscr{K}^\star
    \;.
    \label{eq:NstarKstar}
\end{align}
In fact, one can go so far as to demand \(k = k^\star\), but
we will \emph{not} do so as this would be unnatural if one views
\(\mathscr{K}\) as on-shell ``momentum space'' (which we can
take literally in the \(\Lambda=0\) Bianchi I model --- see \cref{sec:ksBianchiI}).
In that case, it may be more natural to take \(k^\star=-k\) as the action of
\(\star\) on the momentum-space labels. Regardless, because \(\mathsf{T}\) is a
symmetry,
\begin{align}
    \kappa(k) &= \kappa(k^\star) \;.
                \label{eq:starmeasures}
\end{align}

\subsubsection{Antilinear orientation reversal}
\label{sec:orientrev}

Next, let us consider gravitational theories where spacetime (worldline)
configurations carry orientation, and the Lagrangian and Hamiltonian might
depend on this orientation.

This orientation should be regarded as a gravitational degree of freedom like
the metric and should in principle be summed over in the path integral, subject
to boundary conditions. The dynamics of spacetime orientation are trivial,
however: for a given bulk topology and boundaries with compatible orientations,
there is only one possible spacetime orientation which correctly induces the
orientations prescribed on the boundaries. (For one-dimensional worldlines, the
orientation of a zero-dimensional endpoint is just a sign.)

Let us again use \(x\) to denote the induced field configuration of tensors
--- now including the induced orientation --- on a boundary which prepares the
state \(\ket{x}\). The boundary condition \(x^\star\) that prepares the
corresponding bra-state \(\bra{x}\) is then related to \(x\) by a reversal of
orientation, so generically\footnote{For one-dimensional worldline theories,
  \(\mathscr{M}\) decomposes into two mutually disconnected pieces associated
  with the two superselection sectors of \(\mathcal{H}_{\mathrm{kin}}\) with
  opposite boundary orientations; \(\star\) maps between the two pieces, so
  \cref{eq:xstarnex} always holds. In higher dimensions, it is possible for
  \(x^\star\) to be diffeomorphic to \(x\), as described in
  \cref{foot:circleboundary}.}
\begin{align}
  x &\ne x^\star
      \label{eq:xstarnex}
\end{align}
in contrast to \cref{eq:startimerev}.

For one-dimensional worldline theories, the Hilbert space
\(\mathcal{H}_{\mathrm{kin}}\) (as well as the test subspace
\(\mathcal{D}_{\mathrm{test}}\subset\mathcal{H}_{\mathrm{kin}}\)) decomposes
into two superselection sectors according to the orientation (sign) of the ket
boundary (endpoint). It might be tempting therefore to treat each superselection
sector separately as an independent theory, in which case the states \(\ket{x}\)
and \(\ket{x^\star}\) prepared by boundary conditions \(x\) and \(x^\star\)
belong respectively to these two independent theories. Such a decomposition,
however, may not be natural in higher dimensions, where a boundary with a given
orientation can be related by diffeomorphism to a boundary with the opposite
orientation.\footnote{Consider, for example, two-dimensional gravity with
  one-dimensional boundaries. A circular boundary with a given orientation is
  diffeomorphic to a circular boundary with the opposite orientation. (However,
  if the circle is endowed with other features, \eg{} insertions of different
  matter operators, then the ordering of those features relative to the
  orientation is then diffeomorphism-invariant and thus meaningful.) See
  \cref{app:pathintgauging} for a path integral description of gauging spatial
  reflections. See \cref{foot:diffoned} for related comments on the loss of
  spatial diffeomorphisms upon dimensional reduction to just
  time. \label{foot:circleboundary}} In anticipation of future extensions to
higher dimensions, we will therefore only refer to the full worldline theory
whose Hilbert space \(\mathcal{H}_{\mathrm{kin}}\) (as well as
\(\mathcal{D}_{\mathrm{test}}\)) includes states prepared by boundaries with
both orientations, and for which
\begin{align}
  \mathscr{M}
  &= \mathscr{M}^\star
\end{align}
is a label set of boundary conditions including both orientations.

In this full theory where orientation is a degree of freedom, let us consider
the antilinear orientation-reversal transformation\footnote{This is essentially
  ref.~\cite{Witten:2025ayw}'s \(\mathcal{T}\); by including orientation as a
  degree of freedom, we regard \(\mathcal{T}\) as a symmetry of a single
  gravitational theory, as opposed to a map between different QFT Hilbert
  spaces.} \(\mathcal{T}\) defined by the property of being anti-linear and
\begin{align}
  \mathcal{T}:\ket{x} \mapsto \ket{x^\star}
  \;.
  \label{eq:orientrev}
\end{align}
For a general (test) state \(\ket{j}\), the boundary conditions \(j^\star\)
which prepare \(\bra{j}\) are the same as the ones used to prepare
\(\ket{j^\star}=\mathcal{T}\ket{j}\) \cite{Witten:2025ayw}. In terms of the
wavefunction \(j(x)\) description,
\begin{align}
  j^\star(x)
  &= j(x^\star)^*
    \;.
\end{align}
Again, note the distinction between the five-pointed star \(\star\) denoting
spacetime (worldline) orientation reversal here and the six-pointed asterisk
\(*\) used to denote complex conjugation.

The antilinear orientation-reversal transformation \(\mathcal{T}\) is a symmetry,
\begin{align}
  \mathcal{T} H \mathcal{T}
  &= H
    \;,
    \label{eq:orientrevsym}
\end{align}
so again we can choose bases satisfying \cref{eq:NstarKstar,eq:starmeasures}.
Again, we will \emph{not} (though we could in principle) demand
\(k=k^\star\).

\paragraph{In summary,} there is a \(\star\)-involution which acts on boundary
conditions \(j\) preparing a ket-state \(\ket{j}\) and returns boundary
conditions \(j^\star\) which prepare the corresponding bra-state \(\bra{j}\).
Since the latter boundary conditions can also be used to prepare a ket-state
\(\ket{j^\star}\), this is also a \(\star\)-involution on the Hilbert space
\(\mathcal{H}_{\mathrm{kin}}\). In particular, when spacetime (worldline)
configurations do not carry orientation, a spacetime (worldline) time-reversal
symmetry \(\mathsf{T}\) relates \(\ket{j^\star}=\mathsf{T}\ket{j}\). When
spacetime (worldline) configurations carry orientation, by including orientation
as a degree of freedom in the Hilbert space, we can define an antilinear
orientation-reversal symmetry \(\mathcal{T}\) relating
\(\ket{j^\star}=\mathcal{T}\ket{j}\).

In gravity, essentially the only difference between \(\mathsf{T}\) and
\(\mathcal{T}\) is that \(\mathsf{T}\) does not act on orientation while
\(\mathcal{T}\) does. For boundary conditions that prepare states in the
position basis \(\{\ket{x}:x\in\mathscr{M}\}\) by fixing the induced
configurations of ordinary tensor fields on the spacetime boundary (worldline
endpoint), time reversal acts trivially \(x^\star=\mathsf{T}x=x\) while
generically orientation reversal does not \(x^\star=\mathcal{T}x\ne x\). More
generally, however, the distinction between \(\mathsf{T}\) and \(\mathcal{T}\)
is actually somewhat arbitrary. In gravity, orientation is just another degree
of freedom and there is no reason to treat it as special. In fact, if we allow
pseudo-tensor fields, we can have matter fields that behave under time-reversal
\(\mathsf{T}\) similarly to how orientation behaves under \(\mathcal{T}\). For
example, in an unoriented\footnote{A nontrivial example of an unoriented theory
  is a gauge theory with a coupling
  \(\int\dd[4]{X}\sqrt{-g} a \epsilon^{\mu\nu\lambda\rho}\tr
  F_{\mu\nu}F_{\lambda\rho}\) to the pseudoscalar axion \(a\). Neither \(a\) nor
  \(\epsilon^{\mu\nu\lambda\rho}\) are well-defined and single-valued on an
  unoriented manifold, but
  \(\tilde{a}^{\mu\nu\lambda\rho} \equiv a \epsilon^{\mu\nu\lambda\rho}\) is.
  Once written in terms of this better-defined field \(\tilde{a}\), the theory
  is manifestly independent of orientation. Ref.~\cite{Witten:2025ayw} uses this
  particular theory to illustrate how a Hermitian inner product can be
  constructed using \(\mathsf{T}\). \label{foot:axion}} theory with an axion,
time reversal \(\mathsf{T}\) flips the sign of the axion and so
\(\mathsf{T}x\ne x\). The ambiguity between \(\mathsf{T}\) and \(\mathcal{T}\)
becomes even more acute if one views this axionic theory as coming from a
dimensional reduction of an oriented higher-dimensional gauge theory, where
\(\star\) is instead realized as an antilinear orientation reversal
\(\mathcal{T}\).\footnote{In the example of \cref{foot:axion}, if \(y\) denotes
  the compactified fifth dimension and \(a=A_y\), then
  \(\tilde{a}=\ast_5(A_y\,\dd y)\). Reversing the orientation used to define the
  five-dimensional Hodge dual \(\ast_5\) flips the sign of \(\tilde{a}\).}

In this paper, we will mostly write expressions in terms of \(\star\), without
specifying whether this refers to \(\mathsf{T}\) or to \(\mathcal{T}\).
Occasionally, we will discuss unoriented theories where induced field
configurations are generally time-reversal invariant, \(x^\star=\mathsf{T}x=x\),
which we expect to be the case for ordinary tensor fields. We expect
pseudo-tensors to more closely follow discussions of the oriented case where
\(x^\star\ne x\).

%% file: sections/section03_universeFieldTheory.tex
\section{Group-averaged baby universe field theory (GABUFT)}
\label{sec:uft}

In \cref{sec:worldline}, we developed gravitational worldline theories which
live on a single universe with fixed topology --- a worldline with the topology
of an interval, which one might view as representing the minisuperspace
reduction of a higher-dimensional spacetime with cylinder-like
\((\text{time interval})\times(\text{spatial topology})\) topology. With
appropriately chosen contours, the worldline path integral equips the theory
with a \emph{group-averaged} inner product. In this section, we will construct a
``\emph{field theory}'' which describes multiple universes (worldlines),
referred to as \emph{baby universes} following common terminology in the
literature.

Our group-averaged baby universe field theory (GABUFT) for gravitating
worldlines will be a special case of the general baby universe framework
described in relatively recent literature, starting with Marolf and Maxfield
\cite{Marolf:2020xie}. Although GABUFT is superficially analogous to the second
quantization that leads to QFT, we will see that there are important
differences, as already noted in ref.~\cite{Casali:2021ewu}. In this work, we
will aim to be a bit more explicit. In particular, we will construct GABUFT in a
manner that allows the closest possible comparison to older accounts
\cite{Coleman:1988cy,Giddings:1988cx,Giddings:1988wv} phrased in terms of
universe \emph{fields} and baby universe \emph{Fock spaces}. (A starting point
adopted by some of these works, \eg{} \cite{Giddings:1988cx,Giddings:1988wv}, is
an assumed direct analogy between universe field theory and QFT. Because the
term ``third quantization'' was coined in such a context, we will refrain from
calling our construction third quantization.)

In \cref{sec:topologygpi}, we will first present a gravitational path integral
that sums over topologies as the starting point for defining GABUFT. In
\cref{sec:gns}, we review the modern baby universe framework
\cite{Marolf:2020xie}, which amounts essentially to a GNS construction. The
brevity of that review comes at a somewhat steep cost of abstraction, so an
unfamiliar reader may wish to either consult the original literature or refer
immediately to \cref{sec:wick}, which presents a less abstract description of
the GABUFT. There, we begin developing the theory as one of universe field
operators whose Wick contractions correspond to worldline topologies in the
gravitational path integral. In \cref{sec:modeexpansion}, we describe the sense
in which these operators are ``fields'' that satisfy a differential equation of
motion in target space and can be expanded in terms of on-shell modes and
associated creation/annihilation operators. In
\cref{sec:wickuftvsqft,sec:modesuftvsqft}, we will highlight specific
differences between GABUFT and QFT that are immediately visible from our
construction of universe field theory (while leaving a more qualitative
comparison for discussion in \cref{sec:discussqftvsuft}). In
\cref{sec:ufthilbertspace}, we express the baby universe Hilbert space precisely
as a Fock space and re-express it also as a direct (path) integral over
one-dimensional superselection sectors known as \(\alpha\)-sectors.

\subsection{A gravitational path integral summing over worldlines}
\label{sec:topologygpi}

The underlying idea behind the baby universe framework is that the full
gravitational path integral should not only integrate over field configurations
on fixed spacetime topology, but also sum over different spacetime topologies.
The only smooth one-dimensional spacetimes connected to boundaries (endpoints)
are worldlines with the topology of an interval. Thus, in the theory of
worldline baby universes, where multiple boundaries source multiple universes,
we need only include a sum over different ways in which worldlines connect the
boundaries.\footnote{From the perspective of higher-dimensional spacetime, we are
  neglecting many nontrivial topology-changing processes by considering only a
  single connected spacetime topology,
  \((\text{time interval})\times(\text{spatial topology})\), and different ways
  this connects multiple boundaries. In \cref{sec:interactions}, we will discuss
  how nontrivial topology-changing processes might be included.}

\begin{figure}
  \centering
  \includegraphics[width=\textwidth]{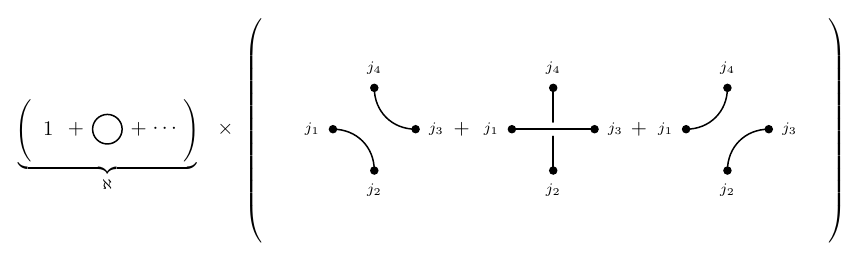}
  \caption{The sum over worldline topologies in
    \(\zeta(j_1\sqcup j_2 \sqcup j_3 \sqcup j_4)\). In
    \cref{sec:gns,sec:wick}, this gravitational path integral will be
    used to define an inner product
    \(\braket{j_4^\star}{j_1\sqcup j_2 \sqcup j_3}_{\mathrm{BU}}\) (as well as
    inner products with other permutations and groupings of \(j_1,\ldots,j_4\))
    in the baby universe Hilbert space.}
  \label{fig:topologygpi}
\end{figure}

In general, the gravitational path integral \(\zeta\) is a function which takes
as input an arbitrary list of boundary conditions and outputs a complex number.
Given everything said so far, we can write an explicit expression for the
gravitational path integral \(\zeta\) that sums over worldline topologies and
integrates over worldline fields \cite{Casali:2021ewu}:
\begin{align}
  \zeta(j_1\sqcup\cdots\sqcup j_m)
  &= \aleph \sum_{\substack{\text{pair partitions}\\\text{\(\pi\) of \(\{1,\ldots,m\}\)}}}
  \prod_{\{a,b\}\in\pi }
  \llangle j_a^\star | j_b \rrangle_{\mathrm{wl}}
  \;.
  \label{eq:topologygpi}
\end{align}
Let us now provide an explanation of the various elements of this formula,
accompanied by an illustration of the sum over topologies in \cref{fig:topologygpi}.

\begin{itemize}
\item{Each \(j_a\) denotes boundary conditions on a single connected piece of
    spacetime boundary, \ie{} a single endpoint. On the LHS,
    \(j_1\sqcup\cdots\sqcup j_m\) denotes boundary conditions on the disjoint
    union of \(m\) endpoints, obtained by imposing boundary conditions \(j_a\)
    on the \(a\)th endpoint.}
\item{Evaluating the path integral \(\zeta\) means summing over spacetime
    (worldline) configurations subject to these boundary conditions
    \(j_1\sqcup\dots\sqcup j_m\). This includes a sum over topologies.}
  \begin{itemize}
  \item{In principle, this might include topologies with pieces of spacetime not
      connected to \(j_1\sqcup\dots\sqcup j_m\), \eg{} closed circular
      worldlines. However, assuming locality of the spacetime (worldline)
      action, these pieces simply contribute a multiplicative constant
      \(\aleph\) to the path integral, by the same accounting as for vacuum
      bubbles in QFT. In the following, we will not need any details about \(\aleph\),
      beyond possibly assuming that it is real and positive:
      \begin{align}
        \aleph \equiv \zeta(\varnothing) > 0 \;.
        \label{eq:alephpos}
      \end{align}
      If \(\aleph\ne0\) but is not real and positive, we will still be able to
      proceed with our construction by replacing \(\zeta\) with the rescaled path integral
      \(\zeta'\equiv\zeta/\aleph\) for which
      \(\aleph'\equiv\zeta'(\varnothing)=1>0\).}
  \item{With \(\aleph\) factored or divided out, we need only sum over different
      ways in which worldline intervals can connect the endpoints, \ie{} sum
      over pair partitions of \(\{1,\ldots,m\}\). For each pair \(\{a,b\}\), the
      path integral on the worldline connecting boundary conditions \(j_a\) and
      \(j_b\) evaluates to a group-averaged inner product
      \(\llangle j_a^\star|j_b\rrangle_{\mathrm{wl}}\), as described in
      \cref{sec:worldline}. (To avoid confusion between states/inner products
      defined in the worldline theory versus those in the baby universe
      theory, we will append subscripts for disambiguation.)

      If \(m\) is odd, then it is not possible to connect all endpoints with
      line segments. The path integral \(\zeta\) then evaluates to zero in our
      current model. (In \cref{sec:interactions}, we will describe turning on
      one-point functions in universe field theory, in which case \(\zeta\) will
      remain nonzero for odd numbers of insertions.)}
  \end{itemize}
\end{itemize}

\subsection{The GNS construction of the baby universe Hilbert space \(\mathcal{H}_{\mathrm{BU}}\)}
\label{sec:gns}

Using the gravitational path integral \(\zeta\), we can construct a Hilbert
space \(\mathcal{H}_{\mathrm{BU}}\) for the baby universe theory. The
procedure for constructing this so-called ``baby universe'' Hilbert space
\(\mathcal{H}_{\mathrm{BU}}\) has already been repeated several times in modern
literature \cite{Marolf:2020xie,Casali:2021ewu,Marolf:2024jze,Chen:2025fwp}, so
we will provide only a formal sketch of it here by invoking a GNS construction.
While this modern baby universe framework and its general construction of the
baby universe Hilbert space may seem somewhat formal, in
\cref{sec:modeexpansion}, we will translate this construction into perhaps more
familiar field theory language.

The Hilbert space \(\mathcal{H}_{\mathrm{BU}}\) for the baby universe theory is
obtained from a GNS construction.

\paragraph{The input of the GNS construction} is the following data: a
\(\star\)-algebra and a positive linear functional on this algebra --- let us
first identify what these ingredients are in our context.

The \(\star\)-algebra of relevance to us is the set of possible boundary
conditions on boundaries with arbitrarily many connected components. At just a
single connected boundary component (an endpoint), we have a vector space
\(\mathcal{D}_{\text{wl test}}\) of test states \(\ket{j}_{\mathrm{wl}}\), \ie{}
boundary conditions \(j\), for the worldline theory. To promote this to an
algebra, we require a multiplication operation --- this is the disjoint union
\(\sqcup\) appearing in \cref{eq:topologygpi} which builds boundary conditions
on multiple boundary components. By allowing the addition of any two such
multi-boundary conditions, we form an algebra
\begin{align}
  \mathcal{J}
  &= \Span\{
    j_1 \sqcup \cdots \sqcup j_m \mid m\in\mathbb{Z}_{\ge 0}, j_i\in\mathcal{D}_{\text{wl test}}
    \}
\end{align}
comprised of polynomials in single-boundary conditions --- their meaning as an
input for the gravitational path integral will be specified by
\cref{eq:zetalinear} below. The algebra \(\mathcal{J}\) is a unital and
commutative algebra, because it has a \(\sqcup\)-multiplicative identity
\(\varnothing\) --- a ``no-boundary condition'' --- and \(\sqcup\) is a
commutative multiplication operation. Moreover, \(\mathcal{J}\) is a
\(\star\)-algebra equipped with a \(\star\)-involution acting on boundary
conditions, as described in \cref{sec:reversal}.

For us, the positive linear functional is given by the gravitational path
integral \(\zeta:\mathcal{J}\to\mathbb{C}\). The gravitational path integral
\labelcref{eq:topologygpi} is manifestly linear in the boundary conditions \(j\)
of each boundary component in a disjoint union; more generally for any
\(c\in\mathbb{C}\) and \(J,J'\in\mathcal{J}\) which themselves could be sums of
disjoint unions, we can define \(\zeta\) to act linearly:
\begin{align}
  \zeta(c J+J')
  &= c \zeta(J) + \zeta(J')
    \;.
    \label{eq:zetalinear}
\end{align}
The positivity
\begin{align}
  \zeta(J^\star\sqcup J)
  &\ge 0
    \label{eq:zetapos1}
\end{align}
of the linear functional \(\zeta\), with expression \labelcref{eq:topologygpi},
follows from the positivity of \(\aleph\) and the worldline group-averaged inner
product \(\llangle\bullet|\bullet\rrangle_{\mathrm{wl}}\). (This follows from
Wick's theorem or, equivalently, Isserlis's theorem which underlie the Fock
space and Gaussian integral constructions described in later subsections.)

\paragraph{The output of the GNS construction} is a Hilbert space
\(\mathcal{H}_{\mathrm{BU}}\), a \(\star\)-representation \(Z\) of the algebra
\(\mathcal{J}\) on \(\mathcal{H}_{\mathrm{BU}}\), and a cyclic vector
\(\ket{\varnothing}_{\mathrm{BU}}\in\mathcal{H}_{\mathrm{BU}}\). An explicit
construction of these objects in \cref{sec:wick,sec:modeexpansion} will clarify
their physical interpretation, but let us first explain what the preceding words
mean mathematically.

Having a cyclic state means that a dense subspace
\(\mathcal{D}_{\mathcal{J}}\subset\mathcal{H}_{\mathrm{BU}}\) of the Hilbert
space is generated by acting on \(\ket{\varnothing}_{\mathrm{BU}}\) with
operators, in this case \(Z(J):J\in\mathcal{J}\) --- let us write the resulting
states as
\begin{align}
  Z(J) \ket{\varnothing}_{\mathrm{BU}}
  &= \ket{J}_{\mathrm{BU}}
    \;,
    &
    Z(J) \ket{J'}_{\mathrm{BU}}
  &= \ket{J\sqcup J'}_{\mathrm{BU}}
    \;,
  &
    \mathcal{D}_{\mathcal{J}}
  &= \{\ket{J} \mid J\in\mathcal{J}\}
    \;.
    \label{eq:ketJstate}
\end{align}
The Hilbert space \(\mathcal{H}_{\mathrm{BU}}\) is obtained by completing
\(\mathcal{D}_{\mathcal{J}}\), with respect to an inner product given by the
positive linear functional \(\zeta\):
\begin{align}
  \braket{J'}{J}_{\mathrm{BU}}
  = \zeta((J')^\star \sqcup J)
  \;.
  \label{eq:buinnerprod}
\end{align}
For a GNS construction from a commutative \(\star\)-algebra \(\mathcal{J}\),
having a \(\star\)-representation \(Z\) means that the operators \(Z(J)\) are
defined on \(\mathcal{D}_{\mathcal{J}}\) (as described above), where they
satisfy
\begin{align}
  Z(J\sqcup J')
  &=Z(J)Z(J') = Z(J')Z(J)
    \;,
    \label{eq:commutativity}
  \\
    Z(J+J')
  &= Z(J) + Z(J')
    \;,
\end{align}
and
\begin{align}
  \braket{J^\star \sqcup J''}{J'}
  &= \braket{J''}{J\sqcup J'}
    \;,
    \label{eq:zhermitian}
\end{align}
\ie{} \(Z(J)\) is symmetric on \(\mathcal{D}_{\mathcal{J}}\) if
\(J^\star=J\).\footnote{Note that any \(J\in\mathcal{J}\) can be decomposed
  \(J=\Re J + i \Im J\) into \(\star\)-invariant ``real''
  \(\Re J \equiv \frac{J+J^\star}{2} = \left( \Re J \right)^\star\) and
  ``imaginary'' \(\Im J \equiv \frac{J-J^\star}{2i} = \left(\Im J\right)^\star\)
  pieces, so there is no loss in generality in restricting \cref{eq:zhermitian}
  to \(J=J^\star\). \label{foot:reimj}}

While initially only defined on \(\mathcal{D}_{\mathcal{J}}\), it is standard
practice \cite{Marolf:2020xie} to assume that these symmetric operators have
natural self-adjoint extensions,
\begin{align}
  Z(J)^\dagger
  &= Z(J)
    \;,
      &
  (J &= J^\star)
    \label{eq:zhermiticity}
\end{align}
that (strongly\footnote{Strong commutativity means all spectral projections of
  the operators commute. This is sometimes referred to as strong commutativity
  \cite{Schm_dgen_2012} (Proposition 5.27) and other times as just commutativity
  \cite{Samoilenko_1991}. We expect physical observables to correspond to
  \emph{bounded} functions of \(Z(J)\), which provides some physical motivation
  for focusing on spectral projections. As stated in
  \cref{sec:alphasector,foot:alphadiag}, strong commutativity also seems to be
  the right hypothesis mathematically for a decomposition of
  \(\mathcal{H}_{\mathrm{BU}}\) into \(\alpha\)-sectors.}) commute with each
other. In \cref{sec:wick,sec:modeexpansion}, we will at first write adjoints
\(\bullet^\dagger\) of operators without much regard for the domains on which
operators and their adjoints are defined. Once we have a better conceptual grasp
of the Hilbert space \(\mathcal{H}_{\mathrm{BU}}\), however, it will become
clear that the operators \(\eval{Z(J)}_{\mathcal{D}_{\mathcal{J}}}\) in our
worldline model are essentially self-adjoint for \(J=J^\star\) and the
self-adjoint extensions
commute (strongly) with each other --- see \cref{app:alphapathint}.

\subsection{Wick's theorem for universe creation and annihilation}
\label{sec:wick}

We will now rephrase the formal GNS construction of \cref{sec:gns} in a way that
more closely resembles the construction of QFT from particle worldline theory.
In particular, we will describe the operators \(Z(J)\) as creating or
annihilating universes (worldlines), analogous to how QFT operators create or
annihilate particle excitations. As we will describe in \cref{sec:wickuftvsqft},
however, this analogy between GABUFT and QFT is in
fact imperfect.

The natural state from which to launch this discussion is the cyclic state
\(\ket{\varnothing}_{\mathrm{BU}}\) of the baby universe theory. The analogue of
this state in QFT is the vacuum state with no particle excitations --- one can
similarly regard \(\ket{\varnothing}_{\mathrm{BU}}\) as the state with no
universes prepared by spacetime boundaries (worldline endpoints). Thus,
\(\ket{\varnothing}_{\mathrm{BU}}\) is often referred to as the Hartle--Hawking
state \cite{Hartle:1983ai}.

In QFT, states with particle excitations are obtained by acting on the vacuum
with field operators. Similarly, the operators \(Z(J)\) prepare universes on
spacetime boundaries (worldline endpoints) with boundary conditions \(J\). For
example, a universe (worldline) with an arbitrary wavefunction \(j(x)\) in the
worldline theory is prepared by boundary conditions \(J=j\) on one endpoint:
\begin{align}
  Z(j)\ket{\varnothing}_{\mathrm{BU}}=\ket{j}_{\mathrm{BU}}
  \;.
\end{align}
When computing inner products for such states ---
\(\braket{j_2}{j_1}_{\mathrm{BU}}\) defined by \labelcref{eq:buinnerprod}
--- the baby universe path integral \labelcref{eq:topologygpi} reduces to the
path integral on a worldline with boundary conditions \(j_1\) and \(j_2^\star\).
Consequently, the one-universe sector of \(\mathcal{H}_{\mathrm{BU}}\) spanned
by the above states can be identified with
\(\aleph^{1/2}\mathcal{H}_{\mathrm{(co)inv}}\), where
\(\mathcal{H}_{\mathrm{(co)inv}}\) is the worldline theory's (co)invariant
Hilbert space.

Analogous to multiparticle excitations in QFT, we can similarly consider more
general states
\begin{align}
  Z(j_1)\cdots Z(j_m)\ket{\varnothing}_{\mathrm{BU}}
  &= \ket{j_1 \sqcup \cdots \sqcup j_m}_{\mathrm{BU}}
    \label{eq:multiunivstate}
\end{align}
understood as \emph{initially} having \(m\)-many universes prepared by boundary
conditions \(j_1,\ldots,j_m\). In fact, the full Hilbert space
\(\mathcal{H}_{\mathrm{BU}}\) is (densely) spanned by such states for
\(m\in\mathbb{Z}_{\ge 0}\) --- recall by definition from the GNS construction
that \(\ket{\varnothing}_{\mathrm{BU}}\) is cyclic and \(Z\) is a
representation of \(\mathcal{J}=\{J\}\), which in turn is
generated by single-boundary components \(\mathcal{D}_{\text{wl test}}=\{j\}\).

The inner product
\(\braket{j_1' \sqcup \cdots \sqcup j_{m'}'}{j_1 \sqcup \cdots \sqcup
  j_m}_{\mathrm{BU}}\) between two such states will involve summing over
different ways to pair boundary components
\(j_1,\ldots,j_m,(j_1')^\star,\ldots, (j_{m'}')^\star\) using line segments, as
expressed in \cref{eq:topologygpi} and illustrated in \cref{fig:topologygpi}. In
particular, the universes initially prepared by \(j_1,\ldots,j_m\) can connect
amongst themselves (and similarly for \((j_1')^\star,\ldots, (j_{m'}')^\star\)),
allowing nonzero overlaps
\(\braket{j_1' \sqcup \cdots \sqcup j_{m'}'}{j_1 \sqcup \cdots \sqcup
  j_m}_{\mathrm{BU}}\) even for \(m\ne m'\).

\begin{figure}
  \centering
  \begin{subfigure}[t]{0.275\textwidth}
    \centering
    \includegraphics[width=\textwidth]{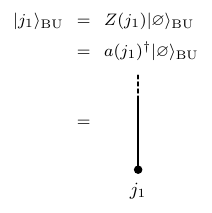}
    \caption{\(Z(j_1)\) acts on the Fock space vacuum
      \(\ket{\varnothing}_{\mathrm{BU}}\) to produce a one-universe state.}
    \label{fig:fockoneuniv}
  \end{subfigure}
  \hfill
  \begin{subfigure}[t]{0.675\textwidth} \centering
    \includegraphics[width=\textwidth]{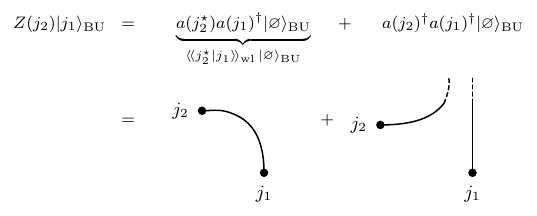}
    \caption{\(Z(j_2)\) acts on the one-universe state prepared in
      \subref{fig:fockoneuniv}. The operators \(Z(j_1)\) and \(Z(j_2)\)
      \emph{initially} prepare two universes with boundary conditions \(j_1\)
      and \(j_2\) respectively. These universes can pair-annihilate or both
      survive to later connect with other boundaries --- the two terms shown
      account for these two possibilities. Note that the LHS, a state
      \(Z(j_2)Z(j_1)\ket{\varnothing}_{\mathrm{BU}}\) with universes initially
      prepared on two boundaries, should not be confused with the state
      \(a(j_2)^\dagger a(j_1)^\dagger\ket{\varnothing}_{\mathrm{BU}}\) with two
      surviving universes.}
  \end{subfigure}
  \par\bigskip
  \begin{subfigure}[t]{\textwidth} \centering
    \includegraphics[width=\textwidth]{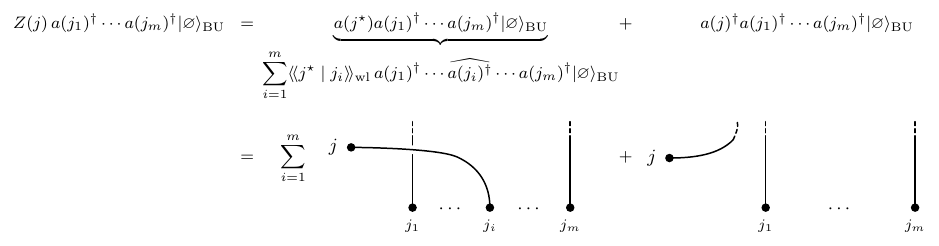}
    \caption{\(Z(j)\) acts on a state
      \(a(j_1)^\dagger \cdots a(j_m)^\dagger \ket{\varnothing}_{\mathrm{BU}}\)
      with \(m\)-many surviving universes. (In the underbrace expression, the
      hat denotes omission from the string of creation operators.)}
  \end{subfigure}
  \caption{The Fock space description of \(\mathcal{H}_{\mathrm{BU}}\) and the
    action of a boundary- (endpoint-)inserting operator
    \(Z(j)=a(j)^\dagger+a(j^\star)\).}
  \label{fig:fock}
\end{figure}

The sum over pairings of boundary components by line segments is reminiscent of
a sum over Feynman diagrams or Wick contractions in free QFT. As we now explain,
the fact that the only connected correlation functions of the operators \(Z(j)\)
are two-point functions implies that the operators \(Z(j)\) can be decomposed
into annihilation \(a(j^\star)\) and creation \(a(j)^\dagger\) operators
\begin{align}
  Z(j)
  &= a(j^\star) + a(j)^\dagger
    \;,
    \label{eq:zjaj}
\end{align}
for which the Hilbert space \(\mathcal{H}_{\mathrm{BU}}\) furnishes a Fock space
representation:
\begin{align}
  a(j) \ket{\varnothing}
  &= 0
    \;,
  &
  [a(j_1),a(j_2)^\dagger]
  &= \llangle j_1 | j_2 \rrangle_{\mathrm{wl}} \;.
    \label{eq:acommutator}
\end{align}
Note that \cref{eq:zjaj} is consistent with the Hermiticity condition
\labelcref{eq:zhermiticity} (formally\footnote{The precise statement is the
  following. Any \(j\) decomposes into ``\(\Re j\)'' and ``\(\Im j\)'' parts
  which are \(\star\)-invariant, as described in \cref{foot:reimj}. For
  \(j=j^\star\), we rigorously prove in \cref{app:multops} that \(Z(j)\) is
  self-adjoint \labelcref{eq:zhermiticity}. Taking complex linear combinations
  of such self-adjoint operators then \emph{formally} gives
  \cref{eq:zsmalljadjoint1} (without carefully specifying operator domains).
  \label{foot:zjadjoint}})
\begin{align}
  Z(j)^\dagger
  &= Z(j^\star)
    \;.
    \label{eq:zsmalljadjoint1}
\end{align}

To see that this reproduces the gravitational path integral
\labelcref{eq:topologygpi}, note that Wick's theorem gives
\begin{align}
  Z(j_1)
  \cdots
  Z(j_m)
  &= \sum_{n=0}^{\lfloor m/2 \rfloor}
    \sum_{\substack{\text{disjoint pairs} \\ \{a_1,b_1\},\ldots,\{a_n,b_n\}}}
  \wick{\c Z(j_{a_1}) \c Z(j_{b_1})}
  \cdots
  \wick{\c Z(j_{a_n}) \c Z(j_{b_n})}
  :\text{uncontracted \(Z(j)\)'s} :
  \label{eq:wick}
\end{align}
where \(:\bullet:\) denotes normal ordering --- placing all creation operators
\(a^\dagger\) to the left of all annihilation operators \(a\) --- and the Wick
contraction is given by
\begin{align}
  \wick{\c Z(j_1) \c Z(j_2)}
  &= Z(j_1) Z(j_2) - :Z(j_1) Z(j_2):
    = [a(j_1^\star),a(j_2)^\dagger]
    = \llangle j_1^\star | j_2 \rrangle_{\mathrm{wl}}
    \;.
    \label{eq:wickcontract}
\end{align}
Wick's theorem \labelcref{eq:wick} immediately implies that the Fock space
description recovers the correlation functions computed by the gravitational
path integral \labelcref{eq:topologygpi}:
\begin{align}
  \mel{\varnothing}{
  Z(j_1)
  \cdots
  Z(j_m)
  }{\varnothing}_{\mathrm{BU}}
  &= \zeta(j_1\sqcup \cdots \sqcup j_m)
    \;.
\end{align}
One can also see pictorially in \cref{fig:fock} how the action of \(Z(j)\) on
the Fock space \(\mathcal{H}_{\mathrm{BU}}\) iteratively produces the worldline
connections that appear in the path integral.

\subsubsection{Contrast with QFT}
\label{sec:wickuftvsqft}

From what we have said so far, it seems that universe field theory is
completely analogous to free QFT. But this analogy is imperfect
\cite{Casali:2021ewu}.

Above, we used a version of Wick's theorem applicable to strings of operators
\emph{ordered as written}, for which the Wick contraction
\labelcref{eq:wickcontract} is the Wightman function
\(\aleph^{-1}\mel{\varnothing}{Z(j_1)Z(j_2)}{\varnothing}_{\mathrm{BU}}\)
identified with the gravitational worldline amplitude
\(\llangle j_1^\star | j_2 \rrangle_{\mathrm{wl}}\). The same version of Wick's
theorem also holds in QFT and the QFT Wightman function can be expressed as a
worldline amplitude --- this is ``Approach 2'' in \cref{tab:qftvsuft}. However,
as described in \cref{app:qftwightman}, that amplitude does not seem to arise
naturally from a gravitational worldline path integral.

Instead, in QFT, one often considers the version of Wick's theorem that applies
to strings of \emph{time-ordered} operators, for which the appropriate Wick
contraction is given by the Feynman propagator, \ie{} time-ordered two-point
function. The latter can be identified with an amplitude
\(\mel{\bullet}{(iH+\epsilon)^{-1}}{\bullet}_{\mathrm{wl}}\) in a worldline
theory that is gravitational in the sense of including an integral over the
worldline lapse, as described around \cref{eq:feynman} --- this is ``Approach
1'' in \cref{tab:qftvsuft}.

In summary, there are two main differences in the construction of universe
field theory versus QFT from a gravitational worldline theory: (i) universe
field theory and QFT respectively identify their Wightman and Feynman
propagators with gravitational worldline amplitudes; (ii) the worldline
amplitudes used as input for universe field theory and QFT respectively are
the group-averaged inner product
\(\llangle \bullet | \bullet \rrangle_{\mathrm{wl}} =
\mel{\bullet}{2\pi\delta(H)}{\bullet}_{\mathrm{wl}}\) and
\(\mel{\bullet}{(iH+\epsilon)^{-1}}{\bullet}_{\mathrm{wl}}\), obtained from integrating the
total worldline lapse \(T=\int\dd{t} N(t)\) over \(\mathbb{R}\) and
\(\mathbb{R}_{>0}\).

\subsection{The universe field, its field equation, and its mode expansion}
\label{sec:modeexpansion}

To demonstrate why one might call this baby universe theory a ``field theory'',
let us now define an operator-valued distribution \(\Phi(x)\) from which the
single-boundary-inserting operators \(Z(j)\) can be obtained by smearing,
\begin{align}
  Z(j)
  &= \int_{\mathscr{M}} \mu(x) j(x) \Phi(x^\star)
    = \int_{\mathscr{M}} \mu(x) j(x^\star) \Phi(x)
    \;.
    \label{eq:zjzx}
\end{align}
With these conventions, \(\Phi(x^\star)\) inserts a boundary (endpoint)
preparing the state \(\ket{x}_{\mathrm{wl}}\) (\cf{} \cref{eq:jsmear}), while
\(\Phi(x)\) inserts \(\ket{x^\star}_{\mathrm{wl}}\) or equivalently
\(\bra{x}_{\mathrm{wl}}\). By analogy to QFT, where the field
\(\Phi_{\mathrm{QFT}}(x)\) measures a particle wavefunction by an overlap with
\(\bra{x}_{\mathrm{wl}}\), we therefore refer to \(\Phi(x)\) as the universe
field. We will see below that \(\Phi(x)\) satisfies a field equation of motion
\(H\Phi(x)=0\) given by the Wheeler--DeWitt Hamiltonian \(H\) acting as a
differential operator on \(\mathscr{M}\) (\eg{} like in \cref{eq:quantumH}).
While \(\Phi(x)\) admits an on-shell mode expansion with creation and
annihilation operators serving as coefficients, we will find that this differs
from the usual QFT mode expansion.

As in \cref{eq:zjaj}, let us decompose the universe field into annihilation and
creation pieces,
\begin{align}
  \Phi(x)
  &= a(x) + a(x^\star)^\dagger
    \;,
\end{align}
where the smeared annihilation (and creation) operators are recovered by
\begin{align}
  a(j)
  &= \int_{\mathscr{M}} \mu(x) j(x)^* a(x)
    \;.
\end{align}
To reproduce \cref{eq:acommutator}, the operator-valued distributions must
satisfy
\begin{align}
[a(x'),a(x)^\dagger]
  &= \llangle x' | x \rrangle_{\mathrm{wl}}
    = \int_{\mathscr{K}} \kappa(k) u_k(x') u_k(x)^* \;.
    \label{eq:axcom}
\end{align}
In the last equality, we used the mode expansion \cref{eq:etaexpansion} of the
group average \(\eta\).

Recall that these on-shell modes \(u_k(x)\) solve the Wheeler--DeWitt equation
\(H u_k(x)=0\), indicating a redundancy
\(|j\rrangle_{\mathrm{wl}}-|e^{-iT H} j\rrangle_{\mathrm{wl}}=0\) in
describing coinvariant worldline states by kinematic test functions \(j\).
Similarly, an annihilation operator that commutes with all creation operators
must vanish,
\begin{align}
  a((1-e^{-i T H})j)
  &=a(j)-a(e^{-iTH}j)=0
  \;,
\end{align}
and so
\begin{align}
  H a(x)
  &= 0
    \;.
\end{align}
In fact, \cref{eq:axcom} tells us that \(a(x)\) admits an on-shell mode
expansion
\begin{align}
  a(x)
  &= \int_{\mathscr{K}} \kappa(k) a_k u_k(x)
    \;,
\end{align}
where the non-redundant content lies in the coefficients \(a_k\), which satisfy
a positive-definite (as opposed to positive-semidefinite) commutation relation
with their creation counterparts,
\begin{align}
  [a_k, a_{k'}^\dagger]
  &= \delta_\kappa(k,k')
    \;.
    \label{eq:aadaggercommut}
\end{align}

Altogether, we find that the universe field \(\Phi(x)\) also admits an on-shell
mode decomposition,
\begin{align}
  \Phi(x)
  &= \int_{\mathscr{K}} \kappa(k) (a_k u_k(x) + a_k^\dagger u_k(x^\star)^*)
    \label{eq:phiexpansion1}
  \\
  &= \int_{\mathscr{K}} \kappa(k) (a_k + a_{k^\star}^\dagger) u_k(x)
    = \int_{\mathscr{K}} \kappa(k) \phi(k) u_k(x)
      \label{eq:phiexpansion2}
\end{align}
where, for later reference, we have named the coefficients
\begin{align}
  \phi(k)
  &= a_k + a_{k^\star}^\dagger
    \;.
    \label{eq:phik}
\end{align}
From the above expressions, it is clear that both the universe field \(\Phi\)
and these coefficients satisfy the Hermiticity condition
\labelcref{eq:zhermiticity},
\begin{align}
  \Phi(x)^\dagger
  &= \Phi(x^\star)
    \;,
  &
    \phi(k)^\dagger
  &= \phi(k^\star)
    \;.
    \label{eq:phihermiticity}
\end{align}
It is also clear that \(\Phi(x)\) satisfies the field equation of motion,
\begin{align}
  H \Phi(x)
  &= 0
    \;,
    \label{eq:ufteom1}
\end{align}
again, indicating a redundancy
\begin{align}
  Z((1-e^{-i T H})j)
  &= Z(j)-Z(e^{-iTH}j)=0
\end{align}
in parameterizing operators by kinematic test functions \(j\).

In general, we see that there is a one-to-one correspondence between nonzero
coinvariant worldline states in \(\mathcal{H}_{\text{wl co}}\) and nonzero
annihilation, creation, or single-boundary-inserting operators of the baby
universe theory,
\begin{align}
  \llangle j | j \rrangle_{\mathrm{wl}}
  &= 0
  &
  &\iff
  &
    a(j)
    &= 0
  &
  &\iff
  &
    Z(j)
  &= 0
    \;.
    \label{eq:nullz}
\end{align}
Equivalently, the baby universe operators can be expressed entirely in terms of
expansion coefficients \(a_k\) and \(\phi(k)\) associated with on-shell modes
spanning \(\mathcal{H}_{\text{wl inv}}\),
\begin{align}
  a(j)
  &= \int_{\mathscr{K}} \kappa(k)
    a_k
    \braket{j}{k}_{\mathrm{wl}}
    \\
  Z(j)
  &= \int_{\mathscr{K}} \kappa(k)
    \phi(k)
    \braket{j^\star}{k}_{\mathrm{wl}}
    \;.
    \label{eq:zphi}
\end{align}

\subsubsection{Contrast with QFT}
\label{sec:modesuftvsqft}

If we ignore the \(\star\) appearing on \(x^\star\) in \cref{eq:phiexpansion1}
--- which we can rightfully do if the worldline theory is unoriented (and
involves only ordinary tensor fields), as described in \cref{sec:timerev} ---
then this equation looks deceptively similar to the mode expansion of a bosonic
field in QFT.

However, an important difference is that every on-shell mode
\(u_k:k\in\mathscr{K}\) appears in the mode expansion of the universe field
multiplying \emph{both} an annihilation operator \(a\) \emph{and} a creation
operator \(a^\dagger\). One consequence of this is that all coefficients
\(\phi(k)=a_k+a_{k^\star}^\dagger\) appearing in the mode expansion
\labelcref{eq:phiexpansion2} mutually commute,
\begin{align}
  [\phi(k), \phi(k')]
  &= 0
    \qquad \forall k,k'\in\mathscr{K}
    \;.
\end{align}
Consequently, all operators that can be built from these basic ingredients also
commute:
\begin{align}
  [\Phi(x), \Phi(x')]
  &= 0
    \qquad
    \forall x,x'\in\mathscr{M}
    \;,
    \label{eq:phicommutes}
  \\
  [Z(j), Z(j')]
  &= 0
    \qquad
    \forall \ket{j},\ket{j'}\in\mathcal{D}_{\text{wl test}}
    \;,
  \\
  [Z(J), Z(J')]
  &= 0
    \qquad
    \forall J,J'\in\mathcal{J}
    \;.
\end{align}
Of course, we should have anticipated this from the GNS construction
reviewed in \cref{sec:gns}: \(Z\) is a representation of the \emph{commutative}
algebra \(\mathcal{J}\) of boundary conditions. More fundamentally, this
commutativity follows from the invariance of the gravitational path integral
under permutations of boundary components in disjoint unions (see \eg{}
\cref{eq:topologygpi}).

In contrast to the universe field \labelcref{eq:phiexpansion1}, the mode
expansion of a real bosonic field in QFT takes the form
\begin{align}
  \Phi_{\mathrm{QFT}}(x)
  &= \int_{\mathscr{K}_+} \kappa(k) (a_k u_k(x) + a_k^\dagger u_k(x)^*)
    \;,
    \label{eq:qftmodeexpansion}
\end{align}
where distinct roles are played by positive-frequency modes
\(\{u_k\mid k\in\mathscr{K}_+\}\) and negative-frequency modes
\(\{u_k\mid k\in\mathscr{K}_-\}=\{(u_k)^*\mid k\in\mathscr{K}_+\}\). For
example, in a Minkowski target spacetime \(\mathscr{M}\ni x\) (and a trivial
potential), the set \(\mathscr{K}\) of all on-shell momenta \(k\) of a massive
field is divided into mutually disconnected hyperboloids
\(\mathscr{K}_+=\{k\in\mathscr{K}\mid k^0 > 0\}\) and
\(\mathscr{K}_-=\{k\in\mathscr{K}\mid k^0 < 0\}\). Annihilation operators \(a\)
are coefficients exclusively for positive-frequency \(k^0>0\) modes, while
creation operators \(a^\dagger\) are coefficients for negative-frequency
\(k^0<0\) modes. In non-stationary settings, there can be many ways to separate
the field \(\Phi_{\mathrm{QFT}}\) into ``positive-frequency'' annihilation and
``negative-frequency'' creation parts,\footnote{Even in the non-stationary
  setting, one does not have \emph{complete} freedom to select ``positive- and
  negative-frequency'' solution spaces arbitrarily. In particular, positivity of
  the QFT inner product requires \([a_k,a_{k'}^\dagger]\) to be a
  positive-definite matrix in \(k,k'\in\mathscr{K}_+\), which is equivalent
  under canonical quantization to the positive- (negative-)definiteness of the
  Klein--Gordon inner product on positive- (negative-)frequency solutions. If
  \(\overline{\Span\{u_k\mid k\in\mathscr{K}_+\}}\) and
  \(\overline{\Span\{u_k\mid k\in\mathscr{K}_-\}}\) are one valid choice of
  positive- and negative-frequency solution spaces, it follows that
  \(\mathscr{K}_+'=\mathscr{K}_-\) and \(\mathscr{K}_-'=\mathscr{K}_+\) is an
  invalid choice. Thus, positivity of the QFT inner product breaks the
  interchangeability of positive- and negative-frequency modes to some degree
  even in non-stationary settings.} corresponding to different choices of the
vacuum; but in any case, each mode is assigned a creation or annihilation
operator, not both. The creation and annihilation operators still satisfy the
same commutation relations \labelcref{eq:aadaggercommut} as in GABUFT, because
the Klein--Gordon inner product on the QFT one-particle Hilbert space is equal to
the inner product \(\lsem\bullet|\bullet\rsem_{\mathrm{wl}}\) on the
positive-frequency subsector
\(L^2(\mathscr{K}_+,\kappa)\subset\mathcal{H}_{\text{wl inv}}\) --- see
\cref{app:qftwightman}. However, an immediate consequence of the difference in
mode expansion is that the QFT field \(\Phi_{\mathrm{QFT}}(x)\) evaluated at
different spacetime points need not commute like in GABUFT, and indeed it does
not commute at timelike separation.

We will review shortly in \cref{sec:alphasector} how the commutativity of the
universe field operators leads to a decomposition of the baby universe Hilbert
space \(\mathcal{H}_{\mathrm{BU}}\) into one-dimensional superselection sectors
known as \(\alpha\)-sectors --- something which does \emph{not} happen in QFT.

\subsection{Revisiting the baby universe Hilbert space
  \(\mathcal{H}_{\mathrm{BU}}\) and \(\alpha\)-sectors}
\label{sec:ufthilbertspace}

Having examined the operators of universe field theory, let us revisit the
baby universe Hilbert space \(\mathcal{H}_{\mathrm{BU}}\) on which these
operators act.

\subsubsection{\(\mathcal{H}_{\mathrm{BU}}\) as a Fock space}
\label{sec:fockspace}

As described in \cref{sec:wick}, one description of
\(\mathcal{H}_{\mathrm{BU}}\) is as a Fock space built from the cyclic
``vacuum'' state \(\ket{\varnothing}_{\mathrm{BU}}\) by acting with
single-boundary-inserting operators \(Z(j)\) (which then generate the algebra of
multi-boundary-inserting operators \(Z(J)\)). We can now be a bit more explicit.
In \cref{eq:nullz}, we concluded that the baby universe operators \(Z(j)\) are
in one-to-one correspondence with states of the worldline theory's physical,
coinvariant Hilbert space \(\mathcal{H}_{\text{wl co}}\); relatedly, in
\cref{eq:zphi}, we expressed these operators as linear combinations of mode
coefficients \(\phi(k)=a_k+a_{k^\star}^\dagger\) associated with on-shell modes
\(u_k\) spanning the invariant Hilbert space \(\mathcal{H}_{\text{wl inv}}\).
The GABUFT Hilbert space can therefore be written explicitly as the symmetric
Fock space over \(\mathcal{H}_{\text{wl co}}\cong \mathcal{H}_{\text{wl inv}}\),
\begin{align}
  \mathcal{H}_{\mathrm{BU}}
  &= \aleph^{1/2}
    \overline{
    \bigoplus_{i=0}^\infty
    \Sym^i(\mathcal{H}_{\text{wl co/inv}})
    }
    \;,
    \label{eq:hbufock}
\end{align}
where, as before, \(\aleph=\braket{\varnothing}{\varnothing}_{\mathrm{BU}}\) is
the proportionality constant relating the baby universe and worldline inner
products in \cref{eq:topologygpi} and the overline indicates a completion, in
this case in norm-topology.

We can carefully check that our answer \labelcref{eq:hbufock} is precisely equal
to the GNS definition of \(\mathcal{H}_{\mathrm{BU}}\) described in
\cref{sec:gns}. By comparing inner products, we readily see that
\begin{align}
  \mathcal{D}_{\mathcal{J}}
  &= \aleph^{1/2} \bigoplus_{i=0}^\infty \Sym^i\left( \frac{\mathcal{D}_{\text{wl test}}}{\ker\eta} \right)
\end{align}
is the dense subspace defined in \cref{sec:gns}, on which the operators
\(Z(J)\), which are sums and products of creation and annihilation operators,
are certainly well-defined. In particular, the above equality is free from any
subtleties associated with trying to extend operators beyond this domain. Taking
the closure then yields \cref{eq:hbufock}.

\paragraph{Contrast with QFT.} Although the Fock space description
\labelcref{eq:hbufock} of \(\mathcal{H}_{\mathrm{BU}}\) evokes an analogy with
the Fock space description of a free bosonic QFT, we should again note that this
analogy is imperfect. In \cref{eq:hbufock}, the invariant Hilbert space of the
worldline theory \(\mathcal{H}_{\text{wl inv}}=L^2(\mathscr{K},\kappa)\) is
spanned by all on-shell modes \(u_k(x)=\braket{x}{k} : k\in\mathscr{K}\). In
contrast, the QFT Hilbert space for a free bosonic field
\labelcref{eq:qftmodeexpansion} is a Fock space
\begin{align}
  \overline{\bigoplus_{i=0}^\infty \Sym^i(L^2(\mathscr{K}_+,\kappa))}
\end{align}
built from only those on-shell modes
\(u_k(x)=\braket{x}{k} : k\in\mathscr{K}_+\subset\mathscr{K}\) that are
``positive-frequency''.

\subsubsection{\(\mathcal{H}_{\mathrm{BU}}\) as a path integral over \(\alpha\)-sectors}
\label{sec:alphasector}

The contrast between universe field theory and QFT becomes even starker when we
consider the structure implied by the \emph{commutative} algebra of operators
\(Z(J)\) in universe field theory.

Let us first review these ideas which arise generally from the baby universe
framework \cite{Marolf:2020xie} described in \cref{sec:gns}. Provided that the
operators \(Z(J)\) for \(J=J^\star\) have natural self-adjoint extensions that
(strongly) commute, these operators can be simultaneously
diagonalized.\footnote{For finite sets of operators, this is
  ref.~\cite{Schm_dgen_2012}'s Theorem 5.23; for countable sets of operators,
  this is ref.~\cite{Samoilenko_1991}'s Theorem 1. Perhaps a similar result for
  uncountable sets also exists. Regardless, the simplicity of our worldline
  GABUFT will allow us to write down an \(\alpha\)-sector decomposition of
  \(\mathcal{H}_{\mathrm{BU}}\) explicitly below. \label{foot:alphadiag}} The
Hilbert space thus decomposes formally,
\begin{align}
  \mathcal{H}_{\mathrm{BU}}
  &= \int_{\mathcal{A}}^\oplus \dd{\alpha} \mathcal{H}_{\mathrm{BU}}^\alpha
    \;,
    \label{eq:alphadecomp}
\end{align}
into simultaneous eigenspaces \(\mathcal{H}_{\mathrm{BU}}^\alpha\) labelled by a
parameter \(\alpha\) where operators \(Z(J)\) take definite eigenvalues
\(Z_\alpha(J)\),
\begin{align}
  Z(J) \ket{\alpha}_{\mathrm{BU}}
  &= Z_\alpha(J) \ket{\alpha}_{\mathrm{BU}}
    \;,
  &
    \ket{\alpha}_{\mathrm{BU}}\in\mathcal{H}_{\mathrm{BU}}^\alpha
    \;.
\end{align}
In \cref{eq:alphadecomp}, \(\mathcal{A}\ni\alpha\) denotes the joint spectrum of
\(Z(J):J\in\mathcal{J}\). The eigenspaces \(\mathcal{H}_{\mathrm{BU}}^\alpha\)
are known as \(\alpha\)-sectors, and they have two notable properties:
\begin{enumerate}
\item{By construction, \(\alpha\)-sectors \(\mathcal{H}_{\mathrm{BU}}^\alpha\)
    are superselection sectors for the operators \(Z(J):J\in\mathcal{J}\)
    (meaning these operators preserve each
    \(\mathcal{H}_{\mathrm{BU}}^\alpha\)). There is a question of whether this
    algebra \(\{Z(J)\mid J\in\mathcal{J}\}\) of boundary-inserting operators
    constitutes the full set of physically interesting operators, but if it does,
    then one might regard each superselection sector
    \(\mathcal{H}_{\mathrm{BU}}^\alpha\) as describing an independent theory of
    quantum gravity.}
\item{Each \(\alpha\)-sector \(\mathcal{H}_{\mathrm{BU}}^\alpha\) is
    one-dimensional. To see this, note that the overlap
    \(\braket{J}{\alpha}=\mel{\varnothing}{Z(J)}{\alpha}=Z_\alpha(J)\braket{\varnothing}{\alpha}\)
    of an \(\alpha\)-state \(\ket{\alpha}\in\mathcal{H}_{\mathrm{BU}}^\alpha\)
    with any state in the dense subspace
    \(\{\ket{J}\mid J\in\mathcal{J}\}\subset\mathcal{H}_{\mathrm{BU}}\) is determined up to a
    \(J\)-independent number \(\braket{\varnothing}{\alpha}\). It follows that
    any two \(\alpha\)-states in the same \(\alpha\)-sector
    \(\mathcal{H}_{\mathrm{BU}}^\alpha\) are proportional to each other.}
\end{enumerate}
In summary, the baby universe Hilbert space \(\mathcal{H}_{\mathrm{BU}}\)
decomposes into \(\alpha\)-sectors \(\mathcal{H}_{\mathrm{BU}}^\alpha\) which
are one-dimensional superselection sectors.\footnote{Each
  \(\mathcal{H}_{\mathrm{BU}}^\alpha\) being one-dimensional is the general
  sense in which closed universes have one-dimensional Hilbert spaces --- a
  statement which has seen reinvigorated attention in recent literature
  \cite{Marolf:2020xie, Usatyuk:2024mzs, Usatyuk:2024isz, Harlow:2025pvj,
    Abdalla:2025gzn, Akers:2025ahe, Blommaert:2025bgd, Chen:2025fwp,
    Nomura:2025whc, Wei:2025guh, Higginbotham:2025dvf, Antonini:2025ioh,
    Higginbotham:2025clp, Harlow:2026hky, Nomura:2026igt}. In the special case
  where \(\alpha\) is unique (which might be the generic case for gravity in
  higher dimensions \cite{McNamara:2020uza}),
  \(\mathcal{H}_{\mathrm{BU}}=\mathcal{H}_{\mathrm{BU}}^\alpha\) is also
  one-dimensional.}

Our relatively simple model of worldline gravity allows us to describe its
\(\alpha\)-sectors slightly more explicitly.

As already mentioned, arbitrary multi-boundary-inserting operators \(Z(J)\) can
be generated by single-boundary-inserting operators \(Z(j)\), and the latter are
given by linear combinations \labelcref{eq:zphi} of universe field mode
coefficients \(\phi(k)=a_k + a_{k^\star}^\dagger\). Restricting to an
\(\alpha\)-sector \(\mathcal{H}_{\mathrm{BU}}^\alpha\) is then equivalent to
restricting to a simultaneous eigenspace where \(\phi(k): k\in\mathscr{K}\) take
definite eigenvalues \(\phi_\alpha(k)\):
\begin{align}
  \phi(k)\ket{\alpha}_{\mathrm{BU}}
  &= \phi_\alpha(k)\ket{\alpha}_{\mathrm{BU}}
    \;.
    \label{eq:phialpha}
\end{align}
Each element \(\alpha\in\mathcal{A}\) thus corresponds to a function --- more
precisely, distribution --- \(\phi_\alpha(k)\) of \(k\) satisfying the reality condition
\begin{align}
  \phi_\alpha(k)^* = \phi_\alpha(k^\star)
\end{align}
implied by the analogous condition \labelcref{eq:phihermiticity} on \(\phi(k)\).

To make practical sense of the formal decomposition \labelcref{eq:alphadecomp},
we can insert a resolution of the identity,
\begin{align}
  1 &= \int_{\mathcal{A}} \dd{\alpha} \ketbra{\alpha}{\alpha}_{\mathrm{BU}}
      \;,
\end{align}
to express correlation functions,
\begin{align}
  \mel{\varnothing}{\phi(k_1)\cdots\phi(k_m)}{\varnothing}_{\mathrm{BU}}
  &= \int_{\mathcal{A}} \dd{\alpha}
    \abs{\braket{\alpha}{\varnothing}_{\mathrm{BU}}}^2
    \phi_\alpha(k_1)\cdots\phi_\alpha(k_m)
  \\
  &\equiv \int_{\mathcal{A}} \mathcal{D}\phi_\alpha
    \,
    e^{-I[\phi_\alpha]}
    \,
    \phi_\alpha(k_1)\cdots\phi_\alpha(k_m)
    \;,
    \label{eq:alphapathint}
\end{align}
as path integrals of \(\phi_\alpha(k)\). This path integral is weighted by the
squared vacuum wavefunction
\(\abs{\braket{\alpha}{\varnothing}_{\mathrm{BU}}}^2\) of decoupled harmonic
oscillators \(\phi(k)=a_k + a_{k^\star}^\dagger\) labelled by \(k\):
\begin{align}
  \mathcal{D}\phi_\alpha
    \,
  e^{-I[\phi_\alpha]}
  &\equiv
  \dd{\alpha}
  \abs{\braket{\alpha}{\varnothing}_{\mathrm{BU}}}^2
    \;.
\end{align}
In particular, setting
\begin{align}
    I[\phi_\alpha]
  &= \frac{1}{2} \int_{\mathscr{K}} \kappa(k) \phi_\alpha(k)^* \phi_\alpha(k)
    \;.
    \label{eq:alphaaction}
\end{align}
and normalizing\footnote{In physicist's language, the
  ``\(\mathcal{D}\phi_\alpha\) part of the measure'' is independent of
  \(\phi_\alpha\), so \cref{eq:alphaaction,eq:measurenorm} completely determine
  the Gaussian measure \(\mathcal{D}\phi_\alpha \, e^{-I[\phi_\alpha]}\). (But,
  more precisely, it is really only the combination
  \(\mathcal{D}\phi_\alpha\,e^{-I[\phi_\alpha]}\) that gives a well-defined
  measure, as described in \cref{foot:precisemeas}.)}
\begin{align}
  \aleph
  &= \int_{\mathcal{A}} \mathcal{D}\phi_\alpha
    \,
    e^{-I[\phi_\alpha]}
    \;,
    \label{eq:measurenorm}
\end{align}
the Gaussian moment path integral \labelcref{eq:alphapathint} correctly
reproduces the correlation functions of the gravitational path integral
\labelcref{eq:topologygpi} by Isserlis's theorem. Readers interested in
mathematical details of the infinite-dimensional Gaussian path integral may wish to
consult \cref{app:alphal2}, where the space \(\mathcal{A}\) of distributions
\(\phi_\alpha(k)\) and the path integral over \(\mathcal{A}\) are defined more
precisely.

Coming back to the Hilbert space, we realize what we meant by the formal
\(\alpha\)-sector decomposition \labelcref{eq:alphadecomp} is more precisely
that
\begin{align}
  \mathcal{H}_{\mathrm{BU}}
  &= L^2\left(
    \mathcal{A}, \dd{\alpha} \abs{\braket{\alpha}{\varnothing}_{\mathrm{BU}}}^2
    \right)
    = L^2\left(
    \mathcal{A}, \mathcal{D}\phi_\alpha \, e^{-I[\phi_\alpha]}
    \right)
    \;.
    \label{eq:alphal2}
\end{align}
In other words, states are wave-functionals of \(\phi_\alpha(k)\) and inner products
are given by path integrals over \(\mathcal{A}\) with the specified measure,
\begin{align}
  \braket{\Psi'}{\Psi}_{\mathrm{BU}}
  &= \int_{\mathcal{A}} \mathcal{D}\phi_\alpha
    \,
    e^{-I[\phi_\alpha]}
    \,
    (\Psi'[\phi_\alpha])^* \Psi[\phi_\alpha]
    \;.
\end{align}
In this language, the dense subspace \(\mathcal{D}_{\mathcal{J}}\)
defined in the GNS construction of \cref{sec:gns} is spanned by
wave-functionals of \(\phi_\alpha\),
\begin{align}
  \ket{\varnothing}_{\mathrm{BU}}
  &\leftrightarrow
    1
    \;,
  \\
  \ket{j}_{\mathrm{BU}}
  &\leftrightarrow
    Z_\alpha(j)
    = \int_{\mathscr{K}} \kappa(k) \phi_\alpha(k) \braket{j^\star}{k}_{\mathrm{wl}}
    \;,
  &
  &(j\in\mathcal{D}_{\text{wl test}})
   \\
  \ket{j_1\sqcup \cdots \sqcup j_m}_{\mathrm{BU}}
  &\leftrightarrow
    Z_\alpha(j_1) \cdots Z_\alpha(j_m)
    \;.
  &
  &(j_i\in\mathcal{D}_{\text{wl test}})
\end{align}
In particular, we see that the operators \(Z(j):\ket{J}\mapsto\ket{j\sqcup J}\)
initially defined on this domain \(\mathcal{D}_{\mathcal{J}}\) correspond to
multiplication by the functional \(Z_\alpha(j)\) of \(\phi_\alpha\) written
above. The universe field \(\Phi(x)\) at each \(\alpha\) likewise takes on a
definite on-shell configuration
\begin{align}
  \Phi_\alpha(x)
  &= \int_{\mathscr{K}} \kappa(k) \phi_\alpha(k) u_k(x)
    \;.
\end{align}

If we did not want to assume anything about extensions of operators beyond
\(\mathcal{D}_{\mathcal{J}}\) to begin with, we could have alternatively
presented this section in reverse. \Cref{app:alphapathint} embellishes this
story with some mathematical details. By comparing overlaps within the dense
subspace \(\mathcal{D}_{\mathcal{J}}\), we first prove that \cref{eq:alphal2} is
equal to \cref{sec:gns}'s GNS construction of the baby universe Hilbert space
\(\mathcal{H}_{\mathrm{BU}}\). Then making use of this path integral
description, we prove that \(Z(J)\) for \(J=J^\star\in\mathcal{J}\) are
essentially self-adjoint once defined on the dense core
\(\mathcal{D}_{\mathcal{J}}\), with the self-adjoint extensions being maximally
defined multiplication operators that strongly commute.

\paragraph{Contrast with the QFT path integral.} There are two main differences
when comparing the path integral over \(\alpha\)-sectors in GABUFT to the path
integral of a QFT: the configurations integrated over, and the Hilbert space
interpretation of the path integral.

As described above, the GABUFT path integral over \(\alpha\)-sectors is
restricted to on-shell configurations, \(H \Phi_\alpha(x)=0\), of the universe
field \(\Phi(x)\). On the other hand, the QFT path integral also includes
off-shell configurations. This stems from the difference in gravitational
worldline amplitudes,
\(\llangle \bullet | \bullet \rrangle_{\mathrm{wl}} =
\mel{\bullet}{2\pi\delta(H)}{\bullet}_{\mathrm{wl}}\) and
\(\mel{\bullet}{(iH+\epsilon)^{-1}}{\bullet}_{\mathrm{wl}}\), used as input for
constructing GABUFT and QFT respectively, as described in
\cref{sec:wickuftvsqft}. In fact, both the GABUFT and QFT path integrals can be
described as Gaussian path integrals with covariance given by these respective
amplitudes. For GABUFT, regulating the \(\delta\)-function as in
\cref{eq:etaisfeynplusafeyn}, one might initially write down an off-shell path
integral with some insertions \(\cdots\) as
\begin{align}
  &\propto \int \mathcal{D}\Phi \exp\left(
    -\frac{1}{4\epsilon} \int_{\mathscr{M}}\mu(x) \Phi(x) (H^2+\epsilon^2) \Phi(x)
    \right)
    \cdots
\end{align}
but this localizes to on-shell configurations in the \(\epsilon\to 0\) limit and
reduces to \cref{eq:alphapathint}. For QFT, we have
\begin{align}
  &\propto \int \mathcal{D}\Phi_{\mathrm{QFT}} \exp\left(
    -\frac{1}{2}\int_{\mathscr{M}}\mu(x) \Phi_{\mathrm{QFT}}(x) (i H+\epsilon) \Phi_{\mathrm{QFT}}(x)
    \right)
    \cdots
    \;.
    \label{eq:qftpathint}
\end{align}

These path integrals also have very different Hilbert space interpretations
under GABUFT and QFT. As emphasized in \cite{Casali:2021ewu}, the baby universe
framework regards its path integral as one might in classical statistical
mechanics: each field configuration \(\Phi_\alpha(x)\) over \(\mathscr{M}\ni x\)
included in the path integral corresponds to one state (an \(\alpha\)-state
\(\ket{\alpha}\)) and the path integral is just computing a classical
statistical average over an ensemble of such states. To have a standard
statistical interpretation, the weighted measure
\(\mathcal{D}\phi_\alpha\,e^{-I[\phi_\alpha]}\) of the path integral must be
positive, and indeed, we see from \cref{eq:alphaaction} that this is equivalent
to the positivity of the inner product in the baby universe Hilbert space. In
contrast, each configuration in a QFT path integral does not correspond to a QFT
state. Instead, one might view QFT states as specifying boundary conditions for
the path integral, say on some codimension-one time slice through
\(\mathscr{M}\) or in the asymptotic past/future. (But, as we will discuss in
\cref{sec:discussqftvsuft}, many reasonable worldline models of gravity don't
seem to have reasonable descriptions as standard QFTs, partly because of issues
related to the notion of ``time'' in \(\mathscr{M}\).)

%% file: sections/section04_discussion.tex
\section{Discussion}

In summary, we have given a field-theoretic description of the baby universe
framework applied to a group-averaged theory of one-dimensional gravity.

We first described the single-universe worldline theory in \cref{sec:worldline}.
The key feature of this theory is a group-averaged inner product
\(\llangle x'|x\rrangle_{\mathrm{wl}}=\mel{x'}{2\pi\delta(H)}{x}_{\mathrm{wl}}\)
computed by the gravitational path integral between two fixed endpoints \(x\)
and \(x^{\prime\star}\) in target space \(\mathscr{M}\). This inner product
leads to a construction of the physical worldline Hilbert space
\(\mathcal{H}_{\text{wl co}}\cong\mathcal{H}_{\text{wl inv}}\) of either
coinvariant states or invariant states annihilated by the Wheeler--DeWitt
Hamiltonian \(H\).

Using this group-averaged worldline theory as input, we then defined in
\cref{sec:uft} a corresponding baby universe field theory (GABUFT) by summing
over worldline topologies pairing arbitrary numbers of boundaries (endpoints).
In this theory, boundary-inserting operators that prepare test states \(\bra{j}\),
\begin{align}
  Z(j^\star)
  &= \int_{\mathscr{M}} \mu(x) j(x)^* \Phi(x)
    \;,
    \label{eq:zsmearing}
\end{align}
are obtained by smearing test wavefunctions \(j\) against a universe field
operator \(\Phi(x)\). Similar to QFT, \(\Phi(x)\) obeys the field equation of motion
\(H \Phi(x)=0\) corresponding to the worldline Wheeler--DeWitt constraint. But in
contrast to QFT, the mode expansion of \(\Phi(x)\) involves each on-shell mode
being multiplied by \emph{both} universe creation \emph{and} annihilation
operators. This allows universe field operators at all target space points to
mutually commute. Consequently, the baby universe Hilbert space
\(\mathcal{H}_{\mathrm{BU}}\) can be described in two ways: as a symmetric Fock
space over the worldline physical Hilbert space \(\mathcal{H}_{\text{wl inv}}\)
(instead of some ``positive-frequency'' subsector like in QFT); and as a
direct path integral over one-dimensional superselection sectors
(\(\alpha\)-sectors) in each of which the universe field takes on a definite
on-shell configuration \(\Phi_\alpha(x)\).

Below in \cref{sec:discussqftvsuft}, we will firstly discuss some reasons why
the GABUFT approach taken in this paper might be preferable
\cite{Casali:2021ewu} to the alternative QFT approach for describing
minisuperspace models, particularly of recollapsing universes. Then in
\cref{sec:interactions}, we will discuss generalizations of GABUFT to include
nontrivial topology change. We also point to directions for future work in
\cref{sec:nextsteps}.

\subsection{QFT and GABUFT approaches to models of recollapsing universes}
\label{sec:discussqftvsuft}

Throughout this paper, we have already described the technical differences
between QFT and GABUFT listed in \cref{tab:qftvsuft}. For the most part, it was
largely assumed that the QFT description exists and makes sense in the first
place. In this section, however, we will describe how the QFT approach can run
into difficulties even in simple minisuperspace models of recollapsing
universes.

At a high level, one might summarize \cref{tab:qftvsuft} by saying that GABUFT
treats directions in target space \(\mathscr{M}\ni x\) more democratically than
QFT. In particular, GABUFT makes no reference to some time direction in
\(\mathscr{M}\). On the other hand, to pass from the worldline theory to QFT,
one can identify worldline amplitudes with \emph{time-ordered} correlation
functions
\(\mel{0}{\mathrm{T}\Phi_{\scriptscriptstyle\mathrm{QFT}}(x')\Phi_{\scriptscriptstyle\mathrm{QFT}}(x)}{0}_{\mathrm{QFT}}\),
or otherwise use worldline amplitudes \(\mel{x'}{P_+\eta}{x}_{\text{wl}}\)
projected onto \emph{``positive frequencies''}. The dichotomous split between
``positive-frequency'' and ``negative-frequency'' solutions subsequently appears
in the QFT mode expansion, and the Hilbert space becomes a Fock space
exclusively over the former.

This begs the question of whether \(\mathscr{M}\) has some natural time
direction with respect to which a reasonable QFT can be defined. The DeWitt
metric \cref{eq:dewittmet} is in fact Lorentzian at each universe point, with
the unique timelike direction being the scale factor. So, at least for
minisuperspace models where the scale factor is included and integration over
homogeneous space just yields an overall volume factor, \(\mathscr{M}\)
is Lorentzian. See, for example, the diagonal minisuperspace models where the
minisuperspace metric \labelcref{eq:misnersupermetric} has a unique timelike
direction, whose coordinate \(x^0\) parametrizes the scale factor \(e^{2x^0}\).
Unfortunately, this variable \(x^0\) can be a poor choice of ``time'', with
respect to which the expected dynamics of many minisuperspace models don't seem
to be recovered in a naive QFT approach.

As foreshadowed at the end of \cref{sec:motivation} and demonstrated by examples
in \cref{app:minisuperspace}, the problem is that the potential term \(V(x)\) in
the Wheeler--DeWitt Hamiltonian \(H\) (or \(\tilde{V}(x)\) appearing in
\(\tilde{H}(x)\) considered in \cref{app:minisuperspace}) can grow arbitrarily
negative over ``time'' \(x^0\). This is perhaps most clearly seen in the simple
recollapsing FLRW, Kantowski--Sachs, and \(\Lambda<0\) Bianchi I models
discussed in \cref{app:flrw,sec:ksBianchiI}. The classical trajectories in these
models describe universes where the scale factor \(e^{2x^0}\) initially grows,
reaches a maximum, and finally recollapses. Because \(x^0\) is timelike, its
kinetic term in the Wheeler--DeWitt Hamiltonian is negative, so the potential
\(V(x)\) that reflects classical trajectories away from large \(x^0\) likewise
grows \emph{negatively} and does so exponentially in the aforementioned models.
Said differently, these models are described by particles propagating on
\(\mathscr{M}\) with a time-dependent mass-squared \(V(x)\) such that they
become exponentially tachyonic at large \(x^0\). This is problematic for a
QFT-like approach \cite{Casali:2021ewu}\footnote{Some older discussions of
  related issues include \cite{Wald:1993kj,Marolf:1994wh}.}.

To see what can go wrong, let us consider trying to write down a QFT-like mode
expansion \labelcref{eq:qftmodeexpansion} for the field
\(\Phi_{\mathrm{QFT}}(x)\) satisfying \(H\Phi_{\mathrm{QFT}}(x)=0\). There is
some freedom in dividing a solution space into ``positive-frequency'' and
``negative-frequency'' sectors corresponding to a choice of vacuum. But if this
vacuum is to resemble something like a nonsingular state in free QFT ---
specifically, a Hadamard state --- then one ought to be able to choose the basis
of modes in the ultraviolet to locally resemble positive- and negative-frequency
plane-waves respectively \cite{Giddings:2025abo}.\footnote{More generally, the
  ``spectrum condition'' of axiomatic QFT \cite{Hollands:2009bke} requires
  singularities of Wightman functions to be governed by wave fronts with
  \emph{future directed} null momenta.} At least in the classically allowed
region, one might be able to choose such positive- and negative-frequency modes
with rapidly oscillating time dependence respectively like
\(e^{i S_{\mathrm{cl}}(x^0)}\) and \(e^{-i S_{\mathrm{cl}}(x^0)}\), where
\(\partial_0 S_{\mathrm{cl}}=(p_{\mathrm{cl}})_0\) and the positive quantity
\(-(p_{\mathrm{cl}})_0\) varies relatively slowly. However, when evolved into
the classically forbidden region where \(-V(x)\) grows without bound, these
modes will stop oscillating in \(x^0\) and will grow without bound.\footnote{In
  fact, because of this behaviour, these modes cannot show up as independent
  eigenfunctions in a spectral decomposition of \(H\). So as described towards
  the end of this subsection, we would not even consider them as independent
  \(u_k\) as defined in \cref{sec:gennotation}.} This is just the familiar fact
in particle quantum mechanics that energy eigenfunctions decaying under a
reflective potential require particular equal-magnitude combinations of left-
and right-moving components, while other combinations grow exponentially in the
potential. Yet, the QFT mode expansion \labelcref{eq:qftmodeexpansion} forces us
to treat positive and negative frequencies independently. A consequence of this
is that states \(\ket{\Psi_{\mathrm{1p}}}_{\mathrm{QFT}}\) in the one-universe
(one-particle) sector of the QFT will have wavefunctions
\(\mel{0}{\Phi_{\mathrm{QFT}}(x)}{\Psi_{\mathrm{1p}}}_{\mathrm{QFT}}\) that grow
without bound in \(x^0\). We thus arrive at a rather bizarre conclusion: in
models where universes classically undergo recollapse, the QFT approach predicts
wavefunctions that grow with universe size.

The above discussion was for a free QFT. One might very optimistically speculate
that including nontrivial topological interactions between universes could
stabilize the potential of the tachyonic field. Even so, we would find it rather
bizarre that, to talk about a simple model where universes are classically
expected to recollapse, we must appeal to interactions that stabilize a tachyon
condensate of large universes. It is not even clear that the resulting quantum
theory describes recollapsing universes at all.

Before moving on, let us also comment briefly on other possible choices of
``time'' in the QFT approach. Above, we talked about using \(x^0\) as a ``time''
coordinate because it is the timelike direction (with negative metric norm) in
Lorentzian minisuperspace. One might have guessed that this is a bad choice for
models of recollapse, because even classical trajectories do not evolve
monotonically in \(x^0\). In some minisuperspace models, it may be possible to
choose another ``time'' variable which does evolve monotonically --- for
example, the spacelike coordinate \(\tilde{x}^1\) in the Kantowski--Sachs model
or \(x^1\) or \(x^2\) in the \(\Lambda<0\) Bianchi I model discussed in
\cref{sec:ksBianchiI}. In other models, such as the Bianchi IX model sketched in
\cref{sec:bianchiIX}, however, no such choice is readily available and one would
have to introduce an extra clock degree of freedom to have a monotonically
evolving time variable. Once an alternative choice of time has been chosen, one
can again try to define a QFT-like theory by writing down a mode expansion
\labelcref{eq:qftmodeexpansion} where one distinguishes ``positive-frequency''
and ``negative-frequency'' modes based on their dependence with respect to this
new time variable. Meanwhile, \(x^0\) is treated like a spatial coordinate; in
particular, one can now restrict to modes that decay in \(x^0\) when the
potential \(-V(x)\) becomes large. It is unclear, however, in what way such an
approach would be advantageous compared to GABUFT. One reason to favour a
QFT-like approach is because it might allow us to import what we already know
about QFT to study gravitating universes. But once we start making
unconventional choices --- treating timelike coordinates \(x^0\) as spatial and
accepting potentials \(V(x)\) that become arbitrarily negative --- the
connection to standard QFT becomes tenuous. In fact, one might argue that a
theory constructed by the procedure just described is more similar to a GABUFT
with an extra (and somewhat arbitrary) ``positive-frequency'' projection of the
Hilbert space.

Finally, let us describe how the GABUFT approach handles minisuperspace models
of recollapsing universes where \(-V(x)\) grows without bound in \(x^0\). The
upshot is that constructing GABUFT using a group-averaged worldline theory
automatically restricts us to wavefunctions that bounce off of perfectly
reflecting potentials. This follows from the fact that any spectral
decomposition of a Sturm--Liouville differential operator \(H\) involves
eigenfunctions \(u_k(x)\) that decay rather than grow under such potentials.
Consequently, each mode \(u_k(x)\) appearing in the mode expansion
of the GABUFT field \(\Phi(x)\) will have precisely the right equal-magnitude
combination of incoming and outgoing components to describe a wave reflecting
off the potential. (See, \eg{} \cref{eq:flrwWKB,eq:kssolnflatmini}.) These are
the usual textbook wavefunctions of a quantum mechanical particle encountering a
perfectly reflecting potential. As there, the classical limits of appropriate
coherent (test) states \(\ket{j}_{\mathrm{wl}}\) of the worldline theory will
evolve under \(H\) exactly like classical trajectories of reflected particles,
here representing recollapsing universes. Two such worldline states \(j_1\) and
\(j_2\) will have a nonzero GABUFT Wick contraction
\begin{align}
  \wick{\c Z(j_1^\star) \c Z(j_2)}
  = \llangle j_1 | j_2 \rrangle_{\mathrm{wl}}
  = \mel{j_1}{\int_{-\infty}^\infty \dd{T} e^{-i T H}}{j_2}_{\mathrm{wl}}
\end{align}
in the classical limit if and only if they are centred (in kinematic phase
space) on the same classical trajectory. In this precise sense, GABUFT recovers
the classical dynamics of minisuperspace models, including
recollapse.\footnote{A closely related Dirac quantization of recollapsing
  minisuperspace models was studied in ref.~\cite{Marolf:1994wh}, where
  gauge-invariant observables were used to show that the resulting quantum
  theory reproduces recollapse.}

\subsection{Generalities, nontrivial topology change, and interactions}
\label{sec:interactions}

In \cref{sec:uft}, we developed free GABUFT for gravitating universes modelled
by non-interacting worldlines. Specifically, the path integral over worldlines
given in \cref{eq:topologygpi} only represents cylinder-like
\((\text{time interval})\times(\text{spatial topology})\) spacetimes, with
constant (dimensionally reduced) spatial topology and a time interval connecting
externally prepared states \(j_a^\star\) and \(j_b\). In some sense, this
already incorporates a set of simple topology-changing processes: universes can
be swapped (\(\ket{j_1\sqcup j_2}=\ket{j_2\sqcup j_1}\)), pair-created, and
pair-annihilated (\(\braket{\varnothing}{j_1\sqcup j_2}_{\mathrm{BU}}\ne0\)).

\begin{figure}
  \centering
  \begin{subfigure}[t]{0.3\textwidth}
    \centering
    \includegraphics[width=\textwidth]{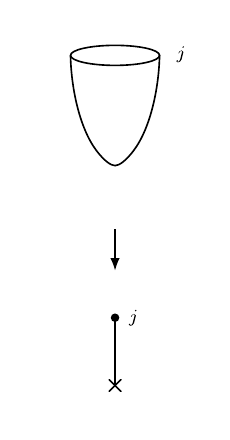}
  \end{subfigure}
  \hfill
  \begin{subfigure}[t]{0.3\textwidth}
    \centering
    \includegraphics[width=\textwidth]{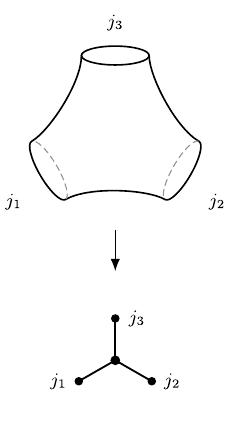}
  \end{subfigure}
  \hfill
  \begin{subfigure}[t]{0.3\textwidth}
    \centering
    \includegraphics[width=\textwidth]{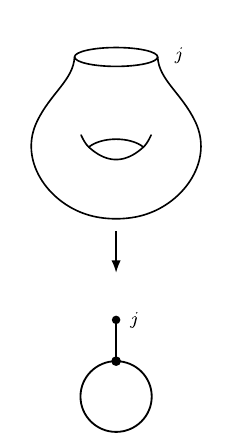}
  \end{subfigure}
  \caption{Dimensional reduction of higher-dimensional (here, two-dimensional)
    universes undergoing topology change resembles Feynman diagrams with sources
    and interaction vertices.}
  \label{fig:feynmandiag}
\end{figure}

More generally, universes can undergo more nontrivial topology-changing
processes that are usually associated with smooth Euclidean spacetime geometries
in dimensions greater than one. Traditionally, gravitational path integrals on
such topologies are performed in Euclidean signature, either ignoring the
conformal factor problem or adopting some ad hoc contour prescriptions for the
path integral \cite{Gibbons:1978ac}. We anticipate that a more principled
Lorentzian starting point should be possible, using a Lorentzian gravitational
path integral generalizing the kind we considered on cylinder-like topologies. To
admit more nontrivial topologies, one may allow this path integral to include,
for example, (almost-)Lorentzian configurations with conical singularities
\cite{Marolf:2022ybi}.

Including a sum over these more nontrivial topologies in the gravitational path
integral \(\zeta\), one can again define a baby universe theory through the GNS
construction described in \cref{sec:gns}. A direction for further development is
to understand how a baby universe theory including nontrivial topologies can be
given a field-theoretic description, generalizing the free GABUFT framework
developed in this paper. In particular, upon dimensional reduction, nontrivial
topologies of higher-dimensional universes reduce to Feynman-like diagrams with
sources and interaction vertices --- see \cref{fig:feynmandiag}. These pictures
evoke the correspondence between string worldsheets and Feynman diagrams of
string field theory, so a field theory description will likely exist and be
useful also in our baby universe context. Yet, we know that taking the
Lorentzian group-averaged approach to gravity on cylinder-like universes leads
to a free GABUFT differing from standard free QFT. So the generalization of
GABUFT in the presence of nontrivial topology change will likewise differ from a
standard interacting QFT or something resembling string field theory. We will
largely leave it to future work to develop the interacting GABUFT framework, but
we'll end this paper with some preliminary comments that might be useful in this
direction.

In \cref{sec:generalprop}, we first discuss some general properties of GABUFT
that we expect to persist, even in the presence of nontrivial topology change.
In \cref{sec:onepoint}, while staying in the free GABUFT framework, we discuss
turning on a one-point function --- the Hartle--Hawking wavefunction
\cite{Hartle:1983ai} --- that describes topologies where universes cap off to
nothing. We then describe in \cref{sec:topmodel} how the topological model of 2D
gravity given by Marolf and Maxfield \cite{Marolf:2020xie} can be given a Fock
space description that is, perhaps surprisingly, exact even when all topological
interactions are included. This model is exceedingly simple and we by no means
claim that the strategy we apply to it will be practically useful more
generally. So, in \cref{sec:nextsteps}, we speculate on what an appropriate set
of next steps would be to understand how topology change in more nontrivial
theories can be given a GABUFT description.

\subsubsection{General comments on reality and on-shell field operators}
\label{sec:generalprop}

When constructing free GABUFT in \cref{sec:uft}, we found that operators in the
theory formally\footnote{See \cref{foot:zjadjoint}.} satisfy a reality condition
--- see \cref{eq:zsmalljadjoint1,eq:phihermiticity} ---
\begin{align}
  Z(j)^\dagger
  &= Z(j^\star)
  &
  &\text{or, equivalently,}
  &
    \Phi(x)^\dagger
  &= \Phi(x^\star)
    \label{eq:zsmalljadjoint2}
\end{align}
and that the GABUFT field operator \(\Phi(x)\) satisfies the Wheeler--DeWitt
\cref{eq:ufteom1}
\begin{align}
  H \Phi(x)
  &= 0
    \;.
    \label{eq:ufteom2}
\end{align}
As we now discuss, on fairly general grounds, we expect these properties of
the GABUFT field \(\Phi(x)\) to survive even with the inclusion of contributions
to the path integral \(\zeta\) representing nontrivial higher-dimensional
topologies. (We will continue to define \(\Phi(x)\) in relation to
single-boundary-inserting operators \(Z(j)\) by \cref{eq:zsmearing}.)

\paragraph{Reality implied by positivity.} Let us recall from \cref{sec:gns} that the
starting point of the baby universe framework \cite{Marolf:2020xie} is the
existence of a gravitational path integral \(\zeta:\mathcal{J}\to\mathbb{C}\)
that is a positive linear map \(J\mapsto \zeta(J)\) from the \(\star\)-algebra
\(\mathcal{J}\) of multi-boundary conditions \(J\) to \(\mathbb{C}\).
Positivity,
\begin{align}
  \zeta(J^\star \sqcup J) \ge 0
  \;,
  \label{eq:zetapos2}
\end{align}
is especially crucial for defining an inner product \labelcref{eq:buinnerprod},
\begin{align}
  \braket{J'}{J}
  &\equiv \zeta((J')^\star \sqcup J)
    \;,
    \label{eq:buinnerprod2}
\end{align}
on the GNS-constructed baby universe Hilbert space. The very definition
\labelcref{eq:buinnerprod2} of the inner product in terms of the gravitational
path integral \(\zeta\) then implies that the reality condition
\labelcref{eq:zsmalljadjoint2} at least formally holds as an equation
\labelcref{eq:zhermitian} of matrix elements within a dense domain of states. In
this formal sense (modulo details related to operator domains), the reality
condition \labelcref{eq:zsmalljadjoint2} is an inevitable consequence of the
positivity axiom \labelcref{eq:zetapos2} that underpins the baby universe
framework. Continuing to define the universe field \(\Phi(x)\) by relation
\labelcref{eq:zsmearing} to \(Z(j)\), the formal reality condition
\labelcref{eq:zsmalljadjoint2} can again be equivalently stated in terms of
\(\Phi(x)\).

As reviewed in \cref{sec:timerev}, when bulk spacetime (worldline)
configurations do not carry orientation, \(\star\) is understood as the
time-reversal symmetry \(\mathsf{T}\) \cite{Witten:2025ayw} that the bulk theory
is expected to have. Because an induced (real) tensor\footnote{Recall from the
  end of \cref{sec:orientrev} that a theory with \emph{pseudo} tensors will
  generically have \(x^\star\ne x\); in these theories, the reality condition of
  \(\Phi\) will therefore be nonlocal, like the oriented case described in the
  next paragraph.} field configuration \(x\) on a Euclidean boundary is
invariant, \(x^\star=x\), \cref{eq:zsmalljadjoint2} then tells us that the
universe field is real in the usual pointwise sense, \(\Phi(x)^*=\Phi(x)\).

When bulk spacetime (worldline) configurations carry orientation, as described
in \cref{sec:orientrev}, \(\star\) reverses the induced orientation that is
included along with other induced tensor fields in the boundary data
\(x\in\mathscr{M}\). Thus, generically in this case,
\(x^\star=\mathcal{T}x \ne x\) and the universe field is not a real field in the
usual sense: \(\Phi(x)^*=\Phi(x^\star)\ne \Phi(x)\). In the simplified worldline
setting, the orientation of an endpoint is given by a sign \(\pm\), which then
distinguishes two mutually disconnected components \(\mathscr{M}_\pm\) of target
space \(\mathscr{M}=\mathscr{M}_+\sqcup\mathscr{M}_-\). Since the reality
condition \labelcref{eq:zsmalljadjoint2} relates the field \(\Phi\) across the
two components \(\mathscr{M}_\pm\), one might view \(\Phi\) as just a complex
field living on one of these components, say \(\mathscr{M}_+\), with no reality
condition imposed. If we want to be more faithful to higher-dimensional gravity,
however, there is a question of whether we are using each point in
\(\mathscr{M}\) to represent (spatially) diffeomorphic boundary data. If so,
then \(\mathscr{M}\) will not separate into two components labelled by
orientation, because orientation can be flipped by spatial
diffeomorphisms.\footnote{See \cref{foot:circleboundary} for a discussion of
  one-dimensional boundaries. Even if one does not initially choose to identify
  points \(x\) of a kinematic target space
  \(\mathscr{M}=\mathscr{M}_+\sqcup\mathscr{M}_-\) that represent boundaries
  related by diffeomorphism, \eg{} an orientation-reversing reflection
  \(\mathcal{R}:\mathscr{M}_+\leftrightarrow\mathscr{M}_-\), the path integral
  will gauge such diffeomorphisms --- this is described in
  \cref{app:pathintgauging}. This enforces \(\Phi(x)=\Phi(\mathcal{R} x)\) and
  therefore a nonlocal reality condition
  \(\Phi(x)^*=\Phi(\mathcal{R} x^\star)=\Phi(\mathcal{R} \mathcal{T} x)\)
  relating the field at points \(x\) and \(\mathcal{R}\mathcal{T} x\) on the
  same component \(\mathscr{M}_\pm\). (Because a reflection diffeomorphism
  \(\mathcal{R}\) will carry along not only orientation but also other boundary
  data --- \eg{} the operator insertions in the example of
  \cref{foot:circleboundary} --- \(\mathcal{R}\mathcal{T}\) will generically map
  a point in \(\mathscr{M}_\pm\) to a different point.)} In that case, \(\Phi\)
might be best thought of as a field on the total space \(\mathscr{M}\) subject
to a nonlocal reality condition \(\Phi(x)^*=\Phi(x^\star)\).

We have described above the reality properties of the GABUFT field operator.
Before proceeding, let us make some related comments about the reality
properties of the gravitational path integral \(\zeta\) and the (lack of)
reality of the baby universe Hilbert space.

First, we note that the reality condition \labelcref{eq:zsmalljadjoint2} on
operators is also closely related to a similar condition
\begin{align}
  \zeta(J)^*
  &= \zeta(J^\star)
    \label{eq:zetareality}
\end{align}
on the gravitational path integral \(\zeta\), as seen by writing
\begin{align}
    \mel{J''}{Z(J)}{J'}^*
  &\equiv
  \zeta((J'')^\star \sqcup J \sqcup J')^*
    =
    \zeta(J'' \sqcup J^\star \sqcup (J')^\star)
    \equiv
    \mel{J'}{Z(J^\star)}{J''}
    \;.
    \label{eq:zsymmetric}
\end{align}
Indeed, the reality \labelcref{eq:zetareality} of \(\zeta\) is itself a
\emph{corollary} of its positivity \labelcref{eq:zetapos2} --- because this
logical connection has perhaps not been emphasized sufficiently in the
literature\footnote{In the literature \cite{Marolf:2022ybi,Colafranceschi:2023moh},
  the reality condition \labelcref{eq:zetareality} for the gravitational path
  integral \(\zeta\) is sometimes stated as a separate, potentially optional,
  axiom rather than a consequence of positivity \labelcref{eq:zetapos2}.}, it is
perhaps worth spelling out more explicitly. A mathematical fact is that any
positive linear functional \(\zeta:\mathcal{J}\to\mathbb{C}\) on a
unital\footnote{Recall unital means the existence of a multiplicative identity
  element \(\varnothing\in\mathcal{J}\).} \(\star\)-algebra \(\mathcal{J}\) must
also be Hermitian (meaning \cref{eq:zetareality} is satisfied).
\begin{proof}
  Positivity \labelcref{eq:zetapos2} of \(\zeta\) implies that
  \begin{align}
    \begin{split}
    0 \le \zeta((c J + c' J')^\star \sqcup (c J + c' J'))
      &= \abs{c}^2 \zeta(J^\star \sqcup J)
        + \abs{c'}^2 \zeta((J')^\star\sqcup J')
        \\
      &\phantom{{}={}}\negmedspace
        + c^* c' \zeta(J^\star \sqcup J')
        + (c')^* c \zeta((J')^\star \sqcup J)
      \end{split}
  \end{align}
  for all \(J,J'\in\mathcal{J}\) and \(c,c'\in\mathbb{C}\). On the RHS, the
  first line is positive again by \cref{eq:zetapos2}, so the second line must
  also be real. Taking \(c=1\) and \(c'=1,i\) tells us that
  \(\zeta(J^\star\sqcup J')\pm\zeta((J')^\star\sqcup J)\) is, respectively, real
  \((+)\) and imaginary (\(-\)). It follows that
  \begin{align}
    \zeta(J^\star\sqcup J') = \zeta((J')^\star\sqcup J)^*
    \;,
    \label{eq:zetahermitianform}
  \end{align}
  \ie{} \(\zeta(\bullet^\star \sqcup \bullet)\) is a Hermitian form on
  \(\mathcal{J}\). Finally, taking \(J'=\varnothing\) to be the identity element
  gives Hermiticity \labelcref{eq:zetareality} of \(\zeta\) as a linear
  functional \(\mathcal{J}\to\mathbb{C}\).
\end{proof}

As a second comment, let us note that any nontrivial unitary \(S\)-matrix, \eg{}
in QFT or string field theory, must fail to be positive. Having a unitary that
is Hermitian is already highly non-generic, as its eigenvalues would be
restricted to \(\pm 1\). Positivity would force the unitary to be the identity.
This is to say that our observation in \cref{sec:fockspace} --- that QFT is
essentially a positive-frequency restriction of GABUFT --- is highly non-generic
and special to free worldlines. With the inclusion of string interactions, the
same worldsheet path integral that computes a nontrivial unitary \(S\)-matrix
does not directly define a positive functional for the GNS construction of a
baby universe Hilbert space (or some subsector). Positivity might be salvageable
with appropriate analytic continuations (\eg{} a Wick rotation to Euclidean
target space) and a suitably modified definition of \(\star\)
\cite{Casali:2021ewu}, but these modifications seem hard to justify from first
principles in defining a (worldsheet) gravitational theory. This is perhaps
another reason to search for a separate, GABUFT approach for describing baby
universes.

A third comment is that, although we first intended ``time-reversal''
\(\mathsf{T}\) to mean a reversal of bulk time, as an aside, let us also note
the following connection to boundary time reversal. For an induced tensor field
configuration \(x\) on a real Euclidean boundary, the property \(x=x^\star\) or,
equivalently,
\(\zeta(x\sqcup J)=\zeta(x^\star \sqcup J)=\zeta(x \sqcup J^\star)^*\) suggests
the existence of a time-reversal symmetry in the boundary theory (or, rather,
theories). In particular, each member of the \(\alpha\)-ensemble is associated
with a boundary theory that has a factorizing path integral satisfying
\(\zeta_\alpha(x)=\zeta_\alpha(x)^*\).\footnote{In general, \(\star\) will flip
  the sign of any nonzero induced configuration for pseudo-tensors like axions,
  so \(x^\star\ne x\); the corresponding sources for these fields also flip sign
  in the Euclidean QFT path integral \(\zeta_\alpha\) under complex conjugation
  (which is expected, even in a time-reversal invariant QFT). To be clear, we
  always use the notion of reality for pseudo-tensors inherited from Lorentzian
  signature.} This should be contrasted with the Euclidean path integral of a
non-time-reversal-symmetric QFT, where complex conjugation
\(\zeta_\alpha(x)^*=\zeta_\alpha(\mathcal{T} x)\) is implemented by the
antilinear orientation-reversal transformation \(\mathcal{T}\) on the boundary
data \(x\) \cite{Witten:2025ayw}. The property
\(\zeta_\alpha(x)=\zeta_\alpha(x)^*\) instead suggests the boundary theory is
independent of such a background orientation that would break boundary time
reversal symmetry.\footnote{But in very special cases, it might instead indicate
  that \(\mathcal{T}x\) is diffeomorphic to \(x\) --- see, \eg{}
  \cref{foot:circleboundary}.}


Finally, while the field satisfies a reality condition
\labelcref{eq:zsmalljadjoint2}, let us emphasize that
\(\mathcal{H}_{\mathrm{BU}}\) is complex when GNS-constructed from an algebra
\(\mathcal{J}\) of boundary conditions over \(\mathbb{C}\) --- see
\cref{sec:gns}. In particular, we do not identify \(|J\rangle_{\mathrm{BU}}\)
with \(|J^\star\rangle_{\mathrm{BU}}\). There have been arguments presented for
\cite{Harlow:2023hjb} and against \cite{Witten:2025ayw} gauging antilinear
symmetries in quantum gravity. We hope to return, in upcoming work
\cite{Chen:2026wip2}, to the question of whether the gravitational path integral
gauges antilinear transformations such as \(\star\) in the sense of identifying
these states, and to the resulting structural consequences and potential
tensions for the baby universe framework.

\paragraph{Wheeler--DeWitt equation.} In \cref{sec:modeexpansion}, by studying
Wick contractions of the free GABUFT field \(\Phi(x)\), we deduced that it
satisfies the Wheeler--DeWitt equation of motion \labelcref{eq:ufteom2} ---
\(H\Phi(x)=0\). This is hardly surprising, because Wick contractions of baby
universe operators were designed in \cref{eq:wickcontract} to give worldline
inner products
\(\llangle j_2 | j_1 \rrangle_{\mathrm{wl}}=\mel{j_2}{\eta}{j_1}_{\mathrm{wl}}\)
including a group average \(\eta\) for which \(\eta H=0\) and thus
\(|H j_1\rrangle_{\mathrm{wl}}=0\) is null. Let us recall again that these free
worldlines represent only cylinder-like universes.

We expect the same \emph{linear} field equation \labelcref{eq:ufteom2} to be
satisfied by the field operator \(\Phi(x)\) (again, still defined by
\labelcref{eq:zsmearing}) even in an \emph{interacting} GABUFT that incorporates
nontrivial topology change. Admittedly, this is perhaps more of an expectation
at the moment, based on some assumptions about the path integral near spacetime
boundaries. Specifically, for any topology that one might want to include in a
gravitational path integral, we still expect each boundary component to be
connected by a locally cylinder-like piece to the rest of the spacetime. Upon
dimensional reduction, these cylinder-like pieces become worldline intervals
that one might have called ``external legs'' of a Feynman diagram in the context
of QFT. In our GABUFT context, each external leg represents the path integral of
a locally cylinder-like universe where integrating lapse over the real line
implements a group average \(\eta\) over time translations, generated again by
the \emph{cylinder} Wheeler--DeWitt Hamiltonian \(H\). Consequently, the
difference \(j- e^{-iT H}j\) between boundary states related by time
translations generated by \(H\) is again a null insertion for the path integral;
phrased in terms of operators that insert boundaries, \(Z(H j)=0\) or
equivalently \(H\Phi(x)=0\).\footnote{What we have just described here is what
  is also depicted in \cref{fig:groupavgeg}, not just for time translations, but
  also spatial diffeomorphisms.}

\begin{figure}
  \centering
  \begin{subfigure}[t]{0.3\textwidth}
    \centering
    \includegraphics[width=\textwidth]{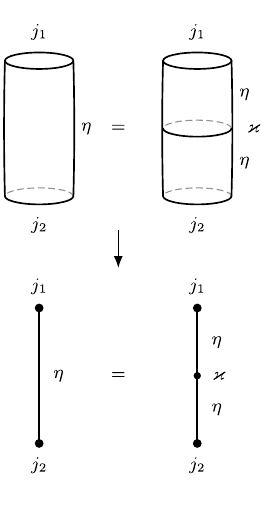}
    \caption{A free worldline representing a higher-dimensional cylinder-like
      universe. Inserting a gauge-fixing condition \(\varkappa\) allows one to
      split the gravitational path integral into two pieces.}
    \label{fig:compwl}
  \end{subfigure}
  \hfill
  \begin{subfigure}[t]{0.65\textwidth}
    \centering
    \includegraphics[width=\textwidth]{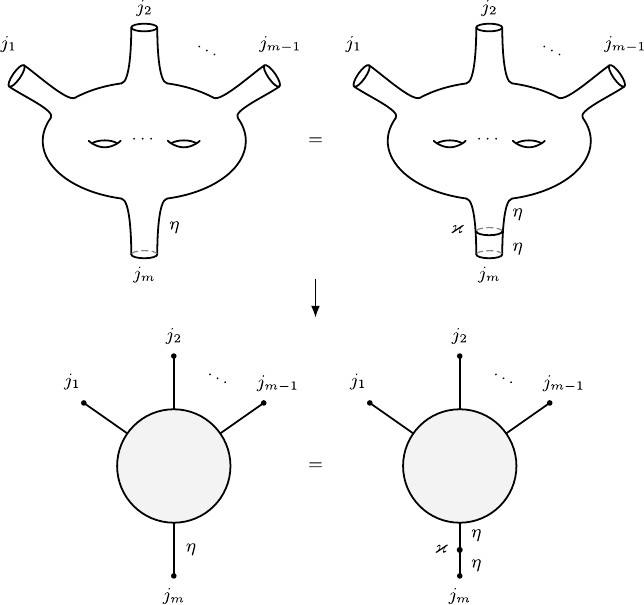}
    \caption{A Feynman-like diagram representing nontrivial higher-dimensional
      topology. We expect each ``external leg'', representing a cylinder-like
      near-boundary spacetime region, to satisfy a similar composition rule.}
    \label{fig:comptop}
  \end{subfigure}
  \caption{A composition rule for the gravitational path integral.}
  \label{fig:composition}
\end{figure}

As an aside, we can formalize the above observation into a composition rule that
we expect the gravitational path integral to satisfy. For free worldlines
representing cylinder-like universes, refs.~\cite{Held:2024rmg,Held:2025mai}
explain that the worldline gravitational path integral satisfies a composition
rule,
\begin{align}
  \llangle j_1 | j_2\rrangle_{\mathrm{wl}}
  &= \int_{\mathscr{M}} \mu(x)
    \int_{\mathscr{M}} \mu(x')\,
    \llangle j_1 | x \rrangle_{\mathrm{wl}}
    \mel{x}{\varkappa}{x'}_{\mathrm{wl}}
    \llangle x' | j_2 \rrangle_{\mathrm{wl}}
    \;,
    \label{eq:compwl}
\end{align}
with the inclusion of a generalized inverse \(\varkappa\) for the group-average
\(\eta\),
\begin{align}
  \eta \varkappa \eta
  &= \eta
    \;.
    \label{eq:geninverse}
\end{align}
The interpretation of \(\varkappa\) is that of a gauge-fixing map from
\(\Image\eta\) back to the test space which picks out an arbitrary
representative \(\ket{j}\) from each equivalence class \(|j\rrangle\).
\Cref{eq:compwl} is illustrated in \cref{fig:compwl}; the two factors
\(\llangle j_1 | x \rrangle_{\mathrm{wl}}\) and
\(\llangle x' | j_2 \rrangle_{\mathrm{wl}}\) are the two group-averaged pieces
of the line segment, sewn together by the insertion
\(\mel{x}{\varkappa}{x'}_{\mathrm{wl}}\). In a Feynman-like diagram representing
nontrivial higher-dimensional topology, as illustrated in \cref{fig:comptop}, we
expect the gravitational path integral on each external leg to behave similarly,
\begin{align}
  \zeta(j_1 \sqcup \cdots \sqcup j_m)
  &= \int_{\mathscr{M}} \mu(x)
    \int_{\mathscr{M}} \mu(x')\,
    \zeta(j_1 \sqcup \cdots \sqcup j_{m-1} \sqcup x) \,
    \mel{x}{\varkappa}{x'}_{\mathrm{wl}}
    \llangle x' | j_m\rrangle_{\mathrm{wl}}
    \;.
    \label{eq:zetacomposition}
\end{align}
This equation is equivalent to saying that \(\zeta\) depends on each test
wavefunction \(j(x)\) only through its invariant image \((\eta j)(x)\); in
particular, the path integral sees no difference between \(j\) and
\(e^{-iT H}j\). This composition property \labelcref{eq:zetacomposition} implies
that \(\Phi(x)\) is not merely a solution to the Wheeler--DeWitt field equation
\labelcref{eq:ufteom2}, \(H\Phi(x)=0\), but one that can be expanded,
\begin{align}
  \Phi(x)
  &= \int_{\mathscr{K}} \kappa(k) \phi(k) u_k(x)
    \;,
    \label{eq:generalphiexp}
\end{align}
in terms of the zero-energy eigenfunctions
\(u_k(x)=\braket{x}{k}_{\mathrm{wl}}\) appearing in the spectral decomposition
\labelcref{eq:etaexpansion} of the group-averaged inner product
\(\llangle\bullet|\bullet\rrangle_{\mathrm{wl}}\).\footnote{Note that these
  modes \(u_k(x)\) are potentially special solutions that may, for example, be
  required to satisfy appropriate boundary conditions in order to appear in the
  spectral decomposition of a self-adjoint \(H\).} The only difference with
\cref{eq:phiexpansion2} is that the operator \(\phi(k)\) need no longer be some
linear combination of creation and annihilation operators.

We acknowledge the stark contrast against QFT or string field theory, where
interactions give rise to nonlinearities in field equations over target space.
It seems that the interactions in GABUFT do not give rise to similar nonlinear
corrections to the Wheeler--DeWitt field equation \(H \Phi(x)=0\) satisfied by
fields \(\Phi(x)\) over (kinematic) target space \(\mathscr{M}\ni x\). Perhaps
in somewhat closer analogy to QFT, nontrivial interactions do correspond to
non-Gaussianities in the path integral
\(\int \mathcal{D}\phi_\alpha\, e^{-I[\phi_\alpha]}\) over \(\alpha\)-parameters
(which are again the eigenvalues \(\phi_\alpha(k)\) of \(\phi(k)\)). We see no
obvious argument either for or against the possibility that this path integral,
possibly with some field redefinition or integral transform of
\(k\in\mathscr{K}\) to another space, could become the path integral of some
standard QFT --- see \cref{sec:topologyperturbation}. If such a description
exists, then the field equations of that QFT would then capture the topological
interactions of baby universes. (But, such a QFT should not be confused with a
field theory on the kinematic target space \(\mathscr{M}\); additionally, as
explained below \cref{eq:qftpathint}, the QFT Hilbert space and baby universe
Hilbert space interpretations of a given path integral differ significantly.)

\subsubsection{One-point functions and the Hartle--Hawking wavefunction}
\label{sec:onepoint}

\begin{figure}
  \centering
  \includegraphics[width=0.75\textwidth]{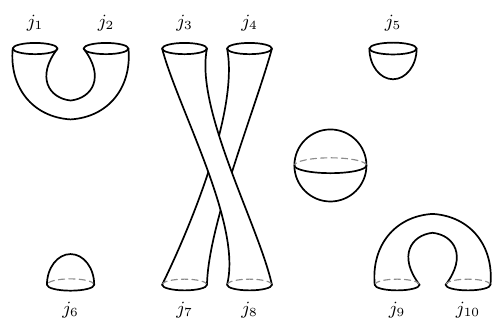}
  \caption{An example contribution to the gravitational path integral
    \(\zeta(j_1\sqcup\cdots\sqcup j_{10})\)
    upon allowing universe topologies that cap off to nothing.}
  \label{fig:onepointfunc}
\end{figure}

Perhaps the simplest generalization of the framework developed in \cref{sec:uft}
is to include contributions in the path integral \(\zeta\) that represent
topologies where universes cap off to nothing --- see \cref{fig:onepointfunc}.
Again defining the baby universe theory by a GNS construction, such topologies
will contribute to an overlap between a state \(\ket{j}_{\mathrm{BU}}\)
initially prepared with one universe and the no-boundary state
\(\ket{\varnothing}_{\mathrm{BU}}\),
\begin{align}
  \braket{\varnothing}{j}_{\mathrm{BU}}
  &= \mel{\varnothing}{Z(j)}{\varnothing}_{\mathrm{BU}}
    \;,
\end{align}
thus giving nonzero one-point functions to the boundary-inserting operators
\(Z(j)\) or, equivalently, to the universe field \(\Phi(x)\) (still defined by
\cref{eq:zsmearing}),
\begin{align}
  \Phi_\varnothing(x)
  &\equiv
    \frac{
    \mel{\varnothing}{\Phi(x)}{\varnothing}_{\mathrm{BU}}
    }{
    \braket{\varnothing}{\varnothing}_{\mathrm{BU}}
    }
    \;.
    \label{eq:onepointfunc}
\end{align}
With disconnected contributions
\(\braket{\varnothing}{\varnothing}=\zeta(\varnothing)\) divided out, this
normalization convention sets \(\Phi_\varnothing(x)\) equal to the connected
part of the overlap \(\braket{x}{\varnothing}_{\mathrm{BU}}\). Thus,
\(\Phi_\varnothing(x)\) can be identified as the usual ``Hartle--Hawking
wavefunction'', for which Hartle and Hawking's original calculation
\cite{Hartle:1983ai} on a disk-like (hemisphere) topology provides the leading
contribution.\footnote{Here, \(\ket{x}_{\mathrm{BU}}\) is not normalized (or
  even \(\delta\)-function normalized) in the baby universe Hilbert space. It is
  defined by the same boundary conditions that would have produced a
  \(\delta\)-function normalized ``state'' in the kinematic Hilbert space of the
  worldline/single cylinder-like universe theory,
  \(\braket{x'}{x}_{\mathrm{wl}}=\delta_\mu(x,x')\). Ref.~\cite{Abdalla:2026mxn}
  highlights the fact that the \emph{normalized} probability
  \(\abs{\braket{j}{\varnothing}_{\mathrm{BU}}}^2/\braket{j}{j}_{\mathrm{BU}}\braket{\varnothing}{\varnothing}_{\mathrm{BU}}\)
  is approximately identically \(1\) (as long as \(j\) is not non-perturbatively
  sharply peaked), in contrast to what one might naively expect from the
  \(\Phi_\varnothing(x)\) whose absolute value depends nontrivially on \(x\).}

Let us consider what the general properties discussed in \cref{sec:generalprop}
have to say about the one-point function \(\Phi_\varnothing(x)\). Firstly,
reality \labelcref{eq:zsmalljadjoint2} tells us that
\begin{align}
  \Phi_\varnothing(x^\star)
  &= \Phi_\varnothing(x)^*
    \;.
    \label{eq:oneptreal}
\end{align}
Secondly, the composition rule \labelcref{eq:zetacomposition} for the
gravitational path integral tells us that \(\Phi_\varnothing(x)\) solves the
Wheeler--DeWitt field equation \labelcref{eq:ufteom2}, \(H\Phi_\varnothing=0\),
and moreover admits an expansion \labelcref{eq:generalphiexp},
\begin{align}
  \Phi_\varnothing(x)
  &= \int_{\mathscr{K}} \kappa(k) \phi_\varnothing(k) u_k(x)
    \;,
    \label{eq:onepointmodes}
\end{align}
in terms of zero-energy eigenfunctions \(u_k(x)=\braket{x}{k}_{\mathrm{wl}}\) of
\(H\). Together, this means the coefficients satisfy
\begin{align}
  \phi_\varnothing(k^\star)
  &= \phi_\varnothing(k)^*
    \;.
    \label{eq:oneptcoeffreal}
\end{align}
The leading, disk-like contribution to \(\Phi_\varnothing(x)\) calculated by Hartle
and Hawking in de Sitter cosmology is indeed real (considering an unoriented
theory where \(x^\star=x\)) and the authors also give a similar path integral
argument for the Wheeler--DeWitt equation \(H\Phi_\varnothing=0\). This should be
contrasted with the Vilenkin tunneling-from-nothing wavefunction
\cite{Vilenkin:1984wp,Vilenkin:1986cy} consisting purely of the branch
describing growing universes --- this wavefunction fails to be real and fails to
satisfy boundary conditions corresponding to a self-adjoint extension of \(H\).
Assuming the properties discussed in \cref{sec:generalprop} are indeed
requirements of GABUFT, either of these failures is enough to preclude the
Vilenkin wavefunction from describing a one-point function in GABUFT; the latter
failure moreover precludes the Vilenkin wavefunction from describing any
correlation function
\(\mel{\varnothing}{\Phi(x)\cdots}{\varnothing}_{\mathrm{BU}}\).

To make further progress in our discussion, let us continue to ignore more
nontrivial topology-changing processes for now, so the baby universe theory has
only the following connected correlators: the same connected two-point function
as before (in \cref{eq:topologygpi}) from cylinder-like
\((\text{time interval})\times(\text{spatial topology})\) topologies; and the
one-point function from disk-like topologies, where the spatial topology
degenerates at one end of a finite time interval. The Gaussian nature of this
theory suggests that it might be the same GABUFT as before, only now with a
shifted \(\ket{\varnothing}_{\mathrm{BU}}\) state. In fact,
\cref{eq:onepointmodes,eq:oneptcoeffreal} imply essentially a unitary
equivalence to the GABUFT constructed in \cref{sec:uft} with vanishing one-point
function. Specifically, up to technicalities when \(\phi_\varnothing\) is
non-normalizable\footnote{\Cref{eq:nobdyshiftedu} defines a unitary map if
  \(\phi_\varnothing\in L^2(\mathscr{K},\kappa)\). If \(\phi_\varnothing\) is
  not normalizable, but \(\delta\)-function normalizable as in the case of dS JT
  \cite{Held:2024rmg}, then \cref{eq:nobdyshiftedu,eq:nobdyshifted} are not,
  strictly speaking, mathematically well-defined objects of the original
  theory. \Cref{eq:nobdyshiftop} morally still stands as the statement that
  correlation functions of the new theory are obtained by replacing
  \(\Phi(x)\to\Phi(x)+\Phi_\varnothing(x)\) in the old theory.}, there is a
unitary map
\begin{align}
  U
  &= \exp\left(
    \frac{1}{2} \int_{\mathscr{K}} \kappa(k) (a_k^\dagger - a_{k^\star}) \phi_\varnothing(k)
    \right)
    \label{eq:nobdyshiftedu}
\end{align}
relating the normalized cyclic state considered in \cref{sec:uft} to the
normalized no-boundary state of our new baby universe theory including disk-like
topologies. Denoting these respectively by
\(\aleph_0^{-1/2} \ket{\varnothing_0}\) and \(\aleph^{-1/2} \ket{\varnothing}\)
for now (and deferring comments on normalization \(\aleph/\aleph_0\) for later),
we have
\begin{align}
  \aleph^{-1/2} \ket{\varnothing}
  &= U\, \aleph_0^{-1/2} \ket{\varnothing_0}
    \;.
    \label{eq:nobdyshifted}
\end{align}
This map \(U\) is designed to shift the GABUFT field by
\(\Phi_\varnothing(x)\),
\begin{align}
  U^\dagger \Phi(x) U
  &= \Phi(x) + \Phi_\varnothing(x)
    \;,
    \label{eq:nobdyshiftop}
\end{align}
thus turning on this one-point function \labelcref{eq:onepointfunc} in the state
\(\aleph^{-1/2} \ket{\varnothing}\).

As one might expect, the baby universe theory with this one-point function
turned on is described by a shifted Gaussian path integral over \(\alpha\)-sectors.
Applying the Baker--Campbell--Hausdorff formula to \(U\) and noting that
\(\ket{\varnothing_0}\) is annihilated by the annihilation operators \(a\),
\cref{eq:nobdyshifted} can be rewritten as
\begin{align}
  \aleph^{-1/2} \ket{\varnothing}
  &= \exp\left(
    \int_{\mathscr{K}} \kappa(k) \left[
    \frac{1}{2} \phi(k)^\dagger \phi_\varnothing(k)
    - \frac{1}{4} \phi_\varnothing(k)^* \phi_\varnothing(k)
    \right]
     \right)
    \, \aleph_0^{-1/2} \ket{\varnothing_0}
    \;,
    \label{eq:nobdyshifted2}
\end{align}
where \(\phi(k)\) is the operator introduced in \cref{eq:phik}. As described in
\cref{eq:alphapathint,eq:alphaaction}, correlation functions in the state
\(\ket{\varnothing_0}\) are given by a centred Gaussian path integral over
\(\alpha\)-sectors labelled by simultaneous eigenvalues \(\phi_\alpha(k)\) of
\(\phi(k)\). The exponentiated insertion introduced by \cref{eq:nobdyshifted2}
merely shifts the Gaussian to be centred on
\(\phi_\alpha(k)=\phi_\varnothing(k)\),
\begin{align}
  \begin{split}
    \MoveEqLeft[2]
    \aleph^{-1} \mel{\varnothing}{\phi(k_1)\cdots\phi(k_m)}{\varnothing}_{\mathrm{BU}}
    \\
    &= \aleph_0^{-1}
      \int_{\mathcal{A}} \mathcal{D}\phi_\alpha
    \,
    \exp\left( 
    -\frac{1}{2}\int_{\mathscr{K}} \kappa(k)
    \left[\phi_\alpha(k)-\phi_\varnothing(k)\right]^*
    \left[\phi_\alpha(k)-\phi_\varnothing(k)\right]
       \right)
    \phi_\alpha(k_1)\cdots\phi_\alpha(k_m)
    \;.
  \end{split}
  \label{eq:shiftedalphapathint1}
\end{align}

Let us say a few words about the normalization
\(\aleph=\braket{\varnothing}{\varnothing}_{\mathrm{BU}}=\zeta(\varnothing)\) of
the no-boundary state. In principle, this should be defined by the gravitational
path integral \(\zeta(\varnothing)\) over spacetimes with no boundary. In a
simplified worldline model, we should consider sums of worldline diagrams that
represent such spacetimes. It is natural therefore to choose
\begin{align}
  \aleph
  &= \aleph_0 e^{\frac{1}{2} \lsem \Phi_\varnothing | \Phi_\varnothing \rsem}
    = \aleph_0 \exp\left(
    \frac{1}{2} \int_{\mathscr{K}} \kappa(k)
    \phi_\varnothing(k)^* \phi_\varnothing(k)
    \right)
    \;.
    \label{eq:oneptaleph}
\end{align}
This includes any previous contribution \(\aleph_0\) already present in the old
model, \eg{} from circular worldlines representing universes pair-created from
nothing and pair-annihilated to nothing. It also includes the factor
\(e^{\frac{1}{2} \lsem \Phi_\varnothing | \Phi_\varnothing \rsem}\) accounting
for individual universes smoothly nucleated from nothing and capping off to
nothing --- see \cref{fig:onepointfunc}. The gravitational path integral on each
such spacetime component yields
\(\frac{1}{2}\lsem \Phi_\varnothing | \Phi_\varnothing \rsem\), where there is a
symmetry factor of \(2\) for the exchange of the two one-point vertices, \ie{}
an exchange of the bra and ket halves of the spacetime.\footnote{Said
  differently, in the connected path integral \(\log\zeta(\varnothing)\), a
  generic configuration and its reflection across a codimension-one surface
  (where we evaluate \(\Phi_\varnothing\)) are counted as the same contribution
  to \(\log\zeta(\varnothing)\) but distinct contributions to
  \(\lsem\Phi_\varnothing|\Phi_\varnothing\rsem\). (To be clear, for oriented
  theories, reflections referred to in this footnote transform not only metric
  and matter fields, but also orientation.) As elsewhere in this paper,
  \(\lsem\bullet|\bullet\rsem\) is the inner product of invariant wavefunctions
  on cylinder-like spacetime topologies;
  \(\lsem\Phi_\varnothing|\Phi_\varnothing\rsem\) is evaluated by viewing
  \(\Phi_\varnothing(x)\) as such a wavefunction. But double-counting then
  occurs, because on a cylinder which has nontrivial boundaries, two bulk
  configurations related by a reflection along the cylinder's axis are
  regarded as gauge-inequivalent.

  It would be interesting to understand how this accounting works in the
  calculations of ref.~\cite{Cotler:2026jdz} in two-dimensional dilaton gravity.
  Comparing the sphere partition function with the norm of the Hartle-Hawking
  wavefunction, the above mentioned factor of \(2\) seems to be part of the
  factor of \(8=2^3\) discrepancy reported by ref.~\cite{Cotler:2026jdz} (v1).
  Another factor of \(2\) could plausibly be due to an incompatible treatment of
  the reflection on the \(S^1\), where the wavefunction was evaluated, relative
  to the treatment of the reflection on \(S^2\) in the partition function.
  (Whether the inner product \(\lsem\bullet|\bullet\rsem\) and the partition
  function should contain explicit divisions by the reflections on \(S^1\) and
  \(S^2\) respectively is related to whether the theory is unoriented and, if
  oriented, whether orientation is explicitly summed over instead of implicitly
  eliminated to already gauge-fix said reflection. See \cref{sec:physhilb} for a
  very brief introduction of \(\lsem\bullet|\bullet\rsem\),
  ref.~\cite{Held:2025mai} for more details, and \cref{app:pathintgauging} for a
  discussion of gauging reflections.) We do not yet see an obvious candidate for
  the final factor of \(2\). \label{foot:symmetryfactor}} Finally, the
one-universe contribution is then exponentiated to give the contribution from
multiple indistinguishable universes. With this choice of normalization,
\cref{eq:shiftedalphapathint1} can be expressed as
\begin{align}
  \begin{split}
    \MoveEqLeft[2]
    \mel{\varnothing}{\phi(k_1)\cdots\phi(k_m)}{\varnothing}_{\mathrm{BU}}
    \\
    &=
      \int_{\mathcal{A}} \mathcal{D}\phi_\alpha
    \,
      \exp\left(
      \int_{\mathscr{K}} \kappa(k)
      \left[
      -\frac{1}{2}\phi_\alpha(k)^* \phi_\alpha(k)
      + \phi_\alpha(k)^* \phi_\varnothing(k)
      \right]
      \right)
    \phi_\alpha(k_1)\cdots\phi_\alpha(k_m)
    \;,
  \end{split}
  \label{eq:shiftedalphapathint2}
\end{align}
where
\begin{align}
  \exp\left(
  \int_{\mathscr{K}} \kappa(k)
  \phi_\alpha(k)^* \phi_\varnothing(k)
  \right)
  =
  \exp\left(
  \int_{\mathscr{K}} \kappa(k)
  \phi_\alpha(k) \phi_\varnothing(k)^*
  \right)
\end{align}
inserts a gas of one-point vertices that seed the nucleation/capping off of
universes from and to nothing. Expanding out this exponential in a special case
of \cref{eq:shiftedalphapathint2} confirms that
\(\aleph=\braket{\varnothing}{\varnothing}_{\mathrm{BU}}\) indeed includes a sum of
indistinguishable vacuum diagrams given by a Wick contraction between two
one-point vertices (as well as the overall normalization \(\aleph_0\) of the
centred Gaussian path integral).

\subsubsection{Exact Fock space description of the topological Marolf--Maxfield model}
\label{sec:topmodel}

We saw in \cref{sec:uft,sec:onepoint} that baby universe theory admits a useful
Fock space description at the level of disk-like and cylinder-like topologies.
Here, we present a very special example of a theory where even higher-point
topology-changing interactions can be resummed exactly in a Fock space
description. This provides a concrete example where we know how to exactly
correct the disk and cylinder approximation of the universe field. We by no
means expect this to be generic, and part of \cref{sec:nextsteps} will be
dedicated to discussing future work studying richer theories.

For completeness, we first briefly define the Marolf--Maxfield model
\cite{Marolf:2020xie} to be considered in this section and then explain how the
general ingredients introduced in this paper are realized in this model. In
particular, the model provides a good opportunity to discuss the interplay of
orientation, gauged spatial reflection, and the (ungauged) \(\star\)
transformation relating boundary conditions for bra- and ket-states. Expert
readers may choose to skip immediately to the Fock space construction at the end
of this section.

\paragraph{The model} we will consider in this section is the topological model of 2D
gravity introduced by Marolf and Maxfield \cite{Marolf:2020xie}, in the closed
cosmology sector without end-of-the-world branes. This theory is purely
topological so there are no local fields; each boundary component is an oriented
topological circle \(S^1\) and each bulk configuration is an oriented
topological 2D spacetime \(M\) interpolating between boundary circles. In fact,
the boundary circles and \(M\) are required to be oriented in this model --- we
will return to this point further below. The path integral for this theory is
just a sum over these oriented topological manifolds \(M\) with the prescribed
boundary circles,
\begin{align}
  \zeta((S^1)^{\sqcup m})
  &= \zeta(\underbrace{S^1 \sqcup \cdots \sqcup S^1}_{\text{\(m\)-many}})
    = \sum_{\substack{\text{\(M\) with}\\ \partial M=(S^1)^{\sqcup m}}} \frac{e^{S_0 \tilde{\chi}(M)}}{\prod_g m_g(M)!}
  \;.
  \label{eq:mmzeta}
\end{align}
Here, \(m_g(M)\) is the number of genus-\(g\) components of \(M\) with no
boundary, so \(\prod_g m_g(M)!\) can be thought of as a symmetry factor
associated with exchanging such components; \(S_0>0\) is a constant; and
\begin{align}
  \tilde{\chi}(M)
  &= \chi(M) + \abs{\partial M}
\end{align}
is the sum of the Euler characteristic \(\chi(M)\) of \(M\) and the number
\(\abs{\partial M}\) of boundary components. 
This provides an exponential suppression \(e^{-2S_0}\) to bulk manifolds of
increasingly high topology, modelling similar behaviour in JT gravity and
worldsheet string theory. The addition of \(\abs{\partial M}\) is just a
convenient normalization convention at the level of the closed cosmology theory
(but is important for ensuring positivity if one considers end-of-the-world
branes or open universes \cite{Marolf:2020xie,Marolf:2024jze}).

This model can be solved exactly by noting that the connected path integral
\(\zeta_{\mathrm{con}}\) --- comprised of contributions to \(\zeta\)
from connected \(M\) --- is simply a geometric series in genus \(g\),
\begin{align}
  \zeta_{\mathrm{con}}((S^1)^{\sqcup m})
  &=\sum_{\substack{\text{Connected \(M\)}\\ \text{with \(\partial M=(S^1)^{\sqcup m}\)}}}
  e^{S_0 \tilde{\chi}(M)}
  = \sum_{g=0}^\infty e^{S_0(2-2g)}
  = \frac{e^{2S_0}}{1-e^{-2S_0}}
  \equiv \lambda
    \;,
    \label{eq:mmconnected}
\end{align}
and in particular is independent of \(\abs{\partial M}=m\). Using this result,
ref.~\cite{Marolf:2020xie} computes the generating function for \cref{eq:mmzeta}
to deduce an exact description of the baby universe Hilbert space in terms of a
Poisson distribution over \(\alpha\)-sectors. We will instead derive an
equivalent, but more intuitive Fock space description of this baby universe
Hilbert space.

\paragraph{One-universe ingredients.} To orient ourselves, let us note how the
general objects introduced elsewhere in this paper are realized in this model.
We can take the target space \(\mathscr{M}=\{x\}=\{S^1\}\) to be comprised of a
single element, a topological circle. The one-universe Hilbert space is
therefore one-dimensional,
\(\mathcal{H}_{\text{wl kin}}=\ell^2(\mathscr{M},e^{-2S_0})=e^{S_0}\mathbb{C}\),
where the second argument denotes the discrete measure \(\mu(S^1)=e^{-2S_0}\)
and, as elsewhere, an overall factor \(e^{S_0}\) multiplying a Hilbert space
\(\mathbb{C}\) denotes a rescaling of its norm relative to the naively canonical
one. Thus,
\begin{align}
  \braket{S^1}{S^1}_{\mathrm{wl}}=\delta_\mu(S^1,S^1)=e^{2S_0}
\end{align}
matches the cylinder amplitude in \cref{eq:mmzeta}. Being topological, the
theory has an identically zero Wheeler--DeWitt Hamiltonian \(H=0\) on cylinder
topologies, and the group average \(\eta=1\) is also trivial as there is no
``lapse'' to integrate over. So, in fact
\begin{align}
  e^{S_0} \mathbb{C}
  &=\mathcal{H}_{\text{wl kin}}
    =\mathcal{D}_{\text{wl test}}
    =\mathcal{H}_{\text{wl co}}
    =\mathcal{H}_{\text{wl inv}}
    \;.
    \label{eq:mmwlhilb}
\end{align}

So far, we have glossed over the fact that the theory is oriented and
configurations carry orientation. In principle, we could have started with a
larger kinematic target space
\(\tilde{\mathscr{M}}=\{\circlearrowleft,\circlearrowright\}\) with two elements
given by circles with opposite orientation, and thus a two-dimensional kinematic
Hilbert space
\(\tilde{\mathcal{H}}_{\text{wl kin}}=\ell^2(\tilde{\mathscr{M}},e^{-2S_0})=e^{S_0}\mathbb{C}^2\). But then
we would need to account for spatial diffeomorphisms; in this topological
theory, the only resulting nontrivial transformation is an orientation-reversing
reflection \(\mathcal{R}\) which maps
\(\ket{\circlearrowleft}_{\mathrm{wl}}\leftrightarrow\ket{\circlearrowright}_{\mathrm{wl}}\)
and extends to a \emph{linear} transformation on
\(\tilde{\mathcal{H}}_{\text{wl kin}}\). As described in
\cref{app:pathintgauging}, \(\mathcal{R}\) is gauged at the cylinder level and
in the full path integral \labelcref{eq:mmzeta}\footnote{By the same reason as
  presented for gauging time evolution in \cref{sec:generalprop}, gauging
  \(\mathcal{R}\) at the cylinder level implies gauging at the level of the full
  path integral.}, which explains why \cref{eq:mmzeta} is independent of the
orientation of the \(S^1\) insertions. In particular, a group average
\begin{align}
  \tilde{\eta}
  &= 1 + \mathcal{R}
    = e^{-2S_0}
    \left[
    (\ketbra{\circlearrowleft}{\circlearrowleft}_{\mathrm{wl}}
    +\ketbra{\circlearrowright}{\circlearrowright}_{\mathrm{wl}})
    + (\ketbra{\circlearrowleft}{\circlearrowright}_{\mathrm{wl}}
    +\ketbra{\circlearrowright}{\circlearrowleft}_{\mathrm{wl}})
    \right]
\end{align}
is included in the cylinder amplitude
\(\llangle \bullet |\bullet\rrangle_{\mathrm{wl}}=\langle
\bullet|\tilde{\eta}|\bullet \rangle_{\mathrm{wl}}\) between states in
\(\tilde{\mathcal{D}}_{\text{wl test}}=\tilde{\mathcal{H}}_{\text{wl kin}}\).
This cylinder amplitude gives an inner product on a one-dimensional
\(\mathcal{R}\)-coinvariant Hilbert space; in \cref{eq:mmwlhilb}, we skipped
these steps and jumped immediately to this coinvariant Hilbert space
\(\mathcal{H}_{\text{wl co}}\), where we now understand \(S^1\) to mean the
equivalence class \(\{\circlearrowleft,\circlearrowright\}\) generated by
\(\mathcal{R}\). To cut down on redundant notation, we will start our
construction of the baby universe theory from \cref{eq:mmwlhilb}.

But first, let us define the operation \(\star\) which relates boundary
conditions preparing ket-states to those preparing bra-states. Because the
theory is oriented, as explained in \cref{sec:orientrev}, \(\star\) should
exchange
\(\ket{\circlearrowleft}_{\mathrm{wl}}\leftrightarrow\ket{\circlearrowright}_{\mathrm{wl}}\),
but (in contrast to \(\mathcal{R}\)) extend to an \emph{antilinear}
transformation on \(\tilde{\mathcal{H}}_{\text{wl kin}}\). Because \(\star\) and
\(\mathcal{R}\) commute, \(\star\) descends to the \(\mathcal{R}\)-quotient
\labelcref{eq:mmwlhilb} where it simply acts as complex conjugation
\(c \ket{S^1}_{\mathrm{wl}}\mapsto c^* \ket{S^1}_{\mathrm{wl}}\).

\paragraph{Fock space construction.} Given the above ingredients, the machinery
developed in \cref{sec:uft,sec:onepoint} allows us to describe a truncation of
the baby universe theory defined by restricting the path integral
\labelcref{eq:mmzeta} to disks and cylinders. Since
\(\mathcal{H}_{\text{wl co}}=\mathcal{H}_{\text{wl inv}}\) is one-dimensional,
the baby universe Hilbert space \(\mathcal{H}_{\mathrm{BU}}\), a Fock space
\labelcref{eq:hbufock} over \(\mathcal{H}_{\text{wl inv}}\), is just the Hilbert
space of a single harmonic oscillator. It will be convenient for us to adopt a
slightly different convention relative to
\cref{eq:nobdyshiftedu,eq:nobdyshifted,eq:nobdyshiftop}: here, we leave
\(\ket{\varnothing}_{\mathrm{BU}}\) proportional\footnote{But, we still maintain
  that the normalization of \(\ket{\varnothing}_{\mathrm{BU}}\) is given by the
  no-boundary path integral,
  \(\aleph\equiv\braket{\varnothing}{\varnothing}_{\mathrm{BU}}\equiv\zeta(\varnothing)\).
  To leading order, the connected path integral
  \(\log\zeta(\varnothing)=\lambda\approx e^{2S_0}\) is dominated by the sphere.
  The astute reader will notice this expression actually disagrees with
  \cref{eq:oneptaleph} which would have predicted
  \(\log\aleph\stackrel{?}{\approx}
  \frac{1}{2}\lsem\Phi_\varnothing|\Phi_\varnothing\rsem \approx
  \frac{1}{2}e^{2S_0}\). This disagreement is an artifact of the topological
  nature of the theory --- the lack of local field fluctuations in the
  gravitational path integral invalidates the argument presented in
  \cref{foot:symmetryfactor}.} to the oscillator ground state,
\(a\ket{\varnothing}_{\mathrm{BU}}=0\), and instead absorb the shift
\labelcref{eq:nobdyshiftop} into the operator,
\begin{align}
  Z(S^1)
  &=\Phi(S^1)
  \approx \Phi_\varnothing(S^1) + e^{S_0} (a + a^\dagger)
    \;,
    \label{eq:mmphiapprox}
  \\
    [a,a^\dagger]
  &=1
    \;,
\end{align}
where again the shift
\begin{align}
  \Phi_\varnothing(S^1)
  &\equiv
    \frac{
    \mel{\varnothing}{\Phi(S^1)}{\varnothing}_{\mathrm{BU}}
    }{
    \braket{\varnothing}{\varnothing}_{\mathrm{BU}}
    }
    \approx e^{2S_0}
    \label{eq:mmdisk}
\end{align}
is approximated by the disk-level Hartle--Hawking wavefunction. Indeed,
\cref{eq:mmphiapprox} already appeared in Marolf and Maxfield's original paper
\cite{Marolf:2020xie}.

\begin{figure}
  \centering
  \begin{subfigure}[t]{0.48\textwidth}
    \centering
    \includegraphics[width=\textwidth]{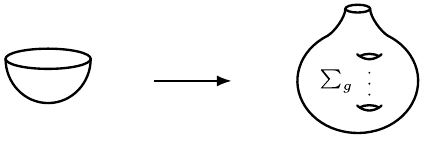}
    \caption{Replacing the disk with the renormalized disk where genus \(g\) has
      been resummed.}
    \label{fig:mmrenormdisk}
  \end{subfigure}
  \hfill
  \begin{subfigure}[t]{0.48\textwidth}
    \centering
    \includegraphics[width=\textwidth]{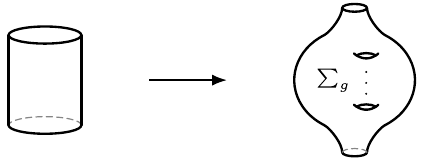}
    \caption{Replacing the cylinder with the renormalized cylinder where genus
      \(g\) has been resummed.}
    \label{fig:mmrenormcylinder}
  \end{subfigure}
  \par\bigskip
  \begin{subfigure}[t]{0.48\textwidth}
    \centering
    \includegraphics[width=\textwidth]{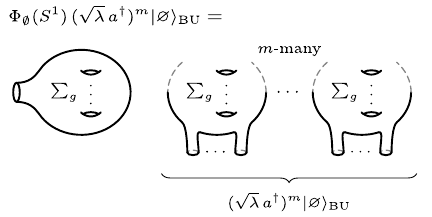}
    \caption{First term \(\Phi_\varnothing(S^1)\) of \cref{eq:mmphiexact} adds a
      connected spacetime that caps off to a renormalized disk.}
    \label{fig:mmdiskterm}
  \end{subfigure}
  \hfill
  \begin{subfigure}[t]{0.48\textwidth}
    \centering
    \includegraphics[width=\textwidth]{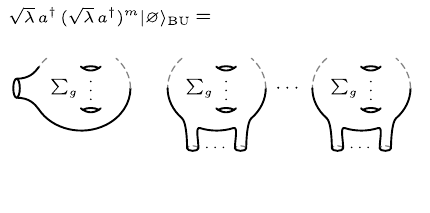}
    \caption{The \(\sqrt{\lambda} a^\dagger\) term of \cref{eq:mmphiexact}
      starts a new connected renormalized spacetime initially with one boundary.
    This connected spacetime does not cap off by itself, the possibility of
    which is already captured by \cref{fig:mmdiskterm}; it will need to be terminated
    later, as described in \cref{fig:mmannihilationterm}.}
    \label{fig:mmcreationterm}
  \end{subfigure}
  \par\bigskip
  \begin{subfigure}[t]{0.48\textwidth}
    \centering
    \includegraphics[width=\textwidth]{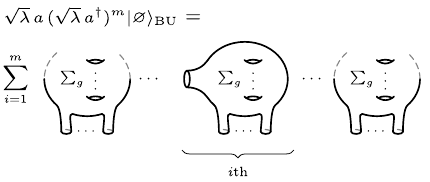}
    \caption{The \(\sqrt{\lambda} a\) term of \cref{eq:mmphiexact} terminates,
      with one last boundary, a surviving connected renormalized spacetime.
      Applied to a state
      \((\sqrt{\lambda}a^\dagger)^m\ket{\varnothing}_{\mathrm{BU}}\) with
      \(m\)-many surviving connected renormalized spacetimes, we would obtain a
      sum iterating through each such survivor to be acted on.}
    \label{fig:mmannihilationterm}
  \end{subfigure}
  \hfill
  \begin{subfigure}[t]{0.48\textwidth}
    \centering
    \includegraphics[width=\textwidth]{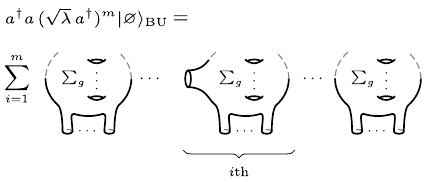}
    \caption{The \(a^\dagger a\) term of \cref{eq:mmphiexact} adds an extra
      boundary to a surviving renormalized spacetime without terminating it. It
      similarly iterates across all surviving spacetime components.}
    \label{fig:mmnumberterm}
  \end{subfigure}
  \caption{Pictorial explanation of the exact formula \labelcref{eq:mmphiexact}
    for the boundary-inserting operator/universe field \(Z(S^1)=\Phi(S^1)\) in
    the Marolf--Maxfield topological model.}
  \label{fig:mmphiexact}
\end{figure}

Slightly more nontrivial is the fact that we can make the Fock space description
exact by renormalizing the terms already in \cref{eq:mmphiapprox} and adding
just \emph{one} additional term,
\begin{align}
  Z(S^1)
  &=\Phi(S^1)
    = \Phi_\varnothing(S^1) + \sqrt{\lambda} (a + a^\dagger)
    + a^\dagger a
    \;,
    \label{eq:mmphiexact}
  \\
  \Phi_\varnothing(S^1)
  &= \lambda
    \;.
\end{align}
The second line is a special case of \cref{eq:mmconnected} and can be understood
as replacing the disk approximation \labelcref{eq:mmdisk} by a ``renormalized
disk'' that resums all handle-disks. The renormalization
\(e^{S_0}\to \sqrt{\lambda}\) in the linear terms of
\cref{eq:mmphiapprox,eq:mmphiexact} can likewise be understood as replacing each
cylinder by a ``renormalized cylinder'' that resums all connected two-boundary
manifolds of arbitrary genus. See \cref{fig:mmrenormdisk,fig:mmrenormcylinder}.

As illustrated in
\cref{fig:mmdiskterm,fig:mmcreationterm,fig:mmannihilationterm,fig:mmnumberterm},
the action of each term in \cref{eq:mmphiexact} on a state has a precise
diagrammatic meaning in terms of such connected renormalized spacetimes, where
the number of handles has been resummed. Namely, the boundary-inserting operator
\(Z(S^1)=\Phi(S^1)\) can: create with this boundary a renormalized disk
(\(\Phi_\varnothing(S^1)\)); create with this boundary a new connected
renormalized spacetime that does not cap off to a renormalized disk
(\(\sqrt{\lambda} a^\dagger\)); terminate at this boundary such a surviving
spacetime (\(\sqrt{\lambda} a\)); or add this boundary to such a surviving
spacetime without terminating it (\(a^\dagger a\)). By acting on the no-boundary
state \(\ket{\varnothing}_{\mathrm{BU}}\) successively with the sum of these
terms and finally selecting with \(\bra{\varnothing}_{\mathrm{BU}}\) terms where
all spacetimes have capped off or have been terminated, it is straightforward to
see that we build up each and every spacetime-with-boundary in the full
gravitational path integral \labelcref{eq:mmzeta} exactly once. The coefficients
in \cref{eq:mmphiexact} have been chosen precisely so that each connected
renormalized spacetime-with-boundary is weighted by \(\lambda\), as required by
\cref{eq:mmconnected}. Provided that the multiplicative contribution from
manifolds with no boundary is supplied by the normalization of
\(\ket{\varnothing}_{\mathrm{BU}}\) itself,
\begin{align}
  \aleph
  &\equiv \braket{\varnothing}{\varnothing}
    \equiv \zeta(\varnothing)
    = e^\lambda
    \;,
\end{align}
we then conclude that indeed
\begin{align}
  \mel{\varnothing}{\Phi(S^1)^m}{\varnothing}_{\mathrm{BU}}
  &= \zeta((S^1)^{\sqcup m})
    \;.
\end{align}

In the disks and cylinders approximation \cref{eq:mmphiapprox}, the occupation
number \(a^\dagger a\) labelled the number of surviving cylindrical universes.
The intuitive proof of the exact formula \cref{eq:mmphiexact} given above shows
that the occupation number \(a^\dagger a\) now labels the number of surviving
connected spacetimes-with-boundary.

In ref.~\cite{Marolf:2020xie}, by first determining the generating function of
the gravitational path integral \labelcref{eq:mmzeta} and subsequently finding
that \(\alpha\)-sectors are labelled by integer eigenvalues of
\(Z(S^1)=\Phi(S^1)\), Marolf and Maxfield then realized that
\(Z(S^1)=\Phi(S^1)\) can itself be identified with a connected-spacetime-number
operator:
\begin{align}
  Z(S^1)
  &=\Phi(S^1)
  = a_0^\dagger a_0
    \;,
  &
    [a_0,a_0^\dagger]
  &= 1
    \;,
    \label{eq:mmunshiftednumber}
  \\
  \ket{\varnothing}_{\mathrm{BU}}
  &= e^{\sqrt{\lambda} a_0^\dagger} \ket{0}_{\mathrm{BU}}
    \;
  &
    a_0 \ket{0}_{\mathrm{BU}}
  &= 0
    \;.
\end{align}
We can easily check that this is related to our formulation by
\begin{align}
  a &= a_0 - \sqrt{\lambda}
      \;.
\end{align}
While Marolf and Maxfield arrived at \labelcref{eq:mmunshiftednumber} by working
backwards from the spectrum of \(Z(S^1)=\Phi(S^1)\), we deduced our exact
expression \labelcref{eq:mmphiexact} by explicitly building up spacetimes using
creation and annihilation operators. This provides one example showing how the
disks and cylinders approximation, \cref{eq:mmphiapprox,eq:mmdisk}, can be
promoted to the exact result.

Unfortunately, as forewarned, our success relied heavily on the simplicity of
this model. In \cref{app:generalizedfock}, we generalize the Fock space
representation \labelcref{eq:mmphiexact} of the universe field \(\Phi(x)\) to
baby universe theories that have Hermitian, conditionally positive connected
path integrals \(\zeta_{\mathrm{con}}\). In the special case of the
Marolf--Maxfield model, the fact that the connected path integral
\labelcref{eq:mmconnected} can be evaluated explicitly is a miracle; the fact
that it happens to be independent of boundary insertions is a second miracle.
Without the former miracle, the Fock space and its Weyl algebra that we write
down in \cref{app:generalizedfock} remain somewhat abstract; without the latter,
the Fock space can be over a highly nontrivial Hilbert space and the algebra
generically closes only with the inclusion of operators labelled by arbitrarily
many boundary components. All we essentially gain is an algebraic way to
organize the combinatorics needed to translate between the connected and full
gravitational path integrals. Moreover, the construction simply doesn't work
when we do not have conditional positivity of \(\zeta_{\mathrm{con}}\) (which is
strictly stronger than positivity of \(\zeta\)), as is the case, for example, in
JT gravity.

It is important therefore to find other ways to incorporate nontrivial topology
change in GABUFT. We discuss some future directions along these lines in
\cref{sec:nextsteps}.

\subsection{Future directions}
\label{sec:nextsteps}

We end with a discussion of some future directions.

\subsubsection{Contours for the gravitational path integral and nontrivial
  topology}
\label{sec:topologycontours}

Traditionally, topology change has been most commonly associated with
``Euclidean'' gravitational path integrals. But, as described in
\cref{sec:interactions}, the conformal factor problem prevents the gravitational
path integral from being straightforwardly defined over real Euclidean
configurations \cite{Gibbons:1978ac}. ``Euclidean'' gravitational path integrals
therefore cannot be truly Euclidean and often involve some nontrivial contour
prescriptions for target space variables \cite{Louko:1995jw}. Consider, for
example, JT gravity. If one were to integrate both the metric and dilaton over
real Euclidean configurations, the JT path integral would be manifestly
divergent. But upon Wick-rotating the dilaton, the dilaton path integral becomes
a \(\delta(R+2)\)-function that localizes the metric path integral onto
geometries with constant curvature \(R=-2\). A related contour issue arises in
worldsheet string theory. In Lorentzian target space, the time-like variable
\(x^0\) has a wrong-signed kinetic term (similar to the conformal mode, in
\cref{eq:miniham,eq:misnersupermetric}). Euclidean worldsheet path integrals ---
where the worldsheet metric is Euclidean --- will often diverge unless the
integration contour for \(x^0\) is Wick rotated so that target space also
becomes Euclidean.

Given various possible choices of contour, which should we pick if our interest
is in describing gravity in closed cosmologies? Since we live in a Lorentzian
spacetime where we see real observables, the most natural choice seems to be a
contour over real Lorentzian configurations. This was the choice we made when
defining the path integrals for free worldline models in
\cref{sec:worldline,sec:uft}. Firstly, this avoids the conformal factor problem.
Secondly, as reviewed in \cref{sec:groupaverage}, integrating the lapse metric
component over all real Lorentzian values gives the gravitational path integral
a desirable interpretation as a physical inner product, group-averaging
\(\eta=\int_{-\infty}^\infty \dd{T} e^{-i T H}=2\pi\delta(H)\) over time
translations. Moreover, as discussed in \cref{sec:discussqftvsuft}, the
resulting GABUFT approach seems to give a more physically reasonable baby
universe theory of recollapsing cosmologies, compared to a QFT-like approach. We
further postulated in \cref{sec:generalprop} that, even on a spacetime manifold
with nontrivial topology, similar lapse integrals on cylindrical near-boundary
regions of the manifold will similarly project boundary data onto solutions
annihilated by the Wheeler--DeWitt Hamiltonian \(H\).

We have not, however, specified how we would like to define a Lorentzian
gravitational path integral on a manifold with nontrivial topology. In fact,
topology change is forbidden on \emph{smooth} Lorentzian manifolds. To allow
topology change, one might instead integrate over (almost) Lorentzian
configurations that can have conical singularities, like a crotch singularity
that appears in a Lorentzian pair-of-pants \cite{Louko:1995jw}. Such
singularities can contribute imaginary parts to the Lorentzian action,
recovering the expected enhancement or suppression of different topologies. This
prescription has been applied in several contexts, including the calculation of
partition functions \cite{Marolf:2022ybi,Dittrich:2024awu}, R\'enyi entropies
\cite{Colin-Ellerin:2020mva,Colin-Ellerin:2021jev,Marolf:2021mbi,Held:2024qcl},
and spectral form factors \cite{Blommaert:2023vbz}. It is shown that desirable
results traditionally obtained from Euclidean methods can be recovered from
Lorentzian gravitational path integrals in this way, at least at the leading
classical order. In ongoing work \cite{Chen:2026wip}, we are also exploring how
the Hartle--Hawking wavefunction can be similarly recovered using a one-boundary
Lorentzian gravitational path integral.

Extending such results beyond leading order, even in the simplest models,
requires care. Analyses in JT gravity, for example, show that recovering
familiar Euclidean disk amplitudes from formulations involving conical
singularities requires a careful treatment of the associated moduli and
path-integral measures \cite{Blommaert:2023vbz,Lin:2023wac}. While some initial
progress has been made in formulating and extracting properties of Lorentzian JT
path integrals on certain higher-topology geometries \cite{Blommaert:2023vbz}, a
controlled organization of the Lorentzian moduli spaces and their measures that
makes higher-topology amplitudes systematically and recursively computable
remains lacking. In particular, it would be valuable to understand whether the
topological recursion governing higher-genus JT amplitudes admits a direct
Lorentzian derivation.


Stepping back from these technical issues, there is a more basic question: why,
and to what extent, should a Lorentzian gravitational path integral reproduce
the standard ``Euclidean'' JT amplitudes in the first place? Let us note that
the standard ``Euclidean'' JT path integral bears close resemblance to a
Lorentzian gravitational path integral if we view the dilaton as giving the
length along an extra, third dimension. Integrating over imaginary values of the
dilaton is then like integrating over values of a timelike length --- a real
Lorentzian lapse --- in the third dimension. But this is not manifestly the same
as a path integral over configurations that have a time-like direction in the
original two dimensions of the JT spacetime. It would be interesting to
understand precisely how calculations obtained from these two path integrals are
related.

A related question concerns the boundary conditions used to prepare states for
closed universes and open universes. In the baby universe framework
\cite{Marolf:2020xie}, closed universe states are prepared by complete boundary
manifolds, while open universe states are prepared by partial boundary manifolds
which can be glued into complete boundary manifolds when computing inner
products. It is often assumed that the set of complete boundary manifolds that
prepare closed universe states and the set that compute inner products of open
universe states are the same. But, is this really true if the gravitational path
integral is formulated over Lorentzian configurations? For closed universes,
states are most naturally prepared by Euclidean boundary manifolds and the bulk
timelike direction runs transverse to the boundary. On the other hand, time for
an open universe naturally runs along the boundary. In particular, when
ref.~\cite{Marolf:2022ybi} evaluates the thermal partition function \(Z(\beta)\)
of an open universe, the Lorentzian gravitational path integral is not
immediately evaluated with a Euclidean boundary condition; instead, the
procedure is to first consider a Lorentzian boundary, with a Lorentzian time
period that is only later replaced by a Euclidean thermal period \(\beta\)
through an integral transform. It would be interesting to better understand
whether this contour choice is really equivalent to a Euclidean boundary
condition that one would use to prepare a closed universe state in a Lorentzian
gravitational path integral.\footnote{See \cite{Held:2026huj} for closely
  related comments in the context of axion wormholes, where it was unclear
  whether Lorentzian formulations of the gravitational path integral in which
  time runs respectively transverse to and along the boundaries are really
  equivalent. The author thanks Don Marolf for discussion on this point, and for
  bringing to my attention a work in progress \cite{Antonini:2026wip} aimed at
  clarifying this issue.}

\subsubsection{Group averaging, topology, and perturbation theory}
\label{sec:topologyperturbation}

In \cref{sec:groupaverage,sec:generalprop}, we discussed how the lapse integral
near each boundary component gives rise to a group average over time
translations that projects onto solutions annihilated by the Wheeler--DeWitt
Hamiltonian \(H\) for cylinder-like universes. In \cref{app:pathintgauging}, we
also explain that, by integrating over shift and summing over inequivalent ways
of attaching a given boundary component to a given near-boundary cylindrical leg
of the bulk, the gravitational path integral also group averages over spatial
diffeomorphisms of the boundary component. Thus, the gravitational path integral
is invariant under the diffeomorphisms of each boundary component. Consequently,
the difference \(j-j'\) between kinematic states related by a diffeomorphism
\(j\stackrel{\mathrm{Diff}}{\sim} j'\) becomes null in the physical Hilbert
space --- this is true for both the single universe theory and the baby universe
theory.

Given multiple boundary components \(j_1 \sqcup \cdots\sqcup j_m\) in baby
universe theory, the gravitational path integral \(\zeta\) will similarly sum
over permutations of the boundary components. That is, for every bulk
configuration appearing in the gravitational path integral
\(\zeta(j_1 \sqcup \cdots\sqcup j_m)\) and every permutation \(\sigma\in S_m\),
there will be another\footnote{For certain topologies in topological theories,
  it may be the case that this second configuration is really the same as the
  original bulk configuration --- this contribution to the path integral then
  satisfies \cref{eq:permutationinv} by itself, without needing to sum over
  images. In theories where local field configurations are path-integrated in
  the bulk, we expect only a measure zero set of the integrated configurations
  to be individually invariant.} bulk configuration obtained by permuting the
components \(j_i\to j_{\sigma(i)}\) before attaching them to the same
bulk.\footnote{It may be the case that the permuted boundary conditions are then
  incompatible with the bulk, \ie{} one or more test functions
  \(j_{\sigma(i)}(x_i)\) vanish on the boundary configuration \(x_i\) supplied
  by the bulk. In this case, the latter bulk configuration with permuted
  boundary conditions will just give a vanishing contribution to the path
  integral. This does not invalidate what is said in the main text.} Because of
this sum over permutations, the path integral becomes invariant under
permutations,
\begin{align}
  \zeta(j_1 \sqcup \cdots\sqcup j_m)
  &= \zeta(j_{\sigma(1)} \sqcup \cdots\sqcup j_{\sigma(m)})
    \;.
  &
    (\sigma\in S_m)
    \label{eq:permutationinv}
\end{align}
This is why the multiplication operation \(\sqcup\) in the algebra
\(\mathcal{J}\) of multi-boundary conditions is taken to be commutative. Even if
we started with a larger ``kinematic'' space with a non-commutative disjoint
union, the path integral \(\zeta\) contains a group average over permutations
that will project out the difference
\(\ket{j_1\sqcup j_2}-\ket{j_2 \sqcup j_1}\).\footnote{See
  \cref{app:dckinematics}, which goes through this exercise for the
  Marolf--Maxfield model.} The permutation of boundary components can be thought
of as a generalization of the single-component spatial diffeomorphisms mentioned
in the previous paragraph.

Is there a similar generalization for time evolution? While the previously
discussed Wheeler--DeWitt Hamiltonian \(H\) generates Lorentzian time evolution
along cylindrical portions of spacetime where topology remains constant, let us
now discuss evolution across nontrivial topology-changing processes.

\paragraph{The Marolf--Maxfield model as a case study.}
For illustrative purposes, let us consider again the topological Marolf--Maxfield
model \cite{Marolf:2020xie} reviewed in \cref{sec:topmodel}, where \(H=0\) is
trivial and the only time evolution is due to topology change.

Let us take as a starting point the truncated theory of only disks and
cylinders. This theory already includes some simple topology-changing
processes --- the swapping, capping off/nucleation, and
pair-creation/-annihilation of universes --- but these are easily captured
by a Gaussian baby universe model of the kind studied in detail in this paper.
In particular, the boundary inserting operator is given exactly by
\cref{eq:mmphiapprox,eq:mmdisk},
\begin{align}
  Z_{\mathrm{DC}}(S^1)
  &= e^{2S_0} + e^{S_0}(a + a^\dagger)
    \;,
  &
    \ket{(S^1)^{\sqcup m}}_{\mathrm{DC}}
  &= Z_{\mathrm{DC}}(S^1)^m \ket{\varnothing}_{\mathrm{DC}}
    \;,
    \label{eq:dcstates}
  &
    a \ket{\varnothing}_{\mathrm{DC}}
  &= 0
    \;.
\end{align}
Here and below where ambiguous, we will label the disk-and-cylinder baby
universe theory with subscript DC. For definiteness, we will take normalization
of \(\ket{\varnothing}_{\mathrm{DC}}\) to be given by the sphere and torus,
\begin{align}
  \log \braket{\varnothing}{\varnothing}_{\mathrm{DC}}
  &= e^{2S_0} + 1
    \;.
\end{align}
It will be helpful to introduce the
vacuum-subtracted and normalized operator
\begin{align}
  z &\equiv \frac{Z_{\mathrm{DC}}(S^1)-e^{2S_0}}{e^{S_0}}
      = a + a^\dagger
      \;.
      \label{eq:zcentredunitvar}
\end{align}
In the following, we will relate the inner product
\(\braket{\bullet}{\bullet}_{\mathrm{DC}}\) of the disk-and-cylinder theory to
that of the full Marolf--Maxfield model. While we focus on this step because it
is nontrivial and conceptually interesting, \cref{app:dckinematics} explains how
the disk-and-cylinder inner product can itself be obtained from a more basic
kinematic inner product where even the aforementioned simple topology-changing
processes are absent.

Denoting the exact inner product of the full Marolf--Maxfield model with
subscript BU as before, we will give an insertion \(\eta_{\mathrm{pants}}\) such
that
\begin{align}
  \braket{(S^1)^{\sqcup m'}}{(S^1)^{\sqcup m}}_{\mathrm{BU}}
  &= \mel{(S^1)^{\sqcup m'}}{\eta_{\mathrm{pants}}}{(S^1)^{\sqcup m}}_{\mathrm{DC}}
    \;.
    \label{eq:pantsip}
\end{align}
The analogy we are trying to evoke here is that
\(\Span\left\{\ket{(S^1)^{\sqcup m}}_{\mathrm{DC}}\mid m\ge 0\right\}\) is
playing the role of a test space of ``kinematic states'' while the insertion
\(\eta_{\mathrm{pants}}\) is a sum over different amounts of evolution generated
by pairs-of-pants. In particular, the first term in \(\eta_{\mathrm{pants}}\)
will be the identity, the following term will insert one pair of pants, the next
will insert two, and so on. This is not obviously a group average over time
translations, because we don't seem to have a group. Nonetheless, there is a
sense in which \(\eta_{\mathrm{pants}}\) is a sum over elapsed times as measured
by counting the number of pair-of-pants processes that have occurred.

Because each pair-of-pants is suppressed by \(e^{-S_0}\) this sum corresponds to
the expansion of \(\eta_{\mathrm{pants}}\) in \(e^{-S_0}\). The first few terms
of \(\eta_{\mathrm{pants}}\) are
\begin{align}
  \eta_{\mathrm{pants}}
  &= 1 + e^{-S_0} \left(
    \frac{1}{3!} z^3
    + \frac{1}{2} z
    \right)
    + e^{-2S_0} \left(
    \frac{1}{72} z^6
    + \frac{3}{8} z^2
    + \frac{5}{12}
    \right)
    + O(e^{-3S_0})
    \;.
\end{align}
To show that these terms are correct (especially the \(\sim e^{-2S_0}\) terms), it
is helpful to organize them into an exponential,
\begin{align}
  \eta_{\mathrm{pants}}
  &= \exp\left[
    e^{-S_0}\left(
    \frac{1}{3!} z^3
    + \frac{1}{2} z
    \right)
    + e^{-2S_0}\left(
    - \frac{1}{12} z^4
    + \frac{1}{4} z^2
    + \frac{5}{12}
    \right)
    + O(e^{-3S_0})
    \right]
    \;,
    \label{eq:pantsexp}
\end{align}
and consider another interpretation of \cref{eq:pantsip}. Namely,
\(\eta_{\mathrm{pants}}\) supplies the necessary interactions needed to turn the
disk-and-cylinder theory's Gaussian integral over \(\alpha\)-sectors into the
integral of the exact theory. Naively, one might have only expected to
see the very first term of \cref{eq:pantsexp},
\begin{align}
  \eta_{\mathrm{pants}}
  &\stackrel{?}{=}
    \exp\left(
    \frac{e^{-S_0}}{3!} z^3
    \right)
    \;,
    \label{eq:pantscubic}
\end{align}
since all interactions can be decomposed into pairs of pants, each corresponding
to a three-point vertex. However, similar to string field theory, the other
terms of \cref{eq:pantsexp} are needed to correct for undercounting and
overcounting. For example, the one-loop tadpole arising from the three-point
vertex \labelcref{eq:pantscubic} is divided by a symmetry factor of \(2\);
adding the additional tadpole term \(\frac{e^{-S_0}}{2}z\) is needed for the
one-boundary genus-one manifold to be counted once, without an extraneous factor
of \(1/2\). The \(\sim e^{-2S_0}\) terms similarly correct the counting of the
four-, two-, and no-boundary manifolds at genus zero, one, and two respectively.

Perhaps working sufficiently hard on these combinatorics, one can show that the
full expression for \(\eta_{\mathrm{pants}}\) is given by\footnote{Note that the
  normal ordering is outside of and does not commute with the exponentiation.
  So, the exponentiation here should not be confused with that appearing in
  \cref{eq:pantsexp}.}
\begin{align}
  \eta_{\mathrm{pants}}
  &=\, :\exp\left(
    \sum_{\substack{h,m\ge 0\\ 2h-2+m>0}}
  \frac{e^{-(2h-2+m)S_0}}{m!} z^m
  \right):
  \;,
  \label{eq:pantsexactexpansion}
\end{align}
where normal ordering\footnote{Upon writing \(z=a+a^\dagger\), normal ordering
  is equivalent to ordering creation operators to the left of annihilation
  operators.} maps
powers of \(z\) to probabilists' Hermite polynomials \(\He_m(z)\),
\begin{align}
  :z^m:
  \,
  &= \He_m(z)
    \;.
    \label{eq:normalhermite}
\end{align}
We will not work that hard. Instead, we note that \cref{eq:pantsexactexpansion}
can be derived by working backwards from the cumulant generating function.

In general, let \(\Delta K(u)=K(u)-K_{\mathrm{G}}(u)\) be the difference between
the cumulant generating function\footnote{We actually don't need the moment or
  cumulant generating function to exist in a neighbourhood of \(u=0\). All we
  need is as many moments or cumulants to exist as the degree of the polynomial
  \(P\) in \cref{eq:relativemeas} --- it is only that many derivatives of
  \(\Delta K\) that will contribute there anyway. If we want
  \cref{eq:relativemeas} to exist for all polynomials, then we need all
  moments/cumulants to exist.} \(K(u)=\log\expval{e^{u z}}\) of a random
variable \(z\) drawn from an arbitrary unnormalized finite measure over
\(\mathbb{R}\) and that for an unnormalized centred unit-variance Gaussian,
\(K_{\mathrm{G}}(u)=\log\expval{e^{uz}}_{\mathrm{G}}\). Then expectation values of
polynomials \(P(z)\) with respect to these measures are related by
\begin{align}
  \expval{
  P(z)
  }
  &=
    \langle
    :e^{\Delta K(z)}:
    P(z)
    \rangle_{\mathrm{G}}
    \;.
    \label{eq:relativemeas}
\end{align}

For the Marolf--Maxfield model, the cumulant-generating function for the variable
\(Z(S^1)\) is known exactly \cite{Marolf:2020xie},
\begin{align}
  \log \mel{\varnothing}{e^{u Z(S^1)}}{\varnothing}_{\mathrm{BU}}
  &= \lambda e^u
    \;,
\end{align}
as easily seen from the connected path integral \cref{eq:mmconnected}. The
cumulant-generating functional for the variable \(e^{-S_0}(Z(S^1)-e^{2S_0})\)
(shifted and rescaled like \(z\) in \cref{eq:zcentredunitvar}) is
\begin{align}
  K_{\mathrm{BU}}(u)
  &= \lambda
    \exp(e^{-S_0} u)
    -e^{S_0} u
    \;.
\end{align}
For the Gaussian disk-and-cylinder theory,
\begin{align}
  K_{\mathrm{DC}}(u)
  =
  e^{2S_0} + 1 + \frac{u^2}{2}
  \;.
\end{align}
The parenthesized expression in
\cref{eq:pantsexactexpansion} is just an expansion of
\(K_{\mathrm{BU}}(z)-K_{\mathrm{DC}}(z)\) in \(e^{-S_0}\). So with
\(\eta_{\mathrm{pants}}\) given by \cref{eq:pantsexactexpansion}, its matrix
elements
\(\mel{(S^1)^{\sqcup m'}}{\eta_{\mathrm{pants}}}{(S^1)^{\sqcup
    m}}_{\mathrm{DC}}\) in the polynomial test space
\(\Span\left\{\ket{(S^1)^{\sqcup m}}_{\mathrm{DC}}\mid m\ge 0\right\}\) converge
and, by design, match the inner product
\(\braket{\bullet}{\bullet}_{\mathrm{BU}}\) of the full Marolf--Maxfield model.

One of the surprises of the Marolf--Maxfield model is that the spectrum of
\(Z(S^1)\) consists of the nonnegative integers,
\(\{Z_\alpha(S^1)\}=\mathbb{Z}_{\ge 0}\) \cite{Marolf:2020xie}; in contrast,
recall that the spectrum of \(Z_{\mathrm{DC}}(S^1)\) in the disk-and-cylinder
theory is \(\{Z_{\mathrm{DC},\alpha}(S^1)\}=\mathbb{R}\). It is as though,
evoking group-averaging, the insertion \(\eta_{\mathrm{pants}}\) has projected
the disk-and-cylinder theory onto constraint surfaces at integer
\(Z_\alpha(S^1)\).

\paragraph{Extensions and next steps.} In contrast to the approach described in
\cref{sec:topmodel,app:generalizedfock}, we expect the above perturbative
approach for adding interactions to remain practically useful in richer baby
universe theories.

While we should not expect to have closed form expressions for the exact
generating function or arbitrary moments of boundaries, we can at least hope to
calculate some lower moments perturbatively in a parameter like \(e^{-S_0}\)
that suppresses higher topologies and higher moments. Taking this first step
requires us to compute or approximate the gravitational path integral on some
basic set of nontrivial topologies. In particular, this will involve addressing
the questions raised in \cref{sec:topologycontours} about how such path
integrals should be defined.

Once one has an approximate expression for the first few moments, a
generalization of \cref{eq:relativemeas} can be used to systematically repackage
those moments as leading interaction terms. Much like in QFT, perturbative
correlation functions of the interacting GABUFT are computed by Feynman
diagrams, where vertices represent insertions of interaction terms and edges
represent Wick contractions of the free GABUFT's Gaussian path integral over
\(\alpha\)-sectors. This can give a useful organization of corrections that are
perturbative in the parameter \(e^{-S_0}\) suppressing higher topologies. However,
short of exact expressions of the moment/cumulant-generating functional, all
moments/cumulants, or all interaction terms, we expect it will be difficult to
discern non-perturbative effects, like the restriction of \(\alpha\)-sectors to
integer \(Z_\alpha(S^1)\) in the Marolf--Maxfield model. Still, there remains much to
be understood even about perturbative GABUFT.

Let us turn to some questions that were absent from the above discussion of the
Marolf--Maxfield model, because of the theory's simple single-element target
space \(\mathscr{M}=\{S^1\}=\mathscr{K}\). In a richer GABUFT, one could explore
different ways of parametrizing the space of field configurations. For example,
one could ask what representation of the path integral over \(\alpha\)-sectors
would make it most resemble the path integral of a local field
theory.\footnote{As reviewed in \cref{sec:alphasector}, a baby universe state is
  more analogous to a state of a classical statistical field theory than a state
  of a QFT \cite{Casali:2021ewu}.} In \cref{sec:alphasector}, we chose to
represent the path integral over \(\alpha\)-sectors as a path integral over
configurations of fields \(\phi_\alpha(k)\) on \(\mathscr{K}\ni k\)
corresponding to an orthonormal basis \(u_k(x)\) of solutions to the
Wheeler--DeWitt equation \(H u_k(x)=0\) on \(\mathscr{M}\ni x\). This was
appealing because it gave the path integral \labelcref{eq:alphapathint} of the
free theory an action \labelcref{eq:alphaaction} that is ultralocal in
\(\mathscr{K}\). However, we see no immediate reason why the topology-changing
processes should produce interactions that are local in the label set
\(\mathscr{K}\) for some orthonormal basis of solutions \(u_k\). Perhaps
interactions will pick out a preferred basis of solutions for which the
interactions are local in \(\mathscr{K}\); perhaps locality will suggest a
completely different way of parametrizing the path integral over
\(\alpha\)-sectors; or, perhaps there is no such way to make interactions local.

One could also search for choices that simplify interactions in other ways. Even
in the simple Marolf--Maxfield model, we saw that the undercounting and
overcounting of two-geometries by Feynman diagrams required correction by
infinitely many interaction terms. In general, we must carefully treat the
mapping class group of the geometries we integrate over, such that we cover
moduli space once in total. As is in string field theory,\footnote{The moduli
  space of string theory will be different, however, from the moduli space of a
  GABUFT, for example, due to differences in the lapse contour, as described in
  \cref{sec:groupaverage,sec:topologycontours}.} this has the potential to
generate a cascade of infinitely many interaction terms. In certain models, it
may be the case that an appropriate choice of field variables can truncate
the series of interaction terms, as happens in Witten's open string field
theory \cite{Witten:1985cc,Zwiebach:1990az}.

In ref.~\cite{Post:2022dfi}, the authors showed that multi-boundary amplitudes
computed by the JT gravitational path integral can be related to the correlation
functions of a QFT, specifically a Kodaira--Spencer theory on the JT spectral
curve. Remarkably, this QFT is local and has a single, three-point interaction
term. The essential idea was to identify the Eynard--Orantin topological
recursion relation for JT gravity with the Schwinger--Dyson equation of the
Kodaira--Spencer theory. It would be interesting to understand how this
Kodaira--Spencer theory fits into the baby universe framework and more
specifically the GABUFT framework of this paper. This involves understanding
precisely how the kinematic, coinvariant, or invariant states of JT closed
universes are related to Kodaira--Spencer fields. It also involves understanding
precisely how the Kodaira--Spencer path integral can be understood as an
integral over \(\alpha\)-sectors.\footnote{Previous studies of
  \(\alpha\)-sectors and factorization in JT gravity include
  refs.~\cite{Saad:2021uzi,Blommaert:2021fob,Blommaert:2022ucs}.} This will
hopefully lay the groundwork for a systematic generalization of the JT
construction of ref.~\cite{Post:2022dfi} to a larger class of theories.

Finally, an interacting GABUFT raises the question of renormalization. One might
expect perturbative calculations in the path integral over \(\alpha\)-sectors to
develop divergences analogous to those encountered in interacting QFT. Such
divergences might become more severe in richer theories, as the intermediate
states summed over on each internal line acquire more independent labels, \eg{}
\(k=(k_0,k_1,\ldots)\in\mathscr{K}\) in a higher dimensional \(\mathscr{K}\).
How should these divergences be treated? In QFT, a Wilsonian approach introduces
a UV regulator and integrates out physics at scales shorter than those one
wishes to resolve. In gravity, it is perhaps natural to integrate out
microscopic wormholes. Regardless of whether this alone cures certain
divergences, it is interesting to study the effective description of observable
physics that emerges from integrating out microscopic wormholes. The wormhole
calculus of Coleman, Giddings, and Strominger
\cite{Coleman:1988cy,Giddings:1988cx,Giddings:1988wv}\footnote{See also
  ref.~\cite{Gesteau:2024gzf} for related discussion. The ``wormhole
  renormalization'' developed there is primarily a BPHZ-like reorganization of
  wormhole contributions motivated by factorization; integrating out wormholes
  below a geometric scale is suggested separately as a possible Wilsonian
  extension.} suggests that the remaining large universes are described by
\(\alpha\)-dependent, renormalized couplings, which become nonlocal couplings
upon averaging. It would be interesting to study the correspondence between the
renormalization of these effective couplings in the gravitational path integral
and the renormalization of terms in the action weighting the path integral over
\(\alpha\)-sectors.

%% file: sections/appendix01_minisuperspace.tex
\section{Examples of minisuperspace models}
\label{app:minisuperspace}

In this appendix, we review some illustrative examples of minisuperspace models.
We will focus on diagonal minisuperspace models in \(D=4\) spacetime dimensions
where the spatial metric can be expressed in terms of a conformal factor
\(e^{2 x^0}\) and anisotropies \(x^1\) and \(x^2\):
\begin{align}
  h_{ij}
  &= e^{2 x^0} \diag\left(
    e^{2x^1 + 2\sqrt{3} x^2},
    e^{2x^1 - 2\sqrt{3} x^2},
    e^{-4x^1}
    \right)
    \;.
\end{align}
Parameterizing the spatial metric in this way leads to a simple Minkowski
metric \(\mathcal{G}_{AB}\):
\begin{align}
  \mathcal{G}_{AB} \dd{x^A}\dd{x^B}
  &= 24\left(
    -(\dd{x^0})^2 + (\dd{x^1})^2 + (\dd{x^2})^2
    \right)
    \;.
\end{align}

Unfortunately, the actual minisuperspace metric that appears in the Hamiltonian
\(H\), given in \cref{eq:miniham}, is not \(\mathcal{G}_{AB}\) but rather
\(\sqrt{h(x)}\mathcal{G}_{AB}=e^{3x^0}\mathcal{G}_{AB}\). Classically, however,
the dynamics generated by \(H\) and the conformally rescaled Hamiltonian
\begin{align}
  \tilde{H}
  &=\sqrt{h} H
    = \mathcal{G}^{AB} p_A p_B + h(x)\,(2\Lambda-R(x))
\end{align}
are equivalent up to time reparametrization.\footnote{While \(H\) generates
  evolution in proper time \(T=\int\dd{t} N(t)\) where \(N\) is the lapse,
  \(\tilde{H}\) generates evolution in \(\tilde{T}=\int\dd{t} N/\sqrt{h}\).} With
appropriate quantum corrections introduced when promoting the Hamiltonian \(H\)
and \(\tilde{H}\) to quantum operators, this equivalence survives in the quantum
theory \cite{Held:2025mai}. More precisely, if \(H\) is promoted to the quantum
operator \labelcref{eq:quantumH} plus a conformal coupling to minisuperspace
curvature \cite{Halliwell:1988wc} (an \(O(\hbar^2)\) quantum correction), then
the group-averaged inner products constructed from \(H\) and
\begin{align}
  \tilde{H}
  &= \Box_{\mathcal{G}} + h(x)\,(2\Lambda-R(x))
  \label{eq:tildeH}
\end{align}
are equal up to a conformal rescaling of wavefunctions.\footnote{See
  ref.~\cite{Held:2025mai} section 6.4.1 for details.}

Because many of the qualitative features we wish to highlight will be visible
already in the WKB approximation of the quantum theory, the equivalence between
\(H\) and \(\tilde{H}\) at the classical level will mostly suffice anyway. We
will therefore focus on the relatively simple dynamics generated by
\(\tilde{H}\) describing a particle propagating in the flat minisuperspace
geometry \(\mathcal{G}_{AB}\) with a potential
\begin{align}
  \tilde{V}(x)
  &= h(x)\,(2\Lambda-R(x))\;.
    \label{eq:tildepotential}
\end{align}

The specific models we will consider are summarized in \cref{tab:minimodels}. In
the following, we will try to highlight properties of some of these
models which arise from the somewhat strange (from a particle perspective)
feature of having potentials \(\tilde{V}(x)\) that are reflective in ``time'' \(x^0\).

\begin{table}
  \centering
  \begin{tabularx}{\linewidth}{|X||c|c|c|c|}
    \hline
    & \makecell{Minisup.~\\coords.}
    & Spatial slice
    & One-forms \(\sigma^i\)
    & \makecell{Spatial curvature\\\(R(x)\)}
    \\
    \hline
    \hline
    FLRW
    & \(x^0\)
    & \(\begin{cases}
      H_3 & k=-1 \\
      \mathbb{R}^3 & k=0 \\
      S^3 & k=1
    \end{cases}\)
    & \(
      \dd{s}_k^2
      = \sum_{i=1}^3(\sigma^i)^2
  \)
    &
      \(6k e^{-2 x^0}\)
    \\
    \hline
    \makecell[l]{Kantowski--\\Sachs}
    & \(x^0,x^1\)
    &
      \(S^2\times\mathbb{R}\)
    &
      \(\begin{aligned}
        \dd{s}_{S^2}^2
        &= (\sigma^1)^2 + (\sigma^2)^2
        \\
        \sigma^3
        &= \dd{z^3}
    \end{aligned}\)
    & \(2 e^{-2x^0-2x^1}\)
    \\
    \hline
    Bianchi I
    & \(x^0,x^1,x^2\)
    &
      \(\mathbb{R}^3\)
    &
      \(\begin{aligned}
        \sigma^1
        &= \dd{z^1}
          \;,
        \\
        \sigma^2
        &= \dd{z^2}
          \;,
        \\
        \sigma^3
        &= \dd{z^3}
    \end{aligned}\)
    & \(0\)
    \\
    \hline
    Bianchi IX
    & \(x^0,x^1,x^2\)
    & \(S^3\)
    &\(\begin{aligned}
      \sigma^1
      &= \cos\chi \dd{\theta}
      \\
      &\phantom{{}={}}\negmedspace
        - \sin\chi \sin\theta \dd{\phi}
        \;,
      \\
      \sigma^2
      &= -\sin\chi \dd{\theta}
      \\
      &\phantom{{}={}}\negmedspace
        - \cos\chi \sin\theta \dd{\phi}
        \;,
      \\
      \sigma^3
      &= \cos\theta \dd{\phi}
        + \dd{\chi}
    \end{aligned}\)
    &\(\begin{aligned}
      &-\textstyle{\frac{1}{2}} e^{-2x^0-8 x^1}
      \\
      &+2 e^{-2x^0-2x^1}
        \\
      &\phantom{{}+{}}\times
        \textstyle{\cosh(2\sqrt{3}x^2)}
      \\
      &- 2 e^{-2x^0+4x^1}
        \\
      &\phantom{{}+{}}\times
        \textstyle{\sinh^2(2\sqrt{3}x^2)}
    \end{aligned}\)
      \\
      \hline
  \end{tabularx}
  \caption{Examples of minisuperspace models with spatial metrics
    \(h_{ij}\sigma^i \sigma^j\) where \(h_{ij}\) is of the diagonal form
    \labelcref{eq:misnermetric} parametrized by minisuperspace coordinates
    \(x^A\). Because of \cref{eq:spacevol} and the neglect of boundary terms at
    spatial boundaries, we implicitly have in mind compactifying space if it is
    not already compact.}
  \label{tab:minimodels}
\end{table}

\subsection{Recollapsing FLRW: null states and oscillatory saddle-point
  contributions}
\label{app:flrw}
Let us start with the simplest examples where we encounter a potential growing
negatively at large \(x^0\), in particular, some recollapsing FLRW models. In
FLRW models, the anisotropies \(x^1\) and \(x^2\) are set to zero (before
quantization) and the spatial slices are hyperbolic (\(k=-1\)), flat (\(k=0\)),
or spherical (\(k=1\)).

\begin{figure}
     \centering
     \begin{subfigure}[t]{0.45\textwidth}
         \centering
         \includegraphics[width=0.925\textwidth]{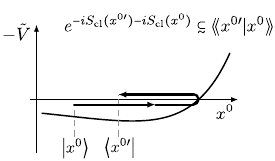}
         \caption{}
         \label{fig:partbounce}
     \end{subfigure}
     \hfill
     \begin{subfigure}[t]{0.45\textwidth}
         \centering
         \includegraphics[width=0.725\textwidth]{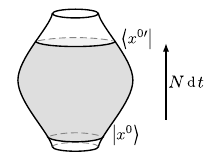}
         \caption{}
         \label{fig:bounce}
     \end{subfigure}
     \par\medskip
     \begin{subfigure}[t]{0.45\textwidth}
         \centering
         \includegraphics[width=0.925\textwidth]{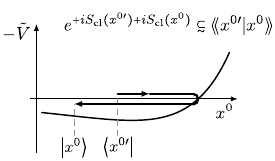}
         \caption{}
         \label{fig:partbouncerev}
     \end{subfigure}
     \hfill
     \begin{subfigure}[t]{0.45\textwidth}
         \centering
         \includegraphics[width=0.725\textwidth]{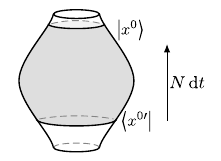}
         \caption{}
         \label{fig:bouncerev}
     \end{subfigure}
     \par\medskip
     \begin{subfigure}[t]{0.45\textwidth}
         \centering
         \includegraphics[width=0.925\textwidth]{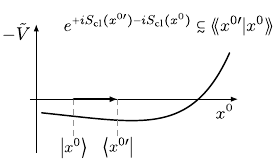}
         \caption{}
         \label{fig:partnobounce}
     \end{subfigure}
     \hfill
     \begin{subfigure}[t]{0.45\textwidth}
         \centering
         \includegraphics[width=0.725\textwidth]{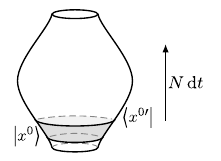}
         \caption{}
         \label{fig:nobounce}
     \end{subfigure}
     \par\medskip
     \begin{subfigure}[t]{0.45\textwidth}
         \centering
         \includegraphics[width=0.925\textwidth]{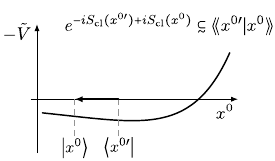}
         \caption{}
         \label{fig:partnobouncerev}
     \end{subfigure}
     \hfill
     \begin{subfigure}[t]{0.45\textwidth}
         \centering
         \includegraphics[width=0.725\textwidth]{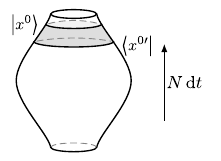}
         \caption{}
         \label{fig:nobouncerev}
     \end{subfigure}
     \caption{The minisuperspace variable \(x^0\) parametrizing the conformal
       factor in FLRW models describes a particle scattering against a potential
       \(-\tilde{V}(x^0)\). In many recollapsing models, this potential takes
       the same qualitative form illustrated in the left panels
       \subref{fig:partbounce}, \subref{fig:partbouncerev},
       \subref{fig:partnobounce}, and \subref{fig:partnobouncerev},
       interpolating between a negative decaying exponential towards negative
       \(x^0\) and a positive growing exponential towards positive \(x^0\).
       These panels show the classical particle trajectories which contribute as
       saddle points to the path integral representation of the group-averaged
       inner product \labelcref{eq:gainnerprod}. The corresponding right panels
       \subref{fig:bounce}, \subref{fig:bouncerev}, \subref{fig:nobounce}, and
       \subref{fig:nobouncerev} illustrate the associated spacetime geometries,
       with the size of the circular sections representing the conformal factor
       of the spatial geometry.}
     \label{fig:flrw}
\end{figure}

Since \(x^0\) is timelike in minisuperspace, the potential felt by this
coordinate is \(-\tilde{V}(x^0)\). Particularly simple examples of recollapsing
models where this potential takes the qualitative form illustrated in
\cref{fig:flrw} include those with: negative cosmological constant
\(\Lambda<0\) and \(k=-1\); \(\Lambda<0\), \(k=0\), and matter; or
\(\Lambda=0\), \(k=1\), and matter. (The inclusion of matter can be modelled by
taking, for example, \(2\Lambda\to2\Lambda+\rho_0 e^{-4 x^0}\) for radiation.) In these
cases, the classical trajectory with zero energy \(\tilde{H}=0\) describes a
particle coming from \(x^0=-\infty\), bouncing off the potential
\(-\tilde{V}(x^0)\) growing towards positive \(x^0\), and rolling back to
\(x^0=-\infty\).


Correspondingly, there is a unique (up to an overall coefficient) zero-energy
eigenstate, \(H\ket{k} =0\). (The following discussion will not qualitatively
change if we instead considered \(\tilde{H}\).) In the classically allowed
region, we can express \(\braket{x^0}{k}\) in the WKB form
\begin{align}
  \braket{x^0}{k}
  &= \chi(x^0) e^{i S_{\mathrm{cl}}(x^0)}
    + \chi(x^0)^* e^{-i S_{\mathrm{cl}}(x^0)}
    \;,
    \label{eq:flrwWKB}
\end{align}
where\footnote{Note that, because the conformal variable \(x^0\) is timelike, a
  classical negative momentum \(p_0<0\) corresponds to increasing \(x^0\) in
  time \(T\) or \(\tilde{T}\).}
\begin{align}
  S_{\mathrm{cl}}(x^0)
  &= \int^{x^0} \dd{y^0}\,(p_{\mathrm{cl}})_0(y^0)
    \;,
  &
  (p_{\mathrm{cl}})_0(x^0)
  &= -\sqrt{24 \tilde{V}(x^0)}
    \;,
\end{align}
and \(\chi(x^0)\) varies much more slowly than
\(e^{i S_{\mathrm{cl}}(x^0)}\) in the WKB approximation. With an appropriate
normalization constant \(\kappa>0\), we have
\begin{align}
  \eta=2\pi\delta(H)
  &= \kappa \ketbra{k}{k}
    \;,
\end{align}
so the group-averaged inner product \labelcref{eq:gainnerprod} is given by
\begin{align}
  \llangle x^{0\,\prime} | x^0 \rrangle
  &= \kappa \braket{x^{0\,\prime}}{k} \braket{k}{x^0}
    \;.
\label{eq:flrwinnerprod}
\end{align}
Substituting the expression \labelcref{eq:flrwWKB} into the two factors on the
RHS, we see that the norm \(\llangle x^0 | x^0 \rrangle\) is non-negative but
oscillates.

In particular, null states \(|x^0\rrangle\) exist at nodes where \(\braket{x^0}{k}\)
vanishes. In general, a state is null if and only if it has vanishing overlap
with \(\ket{k}\). Some obvious null states would be nonzero energy eigenstates
of \(H\). It is perhaps a bit more surprising to find null states at fixed
\(x^0\) --- these are made possible by the destructive interference in
\cref{eq:flrwWKB} between incoming and outgoing waves reflecting off the
potential. This interference and the resulting discrete set of null states will
occur generically for any completely reflective potential. We will see in
\cref{sec:ksBianchiI} that a version of this phenomenon persists even in higher
dimensional minisuperspace.

As an aside, let us remark that each of the four terms resulting from
substituting \cref{eq:flrwWKB} into \cref{eq:flrwinnerprod} corresponds to the
contribution of a saddle-point for the path integral representation of the
group-averaged inner product \labelcref{eq:gainnerprod}. The four saddles,
illustrated in \cref{fig:flrw}, are real and lie on the original Lorentzian
integration contour for the path integral.\footnote{A discussion of a similar
  set of saddles has previously appeared \eg{} in \cite{Araujo-Regado:2022gvw}
  --- compare our \cref{fig:flrw} with their fig.~5.} The original contour can
be deformed into the steepest descent contours of these saddle points, which
therefore contribute to the path integral. More generically (\eg{} if we chose
\(x^0\) or \(x^{0\,\prime}\) to be in the classically forbidden region), we
would have to consider possible contributions from complex saddles which are
necessarily exponentially suppressed.\footnote{The weight with which the
  steepest descent contour of a saddle point contributes to an integral is equal
  to the intersection number of the steepest ascent contour with the original
  contour of integration. (See \eg{} \cite{Witten:2010cx}.) Since the path
  integral is purely oscillatory on the original Lorentzian contour, any saddle
  point on this contour contributes with weight \(\pm 1\)
  \cite{Marolf:2022ybi}. Additionally, the contribution of any saddle point off
  of the original integration contour must be exponentially suppressed.}

\subsection{Kantowski--Sachs and Bianchi I: two- and three-dimensional
  minisuperspace}
\label{sec:ksBianchiI}

Next, let us consider the Kantowski--Sachs and Bianchi I models which
respectively have two- and three-dimensional minisuperspaces, focusing on cases
with enough symmetry to be reduced again to a scattering problem in one
dimension.

\paragraph{The Kantowski--Sachs model} describes \(S^2 \times \mathbb{R}\)
spatial slices with two minisuperspace variables \(x^0\) and \(x^1\) while
\(x^2=0\) in \cref{eq:misnermetric} --- see \cref{tab:minimodels}. Focusing on
the case with zero cosmological constant \(\Lambda=0\), we have a simple
potential \labelcref{eq:tildepotential} which grows in a timelike direction
\(\tilde{x}^0\) in minisuperspace and is constant in an orthogonal spacelike
direction \(\tilde{x}^1\):
\begin{align}
  \tilde{V}(x)
  &= -2 e^{4 x^0 - 2x^1}
    = -2 e^{2\sqrt{3} \tilde{x}^0}
    \;,
    \label{eq:kspotential}
  \\
  \begin{pmatrix}
    \tilde{x}^0
    \\
    \tilde{x}^1
  \end{pmatrix}
  &= \frac{1}{\sqrt{3}}
    \begin{pmatrix}
      2 & -1 \\
      -1 & 2
    \end{pmatrix}
    \begin{pmatrix}
      x^0
      \\
      x^1
    \end{pmatrix}
    \;,
\end{align}
where \(\tilde{x}^0\) and \(\tilde{x}^1\) are related to the original
minisuperspace coordinates \(x^0\) and \(x^1\) by the above Lorentz boost.

For simplicity, let us start by considering the constraint Hamiltonian
\(\tilde{H}\) which commutes with \(\tilde{p}_1 = - i \partial_{\tilde{x}^1}\).
In this case, a basis of gauge-invariant states \(\ket{k}\in\ker \tilde{H}\) is
labelled by the eigenvalue \(k\in\mathbb{R}\) of \(\tilde{p}_1\). For \(x\) in
the classically allowed region (for a given \(k\)), we can express the
wavefunction in the WKB form
\begin{align}
  \braket{x}{k}
  &= e^{ik \tilde{x}^1} \left(
    \chi_k(\tilde{x}^0) e^{i \tilde{S}_k(\tilde{x}^0)}
    + \chi_k(\tilde{x}^0)^* e^{-i \tilde{S}_k(\tilde{x}^0)}
    \right)
    \;,
    \label{eq:kssolnflatmini}
\end{align}
where
\begin{align}
  \tilde{S}_k(\tilde{x}^0)
  &= \int^{\tilde{x}^0} \dd{\tilde{y}^0}\,
    (\tilde{p}_{\mathrm{cl}})_0(\tilde{y}^0;k)
    \;,
  &
  (\tilde{p}_{\mathrm{cl}})_0(\tilde{x}^0;k)
  &= -\sqrt{24 \tilde{V}(\tilde{x}^0)+k^2}
    \;,
\end{align}
and \(\chi_k(\tilde{x}^0)\) varies much more slowly than the explicit
oscillatory factors in the WKB limit. With an appropriate measure \(\kappa(k)\)
over \(\mathbb{R}\ni k\) (determined by the \(\delta\)-function normalization of
\(\ket{k}\)), the group-averaged inner product \labelcref{eq:gainnerprod} can
then be expressed as
\begin{align}
  \llangle x' | x\rrangle
  &= \int \kappa(k) \braket{x'}{k} \braket{k}{x}
    \;.
    \label{eq:ksinnerprod}
\end{align}
For each value of \(k\), there will again be a discrete set of nodal ``times''
\(\tilde{x}^0\) in minisuperspace where \(\braket{x}{k}=0\) is completely
annihilated by the destructive interference of positive and negative frequency
waves reflecting off the potential. For these values of \(\tilde{x}^0\), the
\(\tilde{p}_1=k\) momentum eigenstate
\begin{align}
  |\tilde{x}^0,\tilde{p}_1=k\rangle
  &= \int \dd{\tilde{x}^1} e^{i k \tilde{x}^1}|\tilde{x}^0,\tilde{x}^1\rangle
\end{align}
has vanishing overlap with all \(\ket{k'}\in\ker H\) and consequently,
\(|\tilde{x}^0,\tilde{p}_1=k\rrangle\) is a null state under the inner product
\labelcref{eq:ksinnerprod}.

The symmetry used above makes these null states particularly easy to exhibit,
but is not essential to the underlying mechanism. More generally, consider a
semiclassical wave reflecting perfectly off a potential with a simple turning
hypersurface, for which the usual uniform Airy approximation applies. At least
before additional caustics substantially alter the reflected wave, the Airy
solution oscillates in the classically allowed region and matches onto incoming
and reflected WKB waves of equal flux. Homologous to the turning hypersurface,
one generically expects a sequence of nodal hypersurfaces \(\Sigma\),
approximated by the zeros of the Airy function.

Suppose that a reflected solution \(\psi\in\ker H\) vanishes on such a nodal
hypersurface \(\Sigma\) and consider the state
\begin{align}
  \ket{\delta_\Sigma[\psi]}
  &= \int_\Sigma \dd{\Sigma_A}\,\nabla^A \psi(x)\,\ket{x}
    \;.
\end{align}
For any physical solution \(u_{k'}(x)=\braket{x}{k'}\),
\begin{align}
  \braket{k'}{\delta_\Sigma[\psi]}
  &= \int_\Sigma \dd{\Sigma_A}\,u_{k'}^*\nabla^A \psi
   = \frac{1}{i}\lsem u_{k'} | \psi \rsem_{\mathrm{KG}}
    \;,
\end{align}
where \(\lsem \bullet | \bullet\rsem_{\mathrm{KG}}\) is the Klein--Gordon
``inner product'' and the second equality uses \(\psi|_\Sigma=0\). Conservation of
the Klein--Gordon current allows this flux to be deformed from \(\Sigma\) deep
into the reflecting potential. Complete reflection ensures that \(\psi\),
\(u_{k'}\), and the Klein--Gordon boundary form vanish there, and hence
\(\braket{k'}{\delta_\Sigma[\psi]}=0\) for every \(k'\). Thus
\(\ket{\delta_\Sigma[\psi]}\) is null under the group-averaged inner product.

\paragraph{The Bianchi I model} describes \(\mathbb{R}^3\) spatial slices with
three minisuperspace variables \(x^0\), \(x^1\), and \(x^2\) in
\cref{eq:misnermetric} --- see \cref{tab:minimodels}.

The case with zero cosmological constant \(\Lambda=0\) simply describes a
massless particle propagating in flat \(\mathbb{R}^{1,2}\) minisuperspace (or a
conformal rescaling thereof, if one considers \(H\) instead of \(\tilde{H}\))
with no potential. The case with \(\Lambda>0\) describes a particle with an
\(x^0\)-dependent positive mass-squared. In both cases, we expect a basis of
states \(\ket{k}\in\ker H\) to be labelled by two real numbers, giving the
\(p^{1,2}\) momentum eigenvalues, and a sign, distinguishing positive and
negative frequency solutions: \(k=(p^1,p^2,\pm)\).

The perhaps more interesting case \(\Lambda<0\) features a reflecting potential
which grows exponentially at large positive \(x^0\). The analysis of this model
is qualitatively identical to the Kantowski--Sachs model, except now there are
two spatial minisuperspace directions \(x^1\) and \(x^2\). With the relative
coefficients of positive and negative frequency modes fixed by reflection off
the potential, a basis of solutions can be labelled by just spatial momenta
\(k=(p^1,p^2)\).

\subsection{Bianchi IX: complicated classical trajectories}
\label{sec:bianchiIX}

Using the simple models reviewed so far, we have highlighted some of the features
associated with potentials in the timelike \(x^0\) direction of minisuperspace,
which are somewhat unusual in ordinary particle mechanics (and QFT). In these
models, \(x^0\) does not evolve monotonically along classical trajectories
bouncing off the potential, so one might rightfully conclude that \(x^0\) is not
a good ``time'' variable. Instead, one might consider a ``time'' variable such
as \(x^1\) or \(x^2\) in the Kantowski--Sachs or Bianchi I models, for which a
given time slice picks out a unique instant along a classical trajectory. Thus,
one might regard \(x^1\) or \(x^2\) as time in minisuperspace, and \(x^0\) space
with an ordinary potential.

\begin{figure}
  \centering
  \includegraphics[width=0.75\textwidth]{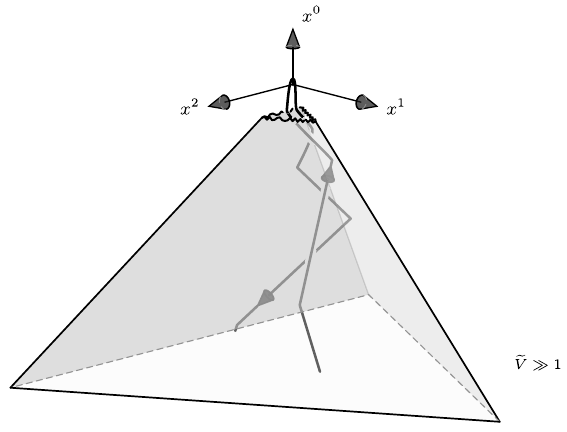}
  \caption{The Bianchi IX model. Deep inside the pyramid, \(\tilde{V}(x)\) is
    small, so classical trajectories are composed of approximately null line
    segments. Crossing each face of the pyramid, \(\tilde{V}(x)\) becomes large
    and positive, causing the trajectory to bounce off the face in a timelike
    manner. Near the origin, \(\tilde{V}(x)\) is negative and becomes stronger
    towards positive \(x^0\), repelling the trajectory back down from that
    region.}
  \label{fig:bianchiIX}
\end{figure}

However, a similar ``time'' variable does not readily exist in the Bianchi IX
model. This model describes \(S^3\) spatial slices with three minisuperspace
variables \(x^0\), \(x^1\), and \(x^2\) in \cref{eq:misnermetric} --- see
\cref{tab:minimodels}. The only comment we will make about this model is that
even its classical trajectories are complicated, as illustrated in
\cref{fig:bianchiIX}, precluding a simple foliation of minisuperspace such that
each leaf is crossed once by each classical trajectory. If desired, one could
add an extra clock degree of freedom, thereby introducing a new variable that
serves as time in minisuperspace, with dynamics designed to make it increase
monotonically along classical trajectories.

This is not a problem for GABUFT, because, as described in \cref{sec:uft}, it is
much more agnostic about the distinction between time and space in the target
space (minisuperspace) compared with the second quantization that leads to QFT.

%% file: sections/appendix02_qftwightman.tex
\section{QFT Wightman function as worldline amplitudes}
\label{app:qftwightman}

In this appendix, we review how the QFT Wightman function of a real
scalar \(\Phi_{\mathrm{QFT}}\), with action
\begin{align}
  I_{\mathrm{QFT}}
  &= - \frac{1}{2} \int_{\mathscr{M}}\mu(x) \Phi_{\mathrm{QFT}} H \Phi_{\mathrm{QFT}}
    \;,
  &
    H
  &= -\Box + V(x)
      \;,
\end{align}
can be expressed in terms of (unoriented) worldline amplitudes. We will present
two ways to do so: one relies on a time-space split of target space
\(\mathscr{M}\) and another that is more covariant. However, in either case, the
worldline amplitude does not seem to arise naturally from a gravitational
worldline path integral.

For the first way, we will focus on QFT in a static Lorentzian spacetime
\(\mathscr{M}=\mathbb{R}\times\mathscr{S}\). Let \(x^0\) be the static time
coordinate, \(x^I\) be coordinates on space \(\mathscr{S}\), and
\begin{align}
  -\mathcal{N}^2(\dd{x^0})^2+ \mathcal{Q}_{IJ} \dd{x}^I \dd{x}^J
\end{align}
be the metric on \(\mathscr{M}\). Then the differential operator appearing in
the QFT action can be expressed as
\begin{align}
  H
  &= \mathcal{N}^{-2} (\partial_0^2 + A)
    \;,
  &
    A
  &\equiv \mathcal{N}^2\left(
    - \frac{1}{\mathcal{N}\sqrt{\mathcal{Q}}}
    \partial_I \left(
    \mathcal{N}\sqrt{\mathcal{Q}}
    \mathcal{Q}^{IJ}
    \partial_J
    \right)
    + V
    \right)
    \;.
\end{align}
With a Killing time symmetry on \(\mathscr{M}\), there is a preferred QFT vacuum
\(\ket{0}_{\mathrm{QFT}}\). For simplicity, we will assume that \(A\) is
strictly positive to avoid zero-mode subtleties. Then, by considering a basis of
on-shell modes normalized with respect to the Klein--Gordon inner product, we can
derive an expression for the vacuum Wightman function
\begin{align}
  \mel{0}{\Phi_{\mathrm{QFT}}(x')\Phi_{\mathrm{QFT}}(x)}{0}_{\mathrm{QFT}}
  &= \mel{x^{I\prime}}{\frac{1}{2\omega} e^{-i \omega (x^{0\prime}-x^0)}}{x^I}_{\mathrm{wl}}
    \label{eq:wightman1}
    \;,
\end{align}
in terms of the worldline amplitude on the RHS described as follows. The
worldline target space in this case is \(\mathscr{S}\ni x^I,x^{I\prime}\)
(rather than \(\mathscr{M}\) as everywhere else in this paper). The worldline
Hilbert space is
\(L^2(\mathscr{S},\mathcal{N}^{-1}\dd{\mathrm{vol}(\mathscr{S})})\) with an
inner product where the volume measure on \(\mathscr{S}\) is rescaled by
\(\mathcal{N}^{-1}\), and the states \(\ket{x^I},\ket{x^{I\prime}}\) are
\(\delta\)-function normalized with respect to this inner product. The operator
\(\omega\) is the square root of \(A\),
\begin{align}
  \omega &\equiv A^{1/2}
           \;.
\end{align}
The worldline amplitude in \cref{eq:wightman1} does not seem to arise from a
worldline path integral that is gravitational in the sense of including a path
integral over a worldline lapse (\cf{}
\cref{eq:gainnerprod,eq:groupavg,eq:feynman,eq:antifeynman}, not to be confused
with the target space lapse \(\mathcal{N}\) above).

Let us consider a second way of expressing the QFT Wightman function in terms of
a worldline amplitude. As elsewhere in this paper, the worldline in this case
propagates in the target space \(\mathscr{M}\) and has a (kinematic) Hilbert
space \(\mathcal{H}_{\text{wl kin}}=L^2(\mathscr{M},\mu)\). We will assume, as
described around \cref{eq:etaisfeynplusafeyn}, that the symmetrized two-point
function of the QFT in some preferred vacuum \(\ket{0}_{\mathrm{QFT}}\) can
indeed be identified with the worldline group-averaged inner product,
\begin{align}
  \mel{0}{\{\Phi_{\mathrm{QFT}}(x'),\Phi_{\mathrm{QFT}}(x)\}}{0}_{\mathrm{QFT}}
  &= \mel{x'}{\eta}{x}_{\mathrm{wl}}
    = \llangle x'|x\rrangle_{\mathrm{wl}}
    \;.
    \label{eq:hadamard}
\end{align}
We will also assume for simplicity that the basis of on-shell modes
\(\{u_k\mid k\in\mathscr{K}\}\) cleanly separates into ``positive-frequency''
modes \(\{u_k\mid k\in\mathscr{K}_+\}\) and ``negative-frequency'' modes
\(\{u_k\mid k\in\mathscr{K}_-\}\), where these designations merely refer to
whether they multiply annihilation or creation operators in the mode expansion
of \(\Phi_{\mathrm{QFT}}\). Then projecting one of the arguments in
\cref{eq:hadamard} onto positive frequencies, we pick out one of the two
operator orderings,
\begin{align}
  \mel{0}{\Phi_{\mathrm{QFT}}(x')\Phi_{\mathrm{QFT}}(x)}{0}_{\mathrm{QFT}}
  &= \mel{x'}{P_+\eta}{x}_{\mathrm{wl}}
    \;,
    \label{eq:wightman2}
\end{align}
where
\begin{align}
  P_\pm
  &= \frac{1\pm S}{2}
    \;,
    &
    S
  &= (+1) \oplus (-1)
\end{align}
on
\begin{align}
  \mathcal{H}_{\text{wl inv}}
  &=L^2(\mathscr{K}_+,\kappa) \oplus L^2(\mathscr{K}_-,\kappa)
    \;.
    \label{eq:invposnegsplit}
\end{align}
(See \cref{sec:gennotation} for the description of
\(\mathcal{H}_{\text{wl inv}}\) as an \(L^2\)-space.) Using the other projection
\(P_-\) would pick out the other ordering from \cref{eq:hadamard}. Whereas the
group-averaged inner product
\(\llangle x'|x\rrangle_{\mathrm{wl}}=\mel{x'}{\eta}{x}_{\mathrm{wl}}\)
naturally arises from a gravitational worldline path integral (as described in
\cref{sec:groupaverage}), the projection \(P_+\) onto positive frequency
solutions of the Wheeler--DeWitt equation seems rather artificial from the
perspective of the gravitational worldline theory. In fact, as discussed in
\cref{sec:discussqftvsuft}, there can be many timelike directions in target
space \(\mathscr{M}\) and, even if we restrict to minisuperspace models where
\(\mathscr{M}\) is Lorentzian with one time-like direction, it is often not
possible to construct a pathology-free QFT with a Hilbert space description in
that direction.

But let us continue assuming we have a model where the QFT approach makes sense,
and end with one more observation\footnote{Ref.~\cite{Hartle:1997dc} derives similar results
  for asymptotically free hyperbolic models, whose Wheeler--DeWitt Hamiltonian
  \(H\) reduces in the asymptotic past or future of target space \(\mathscr{M}\)
  to a Klein--Gordon operator on flat spacetime with constant nonnegative
  mass-squared.}: \cref{eq:hadamard,eq:wightman2} can be used to identify the
Klein--Gordon inner product \(\lsem\bullet|\bullet\rsem_{\mathrm{KG}}\) on
positive-frequency solutions with the inner product
\(\lsem\bullet|\bullet\rsem_{\mathrm{wl}}\) on \(\mathcal{H}_{\text{wl inv}}\)
restricted to the subsector \(L^2(\mathscr{K}_+,\kappa)\). Recall that the
Klein--Gordon inner product is defined in terms of the symplectic form \(\Omega\)
on solution space,
\begin{align}
  \lsem \psi' | \psi \rsem_{\mathrm{KG}}
  &= i\Omega((\psi')^*,\psi)
\end{align}
This is required to be positive definite on the positive-frequency subspace of
solutions, but the above expression extends to a Hermitian form on the full
space of solutions. This Hermitian form behaves like an inverse for
the Pauli--Jordan propagator \cite{Held:2025mai},
\begin{align}
  \Delta(x',x)
  &=[\Phi_{\mathrm{QFT}}(x'),\Phi_{\mathrm{QFT}}(x)]
    \;,
  \\
  i\Omega((\psi')^*,\Delta j)
  &= \braket{\psi'}{j}_{\mathrm{wl}}
    \;,
    \label{eq:symppaulijordan}
\end{align}
analogous to the relation \labelcref{eq:invinnerprod} between
\(\lsem\bullet|\bullet\rsem_{\mathrm{wl}}\) and the group average
\(\eta\).\footnote{This can be phrased in the formalism \cite{Held:2025mai}
  reviewed around \cref{eq:geninverse}. Introduce \(\varkappa_{\mathrm{KG}}\) so
  that the Klein--Gordon inner product can be expressed as a matrix element
  \(\lsem \psi' | \psi \rsem_{\mathrm{KG}} \equiv
  \mel{\psi'}{\varkappa_{\mathrm{KG}}}{\psi}_{\mathrm{wl}}\) in the kinematic
  Hilbert space \(\mathcal{H}_{\text{wl kin}}\). If we call
  \(\eta_{\mathrm{KG}}\equiv \Delta\), then \(\varkappa_{\mathrm{KG}}\) is a
  generalized inverse of \(\eta_{\mathrm{KG}}\) satisfying
  \(\eta_{\mathrm{KG}}\varkappa_{\mathrm{KG}}\eta_{\mathrm{KG}}=\eta_{\mathrm{KG}}\)
  analogous to \cref{eq:geninverse}. The choice of the codimension-one surface
  for the Klein--Gordon inner product is the choice of
  ``gauge-fixing condition'' represented by \(\varkappa_{\mathrm{KG}}\)
  \cite{Held:2025mai}.} The Pauli--Jordan propagator \(\Delta\), like \(\eta\),
maps test functions \(j\) to solutions. In fact, \(\Delta\) and \(\eta\) can be
directly related using \cref{eq:hadamard,eq:wightman2},
\begin{align}
  \Delta
  &= S \eta
    \;.
\end{align}
It then follows by comparison of \cref{eq:invinnerprod,eq:symppaulijordan} that
\begin{align}
  i\Omega((\psi')^*,S \psi)
  &= \lsem \psi' | \psi\rsem_{\mathrm{wl}}
    \;.
\end{align}
In particular, because \(S=1\) on the positive-frequency subspace, the
Klein--Gordon inner product agrees with the inner product
\(\lsem\bullet|\bullet\rsem_{\mathrm{wl}}\) on
\(L^2(\mathscr{K}_+,\kappa)\subset \mathcal{H}_{\text{wl inv}}\).
Ref.~\cite{Held:2025mai} provides further intuition for this relation in terms
of gauge-fixing in the worldline path integral.

%% file: sections/appendix03_alphapathint.tex
\section{Mathematically precise path integral over \(\alpha\)-sectors for free GABUFT}
\label{app:alphapathint}

In this appendix, we retell \cref{sec:alphasector}'s story about the path
integral over \(\alpha\)-sectors, now with mathematical precision. Our starting
point will be the space
\(L^2\left(\mathcal{A}, \Sigma, \mathcal{D}\phi_\alpha \, e^{-I[\phi_\alpha]}
\right)\). In \cref{app:alphal2}, we will firstly define the space
\(\mathcal{A}\), the \(\sigma\)-algebra \(\Sigma\) of measurable subsets of
\(\mathcal{A}\), the Gaussian measure
\(\mathcal{D}\phi_\alpha \, e^{-I[\phi_\alpha]}\) on \((\mathcal{A},\Sigma)\),
and the \(L^2\) space. In \cref{app:alphal2isgns}, we will show that this
Gaussian \(L^2\) space is equal to the baby universe Hilbert space
\(\mathcal{H}_{\mathrm{BU}}\) obtained from the GNS construction described in
\cref{sec:gns}. Finally, in \cref{app:alphal2ops}, we will describe the
boundary-inserting operators \(Z(J)\) as multiplication operators on
\(L^2\left(\mathcal{A},\Sigma,\mathcal{D}\phi_\alpha \, e^{-I[\phi_\alpha]}
\right)\). We will prove that these operators are essentially self-adjoint once
defined on the dense subspace \(\mathcal{D}_{\mathcal{J}}\) that appears in the
GNS construction. Their self-adjoint extensions are multiplication operators
with maximal domains and are strongly commuting.

For simplicity, we assume throughout this appendix that the worldline physical
Hilbert space $\mathcal H_{\mathrm{wl\,inv}}\cong\mathcal H_{\mathrm{wl\,co}}$
is separable.

\subsection{The Gaussian \(L^2\) space of \(\alpha\)-sectors}
\label{app:alphal2}

Let us start by defining the space \(\mathcal{A}\) of \(\alpha\)-parameters, the
\(\sigma\)-algebra \(\Sigma\) of measurable subsets, the Gaussian measure
\(\mathcal{D}\phi_\alpha \, e^{-I[\phi_\alpha]}\) over \((\mathcal{A},\Sigma)\),
and the space
\(L^2\left(\mathcal{A}, \Sigma, \mathcal{D}\phi_\alpha \, e^{-I[\phi_\alpha]}
\right)\). Around \cref{eq:phialpha}, we roughly described \(\mathcal{A}\) as a
space of distributions \(\phi_\alpha\) to path-integrate over. A standard
strategy for evaluating Gaussian path integrals is to express them as integrals
over the real coefficients of countably many modes --- all we have to do is
spell this out explicitly and precisely.

Recall from \cref{sec:worldline} that the physical Hilbert space of the
worldline theory is
\(\mathcal H_{\mathrm{wl\,co}} \cong \mathcal H_{\mathrm{wl\,inv}}\) and has a
\(\star\)-involution (inherited from \(\mathcal{H}_{\text{wl kin}}\)) understood
as antilinear time- or orientation-reversal. It will be useful to consider the
real Hilbert space,
\begin{align}
  \mathcal{H}_{\text{wl }\mathbb{R}}
  &\equiv
    \{
    \psi \in \mathcal{H}_{\text{wl inv}}
    \mid \psi^\star=\psi\}
    \;.
\end{align}
The definition of the measure space
\((\mathcal{A},\Sigma,\mathcal{D}\phi_\alpha \,e^{-I[\phi_\alpha]})\) starting
from \(\mathcal{H}_{\text{wl }\mathbb{R}}\) proceeds along the lines of the
constructive proof of ref.~\cite{MR544188}'s Theorem 2.3A, as we now review.

Given that \(\mathcal{H}_{\text{wl inv}}\) and thus
\(\mathcal{H}_{\text{wl }\mathbb{R}}\) are separable, we can choose a countable
orthonormal basis for \(\mathcal{H}_{\text{wl }\mathbb{R}}\),
\begin{align}
  \mathcal{H}_{\text{wl }\mathbb{R}}
  &= \overline{\Span_{\mathbb{R}}\{\mathsf{e}_1,\mathsf{e}_2,\ldots\}}
    \;.
\end{align}
(As described below, different choices of bases will give constructions that are
equivalent up to isomorphism.) Next, we simply define
\begin{align}
  \mathcal{A}
  &\equiv \mathbb R^{\mathbb{N}}
  = \{\alpha=(\phi_{\alpha,1},\phi_{\alpha,2},\ldots) \mid \phi_{\alpha,i}\in\mathbb{R}\}
  \;.
\end{align}
\Cref{sec:alphasector}'s ``path integral over distributions''
\(\phi_\alpha(k)=\sum_i \phi_{\alpha,i} \mathsf{e}_i(k)\) will now be defined by
integrals over the coefficients \(\phi_{\alpha,i}\).

We equip \(\mathcal{A}\) with the cylinder\footnote{Recall that a
  \emph{cylinder} \(\sigma\)-algebra \(\Sigma\) is one generated by
  \emph{cylinder} sets, meaning product sets where all but a finite number of
  the coordinates \(\phi_{\alpha,i}\) are unrestricted. The countable tensor
  product \labelcref{eq:alphameas} of \emph{probability measures} is well
  defined on \(\Sigma\), because it assigns unit measure to each unrestricted
  direction.} $\sigma$-algebra $\Sigma$ of measurable subsets generated by the
coordinate maps \(\alpha\mapsto \phi_{\alpha,i}\). We define the measure
\(\mathcal{D}\phi_\alpha e^{-I[\phi_\alpha]}\) on \((\mathcal A,\Sigma)\) to be
the tensor product\footnote{We use the notation
  \(\mathcal{D}\phi_\alpha e^{-I[\phi_\alpha]}\) purely for ease of comparison
  with \cref{sec:alphasector}. The individual pieces \(\mathcal{D}\phi_\alpha\)
  and \(e^{-I[\phi_\alpha]}\) don't have mathematically rigorous definitions. In
  particular, the ``\(\mathcal{D}\phi_\alpha\)-measure'' of every cylinder set
  whose finite-dimensional base has positive Lebesgue measure would be infinite,
  and the action \(I[\phi_\alpha]\) written in
  \cref{eq:alphaaction} is infinite almost everywhere on \((\mathcal{A},\Sigma)\)
  with measure \labelcref{eq:alphameas}. \label{foot:precisemeas}}
\begin{equation}
  \mathcal{D}\phi_\alpha e^{-I[\phi_\alpha]}
  \equiv \aleph\bigotimes_{i=1}^\infty \rho_{\mathrm{G}}(\phi_{\alpha,i})
  \label{eq:alphameas}
\end{equation}
of Gaussian measures
\begin{align}
  \rho_{\mathrm{G}}(\phi_{\alpha,i})
  &= \frac{\dd{\phi_{\alpha,i}}}{\sqrt{2\pi}}e^{-\phi_{\alpha,i}^2/2}
    \label{eq:gaussmeas}
\end{align}
for each \(\phi_{\alpha,i}\), where \(\aleph>0\) is the
constant in \cref{eq:topologygpi}.

The complex Hilbert space
\(L^2(\mathcal{A},\Sigma,\mathcal{D}\phi_\alpha e^{-I[\phi_\alpha]})\) with
inner product
\begin{align}
  \int \mathcal{D}\phi_\alpha e^{-I[\phi_\alpha]}
  \Psi'[\phi_\alpha]^* \Psi[\phi_\alpha]
\end{align}
is comprised of equivalence classes of measurable normalizable wavefunctions
\(\Psi:\mathcal{A}\to\mathbb{C}\). As usual, two such wavefunctions
\(\Psi,\Psi'\) represent the same element of the \(L^2\) space if and only if
they are equal almost everywhere (\ie{} except possibly on some measure zero set)
--- this is equivalent to the condition for the \(L^2\)-norm \(\|\Psi-\Psi'\|\) to
vanish.

For later use, let us define a linear map
\begin{align}
  z:
  \mathcal{H}_{\text{wl inv}}
  &\to L^2(\mathcal{A},\Sigma,\mathcal{D}\phi_\alpha e^{-I[\phi_\alpha]})
    \;,
    &
    \psi = \sum_{i=1}^\infty \psi_i \mathsf{e}_i
  &\mapsto
    z_\alpha(\psi)
    \sim \lim_{m\to\infty} \sum_{i=1}^m \psi_i \phi_{\alpha,i}
    \;.
    \label{eq:phimap}
\end{align}
For each \(\psi\in\mathcal{H}_{\text{wl inv}}\), the sequence of partial sums
\(\sum_{i=1}^m \psi_i \phi_{\alpha,i}\) is Cauchy in
\(L^2(\mathcal{A},\Sigma,\mathcal{D}\phi_\alpha e^{-I[\phi_\alpha]})\) ---
choosing \(m,m'\) sufficiently large,
\begin{align}
  \norm{
  \sum_{i=m}^{m'} \psi_i \phi_{\alpha,i}
  }^2
  =
  \aleph \sum_{i=m}^{m'} \abs{\psi_i}^2
\end{align}
can be made arbitrarily small. The \(m\to\infty\) limit in \cref{eq:phimap} thus
selects an element in
\(L^2(\mathcal{A},\Sigma,\mathcal{D}\phi_\alpha e^{-I[\phi_\alpha]})\), which
one can then represent by some measurable normalizable wavefunction
\(z_\bullet(\psi):\mathcal{A}\to\mathbb{C}\).\footnote{This wavefunction need
  not be \emph{pointwise} equal to \(\sum_{i=1}^\infty \psi_i \phi_{\alpha,i}\)
  --- at fixed \(\alpha\) and \(\psi\), this sum need not even converge in
  \(\mathbb{C}\).} By continuity from finite \(m\),
\begin{equation}
  \int_{\mathcal A}
  \mathcal{D}\phi_\alpha e^{-I[\phi_\alpha]}
  e^{i z_\alpha(\psi)}
  = \aleph e^{-\frac{1}{2}\lsem \psi^\star|\psi\rsem_{\mathrm{wl}}}
  \;,
  \label{eq:gausschar}
\end{equation}
\ie{} any countable family of \(z_\alpha(\psi)\)'s are jointly Gaussian random variables with
covariance matrix given by the inner product
\(\lsem \bullet |\bullet\rsem_{\mathrm{wl}}\) on
\(\mathcal{H}_{\text{wl inv}}\). In \cref{app:alphal2isgns}, we will use the
above characteristic function to argue that
\(L^2(\mathcal{A},\Sigma,\mathcal{D}\phi_\alpha e^{-I[\phi_\alpha]})\) is
exactly equal to \(\mathcal{H}_{\mathrm{BU}}\) obtained from the GNS
construction of \cref{sec:gns}.

One might worry about the uniqueness of our construction, which relied on the
choice of basis \(\{\mathsf{e}_i\}\) for \(\mathcal{H}_{\text{wl }\mathbb{R}}\). In fact,
given a real separable Hilbert space \(\mathcal{H}_{\text{wl }\mathbb{R}}\),
ref.~\cite{MR544188}'s Theorem 2.3A guarantees the uniqueness of the probability
measure space
\((\mathcal{A},\Sigma,\aleph^{-1}\mathcal{D}\phi_\alpha e^{-I[\phi_\alpha]})\)
equipped with a linear map \(z\), from \(\mathcal{H}_{\text{wl }\mathbb{R}}\) to
jointly Gaussian random variables with covariance given by
\(\mathcal{H}_{\text{wl }\mathbb{R}}\)'s inner product.\footnote{Uniqueness, of
  course, is meant up to isomorphism. Additionally, the \(\sigma\)-algebra
  \(\Sigma\) is required to be the minimal one for which the random variables
  \(z_\alpha(\psi)\), with
  \(\psi\in\mathcal{H}_{\text{wl }\mathbb{R}}\), are measurable --- this
  precludes useless extensions of the probability measure
  space to directions that are always integrated out.}

\subsection{Identification with the GNS Hilbert space \(\mathcal{H}_{\mathrm{BU}}\)}
\label{app:alphal2isgns}

Recall from the GNS construction of \cref{sec:gns} that the baby universe
Hilbert space
\(\mathcal{H}_{\mathrm{BU}}\equiv \overline{\mathcal{D}_{\mathcal{J}}}\) is
defined to be the completion of a dense domain
\(\mathcal{D}_{\mathcal{J}}=\{\ket{J} \mid J\in\mathcal{J}\}\) with inner
product given by \cref{eq:buinnerprod,eq:topologygpi}. We will now prove that
the baby universe Hilbert space \(\mathcal{H}_{\mathrm{BU}}\) so defined for
free worldline theories is equal to the Gaussian
\(L^2(\mathcal{A},\Sigma,\mathcal{D}\phi_\alpha e^{-I[\phi_\alpha]})\)-space
over \(\alpha\)-parameters \(\alpha\in\mathcal{A}\) described in
\cref{app:alphal2}.

\subsubsection{\(\mathcal{D}_{\mathcal{J}}\) is the polynomial domain of
  \(L^2(\mathcal{A},\Sigma,\mathcal{D}\phi_\alpha e^{-I[\phi_\alpha]})\)}

Recall from \cref{sec:physhilb} that \(\eta\) maps test functions
\(j\in \mathcal{D}_{\text{wl test}}\) to a dense subspace \(\Image\eta\) of
\(\mathcal{H}_{\text{wl inv}}\). Composing with \(z\) defined in
\cref{eq:phimap}, we can construct a linear map
\begin{align}
  Z \equiv z \circ \eta :
  \mathcal{D}_{\text{wl test}}
  &\to
  L^2(\mathcal{A},\Sigma,\mathcal{D}\phi_\alpha e^{-I[\phi_\alpha]})
  \;,
  &
    j
  &\mapsto Z_\alpha(j)\equiv z_\alpha(\eta j)
    \;.
    \label{eq:zonebdymultop}
\end{align}
We can similarly define a linear map from the algebra \(\mathcal{J}\) of
polynomials in \(j\in\mathcal{D}_{\text{wl test}}\),
\begin{align}
  Z : \mathcal{J}
  &\to
  L^2(\mathcal{A},\Sigma,\mathcal{D}\phi_\alpha e^{-I[\phi_\alpha]})
    \;,
    \label{eq:zmultbdymultop1}
  \\
    J
    = \sum_{m=0}^M c_m j_{m,1} \sqcup \cdots \sqcup j_{m,m}
  &\mapsto Z_\alpha(J) \equiv \sum_{m=0}^M c_m Z_\alpha(j_{m,1}) \cdots Z_\alpha(j_{m,m})
    \;.
  &
    (M\in\mathbb{Z}_{\ge 0}, c_m\in\mathbb{C})
    \label{eq:zmultbdymultop2}
\end{align}
Products and sums of finitely many measurable functions are also measurable.
Substituting \(\psi=\eta j\) into the Gaussian characteristic function
\cref{eq:gausschar} and differentiating, we can verify that \(Z_\alpha(J)\) has
finite
\(L^2(\mathcal{A},\Sigma,\mathcal{D}\phi_\alpha e^{-I[\phi_\alpha]})\)-norm.

In fact, we more generally see that the inner products between such
states agree with \cref{eq:buinnerprod,eq:topologygpi}. Thus, the inner product
space \(\mathcal{D}_{\mathcal{J}}\) that arose in the GNS construction of
\cref{sec:gns} can be identified with the subspace
\begin{align}
  \mathcal{D}_{\mathcal{J}}
  &= \Span\{Z_\alpha(J) \mid J\in\mathcal J\} \subset
    L^2(\mathcal{A},\Sigma,\mathcal{D}\phi_\alpha e^{-I[\phi_\alpha]})
\end{align}
comprised of polynomials \(Z_\alpha(J)\) in the Gaussian random variables
\(Z_\alpha(j)\). The GNS construction defines the baby universe Hilbert space to
be the completion
\(\mathcal{H}_{\mathrm{BU}}\equiv\overline{\mathcal{D}_{\mathcal{J}}}\) in
norm-topology, so to prove that
\(\mathcal{H}_{\mathrm{BU}} =L^2(\mathcal{A},\Sigma,\mathcal{D}\phi_\alpha e^{-I[\phi_\alpha]})\), it remains only to show that the polynomial domain
\(\mathcal{D}_{\mathcal{J}}\) is also dense in this Gaussian \(L^2\) Hilbert
space.

\subsubsection{Density of the polynomial domain in
  \(L^2(\mathcal{A},\Sigma,\mathcal{D}\phi_\alpha e^{-I[\phi_\alpha]})\)}
\label{app:polydense}

Let \(\mathsf{j}_1,\mathsf{j}_2,\ldots\in\mathcal{D}_{\text{wl test}}\) be worldline test states
that map to an orthonormal basis for the real section
\(\mathcal{H}_{\text{wl }\mathbb{R}}\) of the invariant Hilbert space
\(\mathcal{H}_{\text{wl inv}}\),
\begin{align}
  \mathcal{H}_{\text{wl }\mathbb{R}}
  =
  \overline{
    \Span_{\mathbb{R}} \{\eta \mathsf{j}_1, \eta \mathsf{j}_2, \ldots\}
  }\;.
\end{align}
Without loss of generality (by the uniqueness described at the end of
\cref{app:alphal2}), we can select \(\mathsf{e}_i=\eta \mathsf{j}_i\) as the
basis used in \cref{app:alphal2}'s construction.\footnote{This simplifying
  choice is not essential. Even if the bases \(\{\mathsf{e}_i\}\) and
  \(\{\eta \mathsf{j}_i\}\) do not coincide, the coordinates
  \(Z_\alpha(\mathsf{j}_i)\) on \(\mathcal{A}\) would still generate
  \(\mathcal{A}\) modulo null sets (which will be sufficient in the following
  argument).} We will now prove that the space of polynomials in
\(Z_\alpha(\mathsf{j}_i)\),
\begin{align}
  \Span\{Z_\alpha(\mathsf{j}_1)^{\ell_1} \cdots Z_\alpha(\mathsf{j}_m)^{\ell_m}
  \mid m,\ell_i\in\mathbb{Z}_{\ge 0}
  \}
  \subset \mathcal{D}_{\mathcal{J}}
  \;,
\end{align}
is dense in \(L^2(\mathcal{A},\Sigma,\mathcal{D}\phi_\alpha
e^{-I[\phi_\alpha]})\) and therefore
\begin{align}
  \mathcal{H}_{\mathrm{BU}}
  \equiv\overline{\mathcal{D}_{\mathcal{J}}}
  =L^2(\mathcal{A},\Sigma,\mathcal{D}\phi_\alpha e^{-I[\phi_\alpha]})
  \;.
  \label{eq:gnsequalsl2}
\end{align}

Our strategy will be to reduce this problem on the infinite dimensional space
\(\mathcal{A}\) to finite dimensions. A useful result for this purpose is proved
by ref.~\cite{MR544188} in the uniqueness part of Theorem 2.1 (Kolmogorov's
theorem): if a family of random variables generates the \(\sigma\)-algebra of a
probability measure space, then bounded functions each of finitely many of those
variables are dense in \(L^2\). In our context, the cylinder \(\sigma\)-algebra
\(\Sigma\) is by definition generated by the coordinates
\(\phi_{\alpha,i}=Z_\alpha(\mathsf{j}_i)\) on \(\mathcal{A}\). Let
\begin{align}
  \Sigma_m
  &= \sigma(Z_\alpha(\mathsf{j}_1),\ldots,Z_\alpha(\mathsf{j}_m))
\end{align}
be the cylinder \(\sigma\)-algebra generated by the first \(m\)-many of these
\(Z_\alpha(\mathsf{j}_i)\). Then ref.~\cite{MR544188}'s result says that
\begin{align}
  \overline{
  \bigcup_{m\ge 0}
  L^2(\mathcal{A},\Sigma_m,\mathcal{D}\phi_\alpha e^{-I[\phi_\alpha]})
  }
  =
  L^2(\mathcal{A},\Sigma,\mathcal{D}\phi_\alpha e^{-I[\phi_\alpha]})
  \;.
  \label{eq:kolmogunique}
\end{align}

It only remains to argue that
\begin{align}
  \overline{\Span\{Z_\alpha(\mathsf{j}_1)^{\ell_1} \cdots Z_\alpha(\mathsf{j}_m)^{\ell_m}
  \mid \ell_i\in\mathbb{Z}_{\ge 0}
  \}}
  &=
    L^2(\mathcal{A},\Sigma_m,\mathcal{D}\phi_\alpha e^{-I[\phi_\alpha]})
    \;.
    \label{eq:polydense}
\end{align}
But this is easy because \((\mathcal{A},\Sigma_m)\) is effectively
\(m\)-dimensional:
\begin{align}
  L^2(\mathcal{A},\Sigma_m,\mathcal{D}\phi_\alpha e^{-I[\phi_\alpha]})
  &\cong L^2(
    \mathbb{R}^m, \aleph \rho_{\mathrm{G}}(x_1)\otimes \cdots \otimes\rho_{\mathrm{G}}(x_m)
    )
    \;,
\end{align}
where \(\rho_{\mathrm{G}}\) is again the unit Gaussian measure \labelcref{eq:gaussmeas}. In particular,
Hermite polynomials span this \(L^2\) space (as energy eigenfunctions span the
Hilbert space of \(m\)-many harmonic oscillators).

Putting together \cref{eq:kolmogunique,eq:polydense}, we have proven
\cref{eq:gnsequalsl2} --- for worldline theories, the baby universe Hilbert
space \(\mathcal{H}_{\mathrm{BU}}\equiv \overline{\mathcal{D}_{\mathcal{J}}}\) defined by
the GNS construction is equal to the Gaussian \(L^2\) Hilbert space over the
space \(\mathcal{A}\) of
\(\alpha\)-parameters.

\subsection{Boundary-inserting operators $Z(J)$}
\label{app:alphal2ops}

The GNS construction described in \cref{sec:gns} not only produces a Hilbert
space \(\mathcal{H}_{\mathrm{BU}}\) but also operators \(Z(J)\) defined
originally on the dense subspace \(\mathcal{D}_{\mathcal{J}}\). We will prove
for worldline theories that, when \(J=J^\star\), these operators are essentially
self-adjoint once defined on \(\mathcal{D}_{\mathcal{J}}\). In particular, they
extend to self-adjoint operators that strongly commute with each other.

Given what we have already developed in \cref{app:alphal2,app:alphal2isgns},
this story is most easily told in reverse. We will first define maximal
multiplication operators \(Z(J)\) obviously satisfying the aforementioned
properties, then later show that their restrictions to
\(\mathcal{D}_{\mathcal{J}}\) are essentially self-adjoint.

\subsubsection{The multiplication operators $Z(J)$}
\label{app:multops}

Previously in \cref{eq:zonebdymultop,eq:zmultbdymultop1,eq:zmultbdymultop2}, we
defined \(Z\) as a map from elements \(J\in\mathcal{J}\) to elements of
\(L^2(\mathcal{A},\Sigma,\mathcal{D}\phi_\alpha e^{-I[\phi_\alpha]})\) which are
(representable) by measurable wavefunctions \(Z_\alpha(J)\) of
\(\alpha\in\mathcal{A}\). We can also view such a wavefunction as defining a
multiplication operator,
\begin{align}
  Z(J):
  \dom(Z(J))
  &\to
    L^2(\mathcal{A},\Sigma,\mathcal{D}\phi_\alpha e^{-I[\phi_\alpha]})
  \;,
  &
    \Psi[\phi_\alpha]
  &\mapsto Z_\alpha(J) \Psi[\phi_\alpha]
    \;,
\end{align}
defined on
\begin{align}
  \dom(Z(J))
  &= \{
    \Psi
    \in
    L^2(\mathcal{A},\Sigma,\mathcal{D}\phi_\alpha e^{-I[\phi_\alpha]})
    \mid
    Z(J)
    \Psi
    \in
    L^2(\mathcal{A},\Sigma,\mathcal{D}\phi_\alpha e^{-I[\phi_\alpha]})
    \}
    \;,
\end{align}
that is, the maximal possible domain such that multiplication by \(Z_\alpha(J)\)
preserves \(L^2\)-normalizability.

In particular, it is clear that \(Z(J)\) so defined acts on the polynomial
domain \(\mathcal{D}_{\mathcal{J}}\subset\dom(Z(J))\) exactly as advertised in
\cref{eq:ketJstate} for the GNS construction. We will show in
\cref{app:polyiscore} that \(\mathcal{D}_{\mathcal{J}}\) is a core for \(Z(J)\),
so the extension from this subdomain to the maximal domain \(\dom(Z(J))\) is
essentially unique.

Before moving on, we note the following properties of \(Z(J)\) that follow
immediately from its definition as a maximal multiplication operator when
\(J=J^\star\).
\begin{itemize}
\item{They are self-adjoint, \(Z(J)=Z(J)^\dagger\). It is clear by definition
    that \(Z(J)\) is closed and symmetric when \(J=J^\star\). By
    ref.~\cite{Schm_dgen_2012}'s Proposition 3.13, self-adjointness is then
    equivalent to the shifted operators \(Z(J)\pm i\) being surjective. This is
    indeed the case, because the inverse maps
    \((Z(J)\pm i)^{-1}: \Psi[\phi_\alpha] \mapsto
    \Psi[\phi_\alpha]/(Z_\alpha(J)\pm i)\) are well-defined on the entire
    Hilbert space.}
\item{They strongly commute --- this means that all spectral projections of any
    two operators \(Z(J)\) and \(Z(J')\) commute. For multiplication operators,
    strong commutativity is trivial because spectral projections are simply
    multiplications by characteristic functions of measurable sets.}
\end{itemize}

\subsubsection{The polynomial domain \(\mathcal{D}_{\mathcal{J}}\) is a core for
  \(Z(J)\)}
\label{app:polyiscore}

A core for an operator \(O\) with domain \(\dom(O)\) is a linear subspace
which is dense in \(\dom(O)\) with respect to the graph norm,
\begin{align}
  \norm{\Psi}_O
  &= \sqrt{\norm{\Psi}^2+\norm{O \Psi}^2}
    \;.
\end{align}
For a closed operator \(O\), a linear subspace is a core if and only if the
original operator \(O\) on \(\dom(O)\) is recovered from its restriction to the
subspace by taking the closure. In our case, for any \(J\in\mathcal{J}\), we
will prove that the polynomial domain \(\mathcal{D}_{\mathcal{J}}\) is a core
for the multiplication operator \(Z(J)\) defined on the maximal domain
\(\dom(Z(J))\) introduced in \cref{app:multops}; for \(J=J^\star\) in
particular, the restriction of \(Z(J)\) to \(\mathcal{D}_{\mathcal{J}}\)
therefore has a unique self-adjoint (and thus closed) extension to
\(\dom(Z(J))\).

For any \(J\in\mathcal{J}\), the relevant squared graph norm is given by
\begin{align}
  \norm{\Psi}_{Z(J)}^2
  &= \int \mathcal{D}\phi_\alpha e^{-I[\phi_\alpha]} (1+\abs{Z_\alpha(J)}^2)
    \abs{\Psi[\phi_\alpha]}^2
  \\
  &= \norm{\Psi}_{L^2(\mathcal{A},\Sigma,\rho_J)}^2
  \;,
\end{align}
which we note is equal to a squared \(L^2\)-norm with the modified measure
\begin{align}
  \rho_J
  &\equiv
    \mathcal{D}\phi_\alpha e^{-I[\phi_\alpha]} (1+\abs{Z_\alpha(J)}^2)
    \;.
\end{align}
Hence, a linear subspace is a core for \(Z(J)\) if and only if it is dense in
\(L^2(\mathcal{A},\Sigma,\rho_J)\).

We now prove that the polynomial domain \(\mathcal{D}_{\mathcal{J}}\) is dense
in \(L^2(\mathcal{A},\Sigma,\rho_J)\) by proceeding analogously to
\cref{app:polydense}'s proof of its density in
\(L^2(\mathcal{A},\Sigma,\mathcal{D}\phi_\alpha e^{-I[\phi_\alpha]})\). Without
loss of generality, let us take the \(J\in\mathcal{J}\) under consideration to
be a polynomial in the first \(m_J\)-many variables considered there,
\(\mathsf{j}_1,\ldots,\mathsf{j}_{m_J}\).\footnote{In general, \(J\in\mathcal{J}\) is a
  polynomial in arbitrary worldline test functions
  \(j\in\mathcal{D}_{\text{wl test}}\). By Gram--Schmidt, we can consider test
  functions that map under \(\eta\) to orthonormal states in
  \(\mathcal{H}_{\text{wl inv}}\). Taking real \(\frac{j+j^\star}{2}\) and
  imaginary \(\frac{j-j^\star}{2i}\) parts, we can further restrict to test
  functions mapping to orthonormal states in \(\mathcal{H}_{\text{wl }\mathbb{R}}\).}
In particular,
\begin{align}
  Z_\alpha(J)
  &= P_J(Z_\alpha(\mathsf{j}_1),\ldots,Z_\alpha(\mathsf{j}_{m_J}))
\end{align}
for some polynomial \(P_J:\mathbb{R}^{m_J}\to \mathbb{C}\).
Similar to before, we have
\begin{align}
  \overline{
  \bigcup_{m\ge m_J}
  L^2(\mathcal{A},\Sigma_m,\rho_J)
  }
  =
  L^2(\mathcal{A},\Sigma,\rho_J)
  \label{eq:unionmodgauss}
\end{align}
and it only remains to argue for the density of polynomials in \(L^2\)-spaces
of finitely many variables,
\begin{align}
  L^2(\mathcal{A},\Sigma_m,\rho_J)
  &\cong L^2\left(
    \mathbb{R}^m, \aleph\,
    (1+\abs{P_J(x_1,\ldots,x_{m_J})}^2)
    \,
    \rho_{\mathrm{G}}(x_1)
    \otimes\cdots\otimes\rho_{\mathrm{G}}(x_m)
    \right)
    \;.
  &
    (m\ge m_J)
    \label{eq:findimmodgauss}
\end{align}
To that end, let us suppose we have a
\begin{align}
  \Psi\in L^2(\mathbb{R}^m,(1+\abs{P_J}^2) \rho_{\mathrm{G}}^{\otimes m})
\end{align}
that is orthogonal to all polynomials \(P:\mathbb{R}^m\to\mathbb{C}\),
\begin{align}
  \int_{\mathbb{R}^m}
  (1+\abs{P_J}^2)\,
  \rho_{\mathrm{G}}^{\otimes m}
  \,
  P^*\, \Psi
  &= 0
    \;.
    \label{eq:polyorthog}
\end{align}
Let us consider the Fourier transform
\begin{align}
  \int_{\mathbb{R}^m}
  (1+\abs{P_J}^2)
  \,
  \rho_{\mathrm{G}}^{\otimes m} \,
  \Psi \,
  e^{i \xi\cdot x}
\end{align}
of the complex measure
\((1+\abs{P_J}^2)\, \rho_{\mathrm{G}}^{\otimes m}\, \Psi\). This Fourier
transform is entire in \(\xi\) because of the exponential suppression of the
integrand by the Gaussian measure \(\rho_{\mathrm{G}}^{\otimes m}\). By
\cref{eq:polyorthog}, all derivatives of the Fourier transform vanish and so the
Fourier transform vanishes identically. By the uniqueness of Fourier transforms,
\begin{align}
  (1+\abs{P_J}^2)\, \rho_{\mathrm{G}}^{\otimes m}\,\Psi
  &=0
\end{align}
and \(\Psi\) is null. This proves that polynomials are dense in the
\(L^2\)-space \labelcref{eq:findimmodgauss}.

Putting this together with \cref{eq:unionmodgauss}, we conclude that
\(\mathcal{D}_{\mathcal{J}}\) is dense in \(L^2(\mathcal{A},\Sigma,\rho_J)\) and
therefore a core for \(Z(J)\).

%% file: sections/appendix04_gaugingstar.tex
\section{How the gravitational path integral may gauge a transformation}
\label{app:pathintgauging}

In various places throughout the main text, we referred to certain
transformations acting on boundary data as being gauged by the gravitational
path integral. Here, we briefly review how this can occur.

Let \(g\) be a linear transformation acting on single-boundary conditions
\(j\in\mathcal{D}_{\text{wl test}}\).\footnote{See
  \cref{sec:topologyperturbation,app:dckinematics} for discussions of
  transformations that intrinsically involve multiple boundary components. In
  upcoming work \cite{Chen:2026wip2}, we hope to address antilinear
  transformations and whether it is reasonable for them to be gauged in the
  sense of \cref{eq:ggaugestate,eq:zetaggauge} in the baby universe framework.}
We say
that \(g\) is gauged if boundary conditions related by \(g\) prepare the same
physical baby universe state,
\begin{align}
  \ket{j}_{\mathrm{BU}}
  \stackrel{?}{=} \ket{g j}_{\mathrm{BU}}
  \;.
  \label{eq:ggaugestate}
\end{align}
In the GNS construction of the baby universe Hilbert space, this is equivalent
to
\begin{align}
  \braket{J'}{j}_{\mathrm{BU}}
  \equiv \zeta((J')^\star\sqcup j)
  &\stackrel{?}{=}
    \zeta((J')^\star\sqcup (g j))
    \equiv \braket{J'}{g j}_{\mathrm{BU}}
    \label{eq:zetaggauge}
\end{align}
for all multi-boundary conditions \(J'\in\mathcal{J}\). Note that if we extend
the action of \(g\) to multi-boundary conditions \(J\in\mathcal{J}\) by
distribution over disjoint unions, \(g(j\sqcup j')=(gj)\sqcup(gj')\), then the
gauging of the single-boundary action above implies also the gauging of the
multi-boundary action --- that is, \(j\in\mathcal{D}_{\text{wl test}}\) above
can be replaced by \(J\in\mathcal{J}\).

Importantly, \cref{eq:zetaggauge} does not require restricting the kinematic
boundary conditions by hand to \(g\)-invariant inputs \(g j=j\). Rather, the
path integral may take general kinematic inputs and automatically project out
their non-invariant components. This can arise dynamically from a sum over
``gauge-field'' configurations that implements an average over the gauge group
\cite{Harlow:2023hjb}.

\begin{figure}
  \centering
  \begin{subfigure}[t]{0.48\textwidth}
    \centering
    \includegraphics[width=0.4\textwidth]{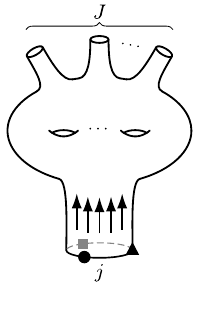}
    \caption{A bulk configuration appearing in the path integral
      \(\zeta(J\sqcup j)\). The wavefunction \(j(x)\) weights configurations
      \(x\) of induced fields on a boundary component. (A particular bulk
      configuration, as intended to be represented here, corresponds to a
      particular \(x\) in the integral
      \(\zeta(J\sqcup j)=\int_{\mathscr{M}}\dd{x} \zeta(J\sqcup x) j(x)\).) In
      particular, each \(x\) prescribes certain field values at the three
      boundary points marked by shapes for comparison in later subfigures. A
      unit normal vector field in the cylinder-like region near \(j\) is also
      shown.}
    \label{fig:groupavgorig}
  \end{subfigure}
  \hfill
  \begin{subfigure}[t]{0.48\textwidth}
    \centering
    \includegraphics[width=0.9\textwidth]{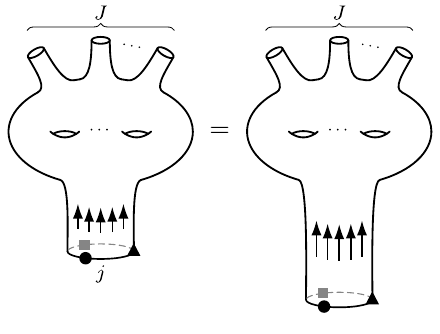}
    \caption{A configuration with a different value of lapse near the \(j\)
      boundary component. On the LHS, a bigger lapse is illustrated by a unit vector field
      with shortened coordinate components. The LHS is diffeomorphic to the RHS
      with a cylinder of greater proper length than \subref{fig:groupavgorig}.
      Integrating lapse over all real values gauges time translations, as
      already described in \cref{sec:groupaverage,sec:generalprop}.}
    \label{fig:groupavglapse}
  \end{subfigure}
  \par\bigskip
  \begin{subfigure}[t]{0.48\textwidth}
    \centering
    \includegraphics[width=0.9\textwidth]{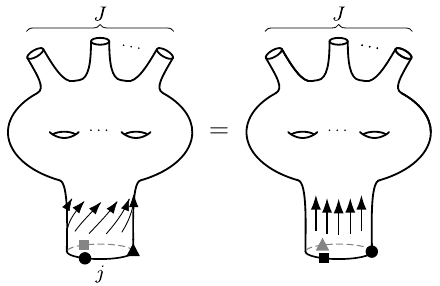}
    \caption{A configuration with a different value of shift near \(j\). On the
      LHS, a nontrivial shift is indicated by a helical normal vector field;
      this can be untwisted on the RHS by a diffeomorphism that rotates the
      region near \(j\) relative to the rest of the spacetime. This
      configuration is equal to the contribution of \subref{fig:groupavgorig} to
      \(\zeta(J\sqcup j')\) where we replace \(j\) with a new wavefunction
      \(j'\), obtained from a spatial rotation of boundary data. The integral
      over shift is thus a group average over diffeomorphisms connected to the
      identity. The path integral depends on \(j\) only through an integral over
      its images under such diffeomorphisms, which are therefore gauged,
      \(j\stackrel{\mathrm{Diff}^+}{\sim}j' \implies \zeta(J\sqcup
      j)=\zeta(J\sqcup j')\).}
    \label{fig:groupavgshift}
  \end{subfigure}
  \hfill
  \begin{subfigure}[t]{0.48\textwidth}
    \centering
    \includegraphics[width=0.9\textwidth]{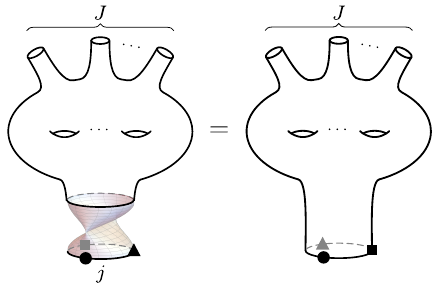}
    \caption{A configuration where the \(j\) boundary component is attached
      through a different cylinder-like bordism to the rest of the spacetime.
      The configuration on the LHS merely looks singular because of limitations
      in pictorially representing the bordism as a two-dimensional surface
      embedded in three dimensions. The LHS is diffeomorphic to the RHS, where
      data on the boundary component is attached with a reflection relative to
      \subref{fig:groupavgorig}. The two terms in the sum
      \subref{fig:groupavgorig}\(+\)\subref{fig:groupavgrefl} exchange under
      reflection, \(j\leftrightarrow \mathsf{R}j\) in an unoriented theory,
      which is thus gauged, \(\zeta(J\sqcup j)=\zeta(J\sqcup \mathsf{R}j)\). In
      an oriented theory, \(\mathsf{R}\) is replaced by \(\mathcal{R}\), which
      also carries along orientation.}
    \label{fig:groupavgrefl}
  \end{subfigure}
  \caption{Contributions to the path integral \(\zeta(J\sqcup j)\) which
    organize into integrals or sums invariant under certain transformations of
    \(j\).}
  \label{fig:groupavgeg}
\end{figure}

The gravitational path integral, for example, naturally integrates over lapse
and shift near each boundary, and sums over other, possibly discrete choices in
attaching the boundary to the bulk --- see \cref{fig:groupavgeg}. We already saw
in \cref{sec:groupaverage} how integrating over the lapse component of the
metric (the ``gauge field'' for diffeomorphisms) leads to a group average
\(\int \dd{T} e^{-i T H}\) over time translations.\footnote{As described in
  \cref{sec:generalprop}, we expect this to continue being true near each
  boundary component, even in the presence of topology change deeper in the
  bulk.} Under the physical inner product defined by the path integral, the
difference between states related by time evolution then becomes null. The path
integral for the shift metric component imposes similar relations between states
related by spatial diffeomorphisms connected to the identity.

Summing also over other, discrete choices in attaching the boundary data to the
bulk can similarly gauge spatial diffeomorphisms disconnected from the identity.
For example, in a theory independent of orientation, \cref{fig:groupavgrefl}
illustrates a configuration where the bulk is connected in a different way to a
boundary component relative to \cref{fig:groupavgorig}. This can be thought of
as the extra insertion of a reflection transformation \(\mathsf{R}\), much like
how configurations with differing lapse have insertions of \(e^{-i T H}\) for
different values of \(T\). The sum of \cref{fig:groupavgorig,fig:groupavgrefl}
thus implements an average over the group \(\{1,\mathsf{R}\}\) generated by
reflections and is invariant under replacing \(j\to \mathsf{R}j\) --- doing so
merely exchanges \cref{fig:groupavgorig,fig:groupavgrefl}. The path integral and
resulting inner product therefore see no difference between \(j\) versus
\(\mathsf{R}j\), so spatial reflection \(\mathsf{R}\) is gauged.

In fact, we expect something similar to happen in theories where bulk
configurations carry orientation. There, one can define a reflection
\(\mathcal{R}\) diffeomorphism on a spacelike boundary which carries along not
only the metric and matter fields, but also the induced orientation field. This
orientation-reversing reflection \(\mathcal{R}\) is gauged in the same way as
described above.\footnote{If a test function \(j(x)\) is supported only on
  boundary data \(x\) with one orientation and not the other, then only one of
  \cref{fig:groupavgorig,fig:groupavgrefl} will be nonzero. More generally
  \(j(x)\) can have support on both orientations and both
  \cref{fig:groupavgorig,fig:groupavgrefl} can be nonzero. In any case, it is
  true that replacing \(j\leftrightarrow\mathcal{R}j\) will swap
  \cref{fig:groupavgorig,fig:groupavgrefl}.}

\paragraph{Comment about spinors/fermions.} In this paper, we have avoided
discussing spinors to sidestep various annoyances. One annoyance is that
spacetime transformations disconnected from the identity are not as universal
for spinor fields. For example, spatial reflections and time reversals need not
square to one and need not commute; in some realizations, these squares and
commutators involve \((-1)^F\) --- the fermion parity operator
\cite{Harlow:2023hjb,Witten:2025ayw}. The only thing we will say about
spinors/fermions is that \((-1)^F\) is gauged in quantum gravity
\cite{Harlow:2023hjb}.

A bulk argument for this is that the gravitational path integral always
group-averages over \(\{1,(-1)^F\}\). In fact, if \(\mathsf{R}^2=(-1)^F\), then
we should really replace \(\{1,\mathsf{R}\}\) in the above discussion with the
actually closed group \(\{1,\mathsf{R},(-1)^F,(-1)^F\mathsf{R}\}\). Even if we
ignore reflections and just consider transformations connected to the identity,
in the vielbein formalism, the path integral will group average over frame
rotations, including \(2\pi\)-rotations which are the same as \((-1)^F\)
\cite{Harlow:2023hjb}.

Let us also give a holographic argument from the boundary side. Suppose a
single-component wavefunction \(j(x)\) is fermion-odd, \((-1)^Fj =-j \), where,
as previously, \((-1)^F\) denotes bulk fermion parity. But this means that
\(j(x)\) prepares a closed universe state using only boundary data \(x\) with
odd numbers of fermionic operator insertions. On such \(x\), the partition
function of a boundary QFT must vanish \(\zeta_\alpha(x)=0\) by \emph{boundary}
fermion-parity symmetry \((-1)^{F_\partial}\): correlation functions ought to be
invariant under conjugation by \((-1)^{F_\partial}\), but an odd product of
fermionic insertions flips sign. A superposition of these correlators will
likewise vanish,
\(\zeta_\alpha(j)=\int_{\mathscr{M}}\mu(x) \zeta_\alpha(x) j(x)=0\). If the
gravitational path integral is equal to an ensemble average
\(\zeta=\int\dd{\alpha}\abs{\braket{\alpha}{\varnothing}_{\mathrm{BU}}}^2\zeta_\alpha\)
of such boundary path integrals \(\zeta_\alpha\) (which each factorizes across
disjoint unions of boundary components), then it must have the property that
\(\zeta(J\sqcup j)=0\) for all fermion-odd \(j\). Equivalently, the difference
between any single-component wavefunction \(j\) and its image \((-1)^Fj\) under
fermion-parity must become null under the gravitational path integral,
\(\ket{j}_{\mathrm{BU}}-\ket{(-1)^Fj}_{\mathrm{BU}}=0\).

%% file: sections/appendix05_generalizedFock.tex
\section{Sch\"urmann Fock space representation}
\label{app:generalizedfock}

We argued in \cref{sec:topmodel} that \cref{eq:mmphiexact} provides an exact
representation of the boundary-inserting operator/universe field
\(Z(S^1)=\Phi(S^1)\) on a Fock space in the Marolf--Maxfield topological model,
by giving each term a diagrammatic interpretation illustrated in
\cref{fig:mmdiskterm,fig:mmcreationterm,fig:mmannihilationterm,fig:mmnumberterm}.
Here, we will generalize this construction to a larger class of baby universe
theories that have Hermitian and conditionally positive \emph{connected}
gravitational path integrals \(\zeta_{\mathrm{con}}\), as described below. In
this context, we will give a representation of boundary-inserting operators
\(Z(j)\) on a Fock space over a ``one-component'' Hilbert space
\(\mathcal{H}_{\mathrm{con}}\). As in the Marolf--Maxfield case, the construction
provides an algebraic way to organize the combinatorics needed to go between the
connected and full path integrals; it does not, however, provide a useful
organization of the connected path integral itself, which is essentially treated
as a black box. We will first describe this as generally as possible, then later
point out in the context of GABUFT that the analogous Fock space
expression for the universe field \(\Phi(x)\) can be expanded into on-shell
modes.

What we will present below is an application of the Sch\"urmann Fock
representation of infinitely divisible states on \(\star\)-bialgebras. For
readers seeking mathematical details, we refer to Sch\"urmann's textbook
\cite{Schurmann:1993wnb}, where Theorems 4.5.6 and 4.5.7 are particularly
relevant.\footnote{The \(\star\)-bialgebra in our case will be the algebra
  \(\mathcal{J}\) of possible (multi-)boundary conditions; the infinitely
  divisible state will be the normalized gravitational path integral
  \(\aleph^{-1}\zeta\) ``generated'' by the connected gravitational path
  integral \(\zeta_{\mathrm{con}}\). (Strictly speaking, a generator \(\psi\) in
  Sch\"urmann's terminology must vanish on the identity
  \(\varnothing\in\mathcal{J}\), so the generator is actually
  \(\psi=\zeta_{\mathrm{con}}\circ(1-\varnothing\varepsilon)\), where
  \(\varepsilon\) is the counit described below.)} Ref.~\cite{Chen:2025uzn}
discusses Poissonization, and the authors'
application\footnote{Ref.~\cite{Chen:2025uzn} also tries to motivate a
  connection between these calculations and the more general idea that the
  emission and absorption of baby universes can be approximated by Poisson
  processes. Let us emphasize that our discussion is completely logically
  independent from this point.} to 2D topological baby universe models like the
Marolf--Maxfield model is closely related to what we will describe below.
However, Poissonization --- as defined by ref.~\cite{Chen:2025uzn} in their
appendix H --- is strictly a special case of the more general Sch\"urmann Fock
representation. Gaussian baby universe models, as considered in
\cref{sec:uft,sec:onepoint}, do not admit a description in terms of
Poissonization (at least, not obviously), because their connected path integrals
\(\zeta_{\mathrm{con}}\) are not positive linear functionals on the set
\(\mathcal{J}\) of (multi-)boundary conditions\footnote{The defining property of
  a Gaussian model is that \(\zeta_{\mathrm{con}}\) vanishes on disjoint unions
  of more than two nontrivial boundary components. We note that the
  Cauchy--Schwarz inequality
  \begin{align}
    \abs{\zeta_{\mathrm{con}}(j_1 \sqcup j_2)}^2
    &=
      \abs{\zeta_{\mathrm{con}}(\varnothing^\star \sqcup j_1 \sqcup j_2)}^2
      \\
    &\stackrel{?}{\le}
    \zeta_{\mathrm{con}}(\varnothing^\star\sqcup \varnothing)
      \zeta_{\mathrm{con}}((j_1 \sqcup j_2)^\star \sqcup (j_1 \sqcup j_2))
      \\
    &= \zeta_{\mathrm{con}}(\varnothing)
    \zeta_{\mathrm{con}}(j_1^\star \sqcup j_2^\star \sqcup j_1 \sqcup j_2)
  \end{align}
  is violated if \(\zeta_{\mathrm{con}}(j_1 \sqcup j_2)\ne 0\) while
  \(\zeta_{\mathrm{con}}(j_1^\star \sqcup j_2^\star \sqcup j_1 \sqcup j_2)=0\),
  regardless of how one might choose to define
  \(\zeta_{\mathrm{con}}(\varnothing)\).}; nonetheless, they admit a Sch\"urmann
Fock representation coinciding with the Fock space representation already
presented in the main text.

The Sch\"urmann Fock space representation requires a Hermitian,
\emph{conditionally} positive linear functional. In our case, this will be the
connected gravitational path integral \(\zeta_{\mathrm{con}}\) which is obtained
by restricting the full gravitational path integral \(\zeta\) to connected
topologies. Like \(\zeta\), \(\zeta_{\mathrm{con}}:\mathcal{J}\to\mathbb{C}\) is
a linear functional on the algebra \(\mathcal{J}\) of multi-boundary conditions.
To construct the Sch\"urmann Fock space, we will require
\(\zeta_{\mathrm{con}}\) to be Hermitian (like \(\zeta\) in
\cref{eq:zetareality}), meaning
\begin{align}
  \zeta_{\mathrm{con}}(J^\star)
  &= \zeta_{\mathrm{con}}(J)^*
    \;,
\end{align}
and be \emph{conditionally} positive, meaning
\begin{align}
  J\in \ker\varepsilon
  \implies \zeta_{\mathrm{con}}(J^\star \sqcup J)
  \ge 0
  \;,
\end{align}
where \(\varepsilon:\mathcal{J}\to\mathbb{C}\) is the linear
functional\footnote{Viewing \(\mathcal{J}\) as a bialgebra, \(\varepsilon\) is
  its counit.}
\begin{align}
  \varepsilon(\varnothing)
  &= 1
    \;,
  &
    \varepsilon(j_1 \sqcup \cdots \sqcup j_m)
  &= 0
    \;.
  &
    (m\ge 1)
\end{align}
Conditional positivity of \(\zeta_{\mathrm{con}}\) implies positivity of
\(\zeta\), as the construction below will show that the baby universe Hilbert
space has a positive inner product coinciding with that of the Sch\"urmann Fock
space. In fact, conditional positivity of \(\zeta_{\mathrm{con}}\) is equivalent
to the positivity of an \(\mathbb{R}_+\)-parameter family of path integrals
\(\zeta_t\) constructed by rescaling the connected part
\(\zeta_{\mathrm{con}}\to t\zeta_{\mathrm{con}}\). Thus, conditional positivity
of \(\zeta_{\mathrm{con}}\) is strictly stronger than the minimal condition of
positivity of \(\zeta=\zeta_{t=1}\) needed for baby universe theory. The
Gaussian models and the Marolf--Maxfield model considered in
\cref{sec:uft,sec:onepoint,sec:topmodel} satisfy this stronger condition.

Conditional positivity here essentially means positivity after projecting out
the identity element \(\varnothing\in\mathcal{J}\). In particular,
\begin{align}
  \braket{J'}{J}_{\mathrm{con}}
  &\equiv
    \zeta_{\mathrm{con}}(
    (J'-\varepsilon(J') \varnothing)^\star
    \sqcup
    (J-\varepsilon(J) \varnothing)
    )
\end{align}
is a positive-semidefinite Hermitian form on \(\mathcal{J}\). Quotienting by the
null subspace and taking a completion, we construct a Hilbert space
\(\mathcal{H}_{\mathrm{con}}\) with this form as the inner product. In the
Gaussian models of \cref{sec:uft,sec:onepoint} where \(\zeta_{\mathrm{con}}\)
vanishes on the disjoint union of more than two nontrivial boundary components,
\(\mathcal{H}_{\mathrm{con}}=\mathcal{H}_{\text{wl co/inv}}\) is just the
physical worldline/one-universe Hilbert space. In the context of the
Marolf--Maxfield model, where \(\zeta_{\mathrm{con}}\) is the constant number
\labelcref{eq:mmconnected} and \(\braket{\bullet}{\bullet}_{\mathrm{con}}\) is
therefore highly degenerate \(\ket{(S^1)^{\sqcup m}}_{\mathrm{con}}=\ket{S^1}_{\mathrm{con}}\),
\(\mathcal{H}_{\mathrm{con}}\) is one-dimensional and spanned by \(\ket{S^1}_{\mathrm{con}}\).

The Sch\"urmann Fock space is the Fock space
\begin{align}
  \mathcal{F}_{\mathrm{Sch}}
  &\equiv
    \overline{
      \bigoplus_{i=0}^{\infty}
      \Sym^i(\mathcal{H}_{\mathrm{con}})
    }
    \;,
    \label{eq:schurmannfock}
\end{align}
over \(\mathcal{H}_{\mathrm{con}}\). The claim is that the baby universe Hilbert
space \(\mathcal{H}_{\mathrm{BU}}\) GNS-constructed from \(\zeta\) embeds into
this Fock space,
\begin{align}
  \mathcal{H}_{\mathrm{BU}}
  \subset \mathcal{F}_{\mathrm{Sch}}
  \;,
  \label{eq:schurmannembed}
\end{align}
where the cyclic no-boundary state \(\ket{\varnothing}_{\mathrm{BU}}\) is
identified with the Fock space vacuum and boundary-inserting operators are
represented as follows.

Each single-boundary-inserting \(Z(j)\) takes the form
\begin{align}
  Z(j)
  &= Z_\varnothing(j)
    + a(j^\star)
    + a(j)^\dagger
    + \Lambda(j)
    \;.
    \label{eq:schurmannzj}
\end{align}
The first term is a c-number one-point function
\begin{align}
  Z_\varnothing(j)
  &\equiv
    \frac{
    \mel{\varnothing}{Z(j)}{\varnothing}_{\mathrm{BU}}
    }{
      \braket{\varnothing}{\varnothing}_{\mathrm{BU}}
    }
    \;.
\end{align}
The second and third terms are annihilation and creation operators of the Fock
space \(\mathcal{F}_{\mathrm{Sch}}\). They are part of a larger algebra of such
operators, labelled in general by multi-boundary conditions \(J\in\mathcal{J}\),
\begin{align}
  [a(J'), a(J)^\dagger]
  &= \braket{J'}{J}_{\mathrm{con}}
    \;,
  &
    a(J)\ket{\varnothing}_{\mathrm{BU}}
  &= 0
    \;.
    \label{eq:schaadag}
\end{align}
Finally, the fourth term is a term preserving the number of spacetime
components, satisfying
\begin{align}
  [\Lambda(j), a(J)^\dagger]
  &= a(j\sqcup J)^\dagger
    \;,
  &
    \Lambda(j)\ket{\varnothing}_{\mathrm{BU}}
    &= 0
    \;.
    \label{eq:schpres}
\end{align}
Acting on the Fock space vacuum \(\ket{\varnothing}_{\mathrm{BU}}\) with creation operators,
\begin{align}
  a(J_1)^\dagger \cdots a(J_m)^\dagger \ket{\varnothing}_{\mathrm{BU}}
  &\equiv \sqrt{\frac{\braket{\varnothing}{\varnothing}_{\mathrm{BU}}}{m!}}\sum_{\sigma\in S_m}
    \ket{J_{\sigma(1)}}_{\mathrm{con}} \otimes \cdots \otimes \ket{J_{\sigma(m)}}_{\mathrm{con}}
    \\
    &\in \Sym^m(\mathcal{H}_{\mathrm{con}})
    \;,
\end{align}
generates a dense subspace of the Fock space \(\mathcal{F}_{\mathrm{Sch}}\).
\Cref{eq:schaadag,eq:schpres} also determine how \(a\) and \(\Lambda\) act on
this dense subspace. We will refrain from specifying the precise domains of
these operators.

The geometric interpretation of each term in \cref{eq:schurmannzj} is again that
illustrated in
\cref{fig:mmdiskterm,fig:mmcreationterm,fig:mmannihilationterm,fig:mmnumberterm}
respectively. The argument presented there for the special case
\labelcref{eq:mmphiexact} can be repeated for the general formula
\labelcref{eq:schurmannzj}, with only one noteworthy modification. Namely,
\(\mathcal{H}_{\mathrm{con}}\) need not be one-dimensional and the
one-spacetime-component states
\(\ket{J}_{\mathrm{con}}\in\mathcal{H}_{\mathrm{con}}\) can depend on the number
of boundaries as well as the data on those boundaries. Correspondingly, the
creation and annihilation operators are labelled by these multi-boundary
conditions \(J\) and the action of attaching a boundary component \(j\) to a
pre-existing spacetime component --- described by
\cref{fig:mmnumberterm,eq:schpres} --- appends \(j\) to this label. With this
generalization, it is again easy to see that acting on
\(\ket{\varnothing}_{\mathrm{BU}}\) recursively with \cref{eq:schurmannzj} will
generate all spacetimes needed to recover normalized correlation functions
\begin{align}
  \frac{
  \mel{\varnothing}{Z(j_1)\cdots Z(j_m)}{\varnothing}_{\mathrm{BU}}
  }{
  \braket{\varnothing}{\varnothing}
  }
  &= \frac{\zeta(j_1 \sqcup \cdots \sqcup j_m)}{\zeta(\varnothing)}
    \;,
\end{align}
determined by the gravitational path integral \(\zeta\). Again, we can choose
the normalization of \(\ket{\varnothing}_{\mathrm{BU}}\) to coincide with the
no-boundary path integral (provided it is positive),
\begin{align}
   \aleph
  &\equiv \braket{\varnothing}{\varnothing}
    \equiv \zeta(\varnothing)
    \;.
\end{align}
Thus, the baby universe Hilbert space \(\mathcal{H}_{\mathrm{BU}}\)
GNS-constructed from \(\zeta\) is embedded \labelcref{eq:schurmannembed} in the
Sch\"urmann Fock space \(\mathcal{F}_{\mathrm{Sch}}\), as claimed.

Finally, let us note that the above construction can also be phrased in terms of
the universe field \(\Phi(x)\), defined as the operator-valued distribution such
that \(Z(j)\) is given by \labelcref{eq:zsmearing},
\begin{align}
  Z(j)
  &= \int_{\mathscr{M}} \mu(x) j(x) \Phi(x^\star)
    \;.
\end{align}
For GABUFT, as explained in \cref{sec:generalprop}, the universe field admits a
mode expansion \labelcref{eq:generalphiexp} in terms of on-shell modes,
\begin{align}
  \Phi(x)
  &= \int_{\mathscr{K}} \kappa(k) \phi(k) u_k(x)
    \;.
    \label{eq:generalizedfockphiexp}
\end{align}
Each mode coefficient \(\phi(k)\) admits a Fock space representation,
\begin{align}
  \phi(k)
  &= \phi_\varnothing(k)
    + a_k + a_{k^\star}^\dagger
    + \lambda_k
    \;.
    \label{eq:generalizedfockphik}
\end{align}
Respective terms in \cref{eq:schurmannzj,eq:generalizedfockphik} are related by
\begin{align}
  Z_\varnothing(j)
  &= \int_{\mathscr{K}} \kappa(k)
    \phi_\varnothing(k)
    \braket{j^\star}{k}_{\mathrm{wl}}
    \;,
  \\
  a(j)
  &= \int_{\mathscr{K}} \kappa(k)
    a_k
    \braket{j}{k}_{\mathrm{wl}}
    \;,
  \\
  \Lambda(j)
  &= \int_{\mathscr{K}} \kappa(k)
    \lambda_k
    \braket{j^\star}{k}_{\mathrm{wl}}
    \;.
\end{align}
The algebraic relations \labelcref{eq:schaadag,eq:schpres} become
\begin{align}
  [a_{k_1^\star\sqcup \cdots \sqcup k_i^\star},a_{k_{i+1}\sqcup \cdots \sqcup k_m}^\dagger]
  &= V(k_1 \sqcup \cdots \sqcup k_m)
    \;,
  &
    a_{k_1 \sqcup \cdots \sqcup k_m}\ket{\varnothing}_{\mathrm{BU}}
  &= 0
    \;,
  \\
  [\lambda_k, a_{k_1 \sqcup \cdots \sqcup k_m}^\dagger]
  &= a_{k \sqcup k_1 \sqcup \cdots \sqcup k_m}^\dagger
    \;,
  &
    \lambda_k \ket{\varnothing}_{\mathrm{BU}}
    &= 0
    \;,
\end{align}
where \(V\) is an equivalent formulation of the connected gravitational path
integral \(\zeta_{\mathrm{con}}\) in terms of \emph{invariant} states associated
to boundary components,
\begin{align}
  \zeta_{\mathrm{con}}(j_1\sqcup \cdots \sqcup j_m)
  &= \int_{\mathscr{K}^m}
    \left(
    \prod_{i=1}^m \kappa(k_i)
    \braket{j_i^\star}{k_i}_{\mathrm{wl}}
    \right)
    V(k_1 \sqcup \cdots \sqcup k_m)
    \;.
\end{align}

In the language commonly used in models like JT gravity,
\(V(k_1 \sqcup \cdots \sqcup k_m)\) might be called a nonperturbatively
genus-resummed ``volume''. It is a function of invariant states, which might be
labelled, for example, by parameters \(k_i=b_i\) that have the interpretation of
geodesic lengths upon gauge-fixing \cite{Held:2024rmg}. It is important to
remember, however, that the Sch\"urmann Fock space representation is only valid
if \(V\) or, equivalently, \(\zeta_{\mathrm{con}}\) is conditionally positive.
In JT gravity, the genus-zero four-boundary Weil--Petersson volume
\cite{Mirzakhani:2006eta},
\begin{align}
  V_{g=0}(b_1\sqcup\cdots\sqcup b_4)
  &= \frac{1}{2}
    \left(
      4\pi^2 + \sum_{i=1}^4 b_i^2
    \right)
    \;,
\end{align}
fails to give a positive-semidefinite Gram matrix
\(V_{g=0}((b_1'\sqcup b_2')\sqcup (b_1\sqcup b_2))\) on two-boundary
states, so \(V\) fails to be conditionally positive even at this leading order
in its genus expansion. (But, let us emphasize again that a failure of
conditional positivity of \(\zeta_{\mathrm{con}}\) does not imply a failure of
the strictly weaker positivity of \(\zeta\) needed for the GNS construction of a
baby universe Hilbert space.)

%% file: sections/appendix06_diskCylinder.tex
\section{The disk-and-cylinder Marolf--Maxfield model, from a
  kinematic Hilbert space with no topology change}
\label{app:dckinematics}

In \cref{sec:topologyperturbation}, we highlighted the relation \cref{eq:pantsip} between
the inner products of the full Marolf--Maxfield model and its disk-and-cylinder
truncation. Here, we explain how the disk-and-cylinder inner product can itself
be obtained from a more basic kinematic inner product in which even the
relatively trivial topology-changing processes in the disk-and-cylinder theory
--- the swapping, capping off/nucleation, and pair-creation/-annihilation of
universes --- are absent.

We begin with kinematic states
\begin{align}
  \ket{(S^1)^{\otimes m}}_0
  &= \ket{S^1}_{\mathrm{wl}}^{\otimes m}
    \;,
\end{align}
which live in the ordered tensor Hilbert space
\begin{align}
  \mathcal{H}_0
  &= \overline{
    \bigoplus_{m=0}^\infty
    \mathcal{H}_{\text{wl co/inv}}^{\otimes m}
    }
    \;.
\end{align}
Recall that
\(\mathcal{H}_{\text{wl co/inv}}=\mathcal{H}_{\text{wl}}
=\Span\{\ket{S^1}_{\mathrm{wl}}\}\) is the one-universe Hilbert space. The inner
product \(\braket{\bullet}{\bullet}_0\) in \(\mathcal{H}_0\) is given by
cylinders that connect ordered boundary components in ket-states to the
corresponding ordered boundary components in bra-states.

Allowing topologies that swap the universes in \(\ket{(S^1)^{\otimes m}}_0\)
before contracting with the bra-state introduces a sum \(\eta_{\mathrm{Sym}}\)
over permutations \(S_m\) within each tensor product
\(\mathcal{H}_{\text{wl co/inv}}^{\otimes m}\). The group-averaged inner product
\begin{align}
  \braket{(S^1)^{\sqcup m'}}{(S^1)^{\sqcup m}}_1
  &= \mel{(S^1)^{\otimes m'}}{\eta_{\mathrm{Sym}}}{(S^1)^{\otimes m}}_0
    \;,
    \label{eq:symip}
\end{align}
is the inner product on the symmetric Fock space
\begin{align}
  \mathcal{H}_1
  &= \overline{
    \bigoplus_{m=0}^\infty
    \Sym^m\mathcal{H}_{\text{wl co/inv}}
    }
    \;.
\end{align}
This can be identified with the same Hilbert space as that of the
disk-and-cylinder theory, \(\mathcal{H}_1\propto\mathcal{H}_{\mathrm{DC}}\),
where \(\ket{\varnothing}_1\propto\ket{\varnothing}_{\mathrm{DC}}\) is the Fock
space vacuum. (Above, we have implicitly adopted the normalization
\(\braket{\varnothing}{\varnothing}_1=\braket{\varnothing}{\varnothing}_0=1\),
but in the disk-and-cylinder theory, it is perhaps more natural to adopt another
normalization, \eg{} where
\(\log\braket{\varnothing}{\varnothing}_{\mathrm{DC}}=e^{2 S_0}+1\) is given by
the sum of the sphere and the torus.) The states appearing in \cref{eq:symip},
however
\begin{align}
  \ket{(S^1)^{\sqcup m}}_1
  &\propto (e^{S_0}a^\dagger)^m
    \ket{\varnothing}_{\mathrm{DC}}
    =\, :(Z_{\mathrm{DC}}(S^1)-e^{2S_0})^m:
    \ket{\varnothing}_{\mathrm{DC}}
    \label{eq:symstates}
\end{align}
are not the same as the disk-and-cylinder states \labelcref{eq:dcstates},
\begin{align}
  \ket{(S^1)^{\sqcup m}}_{\mathrm{DC}}
  &= Z_{\mathrm{DC}}(S^1)^m\ket{\varnothing}_{\mathrm{DC}}
    \;,
\end{align}
because universes in \(\ket{(S^1)^{\sqcup m}}_1\) are not allowed to cap off
into or nucleate from disks, nor are they allowed to be pair-created or
pair-annihilated.

Including these processes amounts, respectively, to shifting
\(Z_{\mathrm{DC}}(S^1)-e^{2S_0}\to Z_{\mathrm{DC}}(S^1)\) and removing the
normal ordering from \cref{eq:symstates}. These operations are implemented by
\begin{align}
  \ket{(S^1)^{\sqcup m}}_{\mathrm{DC}}
  &\propto V\ket{(S^1)^{\sqcup m}}_1
    \;,
  &
    V
  &= \exp\left(
    e^{S_0}a+\frac{1}{2}a^2
    \right)
    \;.
\end{align}
Hence,
\begin{align}
  \braket{(S^1)^{\sqcup m'}}{(S^1)^{\sqcup m}}_{\mathrm{DC}}
  &=
    \mel{(S^1)^{\sqcup m'}}{\eta_V}{(S^1)^{\sqcup m}}_1
    \;,
  &
    \eta_V
  &\equiv \braket{\varnothing}{\varnothing}_{\mathrm{DC}} V^\dagger V
    \;.
\end{align}
Because \(V\) is injective on the domain
\(\bigoplus_{m=0}^\infty \Sym^m\mathcal{H}_{\text{wl co/inv}}\), the insertion
\(\eta_V\) does not introduce any new null states.